\documentclass[iop,numberedappendix,twocolappendix]{emulateapj}
\usepackage{mathrsfs}

\begin{document}

\newcommand{\mirlum}{L_{\rm 8}}
\newcommand{\ebmvstars}{E(B-V)_{\rm stars}}
\newcommand{\ebmvgas}{E(B-V)_{\rm gas}}
\newcommand{\lha}{L(H\alpha)}
\newcommand{\lir}{L_{\rm IR}}
\newcommand{\lbol}{L_{\rm bol}}
\newcommand{\luv}{L_{\rm UV}}
\newcommand{\rs}{{\cal R}}
\newcommand{\ugr}{U_{\rm n}G\rs}
\newcommand{\ks}{K_{\rm s}}
\newcommand{\gmr}{G-\rs}
\newcommand{\hi}{\text{\ion{H}{1}}}
\newcommand{\nhi}{N(\text{\ion{H}{1}})}
\newcommand{\lognhi}{\log[\nhi/{\rm cm}^{-2}]}
\newcommand{\molh}{\text{H}_2}
\newcommand{\nmolh}{N(\molh)}
\newcommand{\lognmolh}{\log[\nmolh/{\rm cm}^{-2}]}
\newcommand{\oii}{\text{[\ion{O}{2}]}}
\newcommand{\neiii}{\text{[\ion{Ne}{3}]}}
\newcommand{\oiii}{\text{[\ion{O}{3}]}}
\newcommand{\nii}{\text{[\ion{N}{2}]}}
\newcommand{\ha}{\text{H$\alpha$}}
\newcommand{\hb}{\text{H$\beta$}}
\newcommand{\sii}{\text{[\ion{S}{2}]}}
\newcommand{\delw}{\Delta\log[W]}
\newcommand{\delsfr}{\Delta\log[{\rm SFR(\ha)}]}
\newcommand{\delz}{\Delta[12+\log({\rm O/H})]}
\newcommand{\delion}{\Delta\log({\rm O32})}
\newcommand{\sfrha}{\text{SFR[H$\alpha$]}}
\newcommand{\sfrsed}{\text{SFR[SED]}}
\newcommand{\ssfrha}{\text{sSFR[H$\alpha$]}}
\newcommand{\ssfrsed}{\text{sSFR[SED]}}

\title{The MOSDEF Survey: Significant Evolution in the Rest-Frame Optical Emission Line 
Equivalent Widths of Star-Forming Galaxies at $\lowercase{z}=1.4-3.8$}
\author{\sc Naveen A. Reddy\altaffilmark{1}, 
Alice E. Shapley\altaffilmark{2}, 
Ryan L. Sanders\altaffilmark{2}, 
Mariska Kriek\altaffilmark{3}, 
Alison L. Coil\altaffilmark{4},
Irene Shivaei\altaffilmark{5},
William R. Freeman\altaffilmark{1},
Bahram Mobasher\altaffilmark{1},
Brian Siana\altaffilmark{1},
Mojegan Azadi\altaffilmark{6},
Tara Fetherolf\altaffilmark{1},
Francesca M. Fornasini\altaffilmark{6},
Gene Leung\altaffilmark{4},
Sedona H. Price\altaffilmark{7},
Tom Zick\altaffilmark{3},
Guillermo Barro\altaffilmark{8}}

\altaffiltext{1}{Department of Physics and Astronomy, University of California, 
Riverside, 900 University Avenue, Riverside, CA 92521, USA; naveenr@ucr.edu}
\altaffiltext{2}{Department of Physics \& Astronomy, University of California,
Los Angeles, 430 Portola Plaza, Los Angeles, CA 90095, USA}
\altaffiltext{3}{Astronomy Department, University of California, Berkeley,
Berkeley, CA 94720, USA}
\altaffiltext{4}{Center for Astrophysics and Space Sciences, University of
California, San Diego, 9500 Gilman Drive, La Jolla, CA 92093-0424, USA}
\altaffiltext{5}{Steward Observatory, University of Arizona, 933 North 
Cherry Avenue, Tucson, AZ 85721, USA}
\altaffiltext{6}{Harvard-Smithsonian Center for Astrophysics, 60 Garden Street, Cambridge, MA 02138, USA}
\altaffiltext{7}{Max-Planck-Institut f\"{u}r Extraterrestrische Physik, Postfach 1312, Garching, D-85741, Germany}
\altaffiltext{8}{University of the Pacific, Stockton Campus, 3601 Pacific Avenue, Stockton, CA 95211, USA}

\slugcomment{submitted to ApJ on 2018 August 23, accepted 2018 October 29}

\begin{abstract}

We use extensive spectroscopy from the MOSFIRE Deep Evolution Field
(MOSDEF) survey to investigate the relationships between rest-frame
optical emission line equivalent widths ($W$) and a number of galaxy
and ISM characteristics for a sample of $1134$ star-forming galaxies
at redshifts $1.4\la z\la 3.8$.  We examine how the equivalent widths
of $\oii\lambda\lambda 3727, 3730$, $\hb$, $\oiii\lambda\lambda 4960,
5008$, $\oiii+\hb$, $\ha$, and $\ha+\nii\lambda\lambda 6550, 6585$,
depend on stellar mass, UV slope, age, star-formation rate (SFR) and
specific SFR (sSFR), ionization parameter and excitation conditions
(O32 and $\oiii/\hb$), gas-phase metallicity, and ionizing photon
production efficiency ($\xi_{\rm ion}$).  The trend of increasing $W$
with decreasing stellar mass is strongest for \oiii\, (and \oiii+\hb).
More generally, the equivalent widths of all the lines increase with
redshift at a fixed stellar mass or fixed gas-phase metallicity,
suggesting that high equivalent width galaxies are common at high
redshift.  This redshift evolution in equivalent widths can be
explained by the increase in SFR and decrease in metallicity with
redshift at a fixed stellar mass.  Consequently, the dependence of $W$
on sSFR is largely invariant with redshift, particularly when examined
for galaxies of a given metallicity.  Our results show that high
equivalent width galaxies, specifically those with high $W(\oiii)$,
have low stellar masses, blue UV slopes, young ages,
high sSFRs, ISM line ratios indicative of high ionization
parameters, high $\xi_{\rm ion}$, and low metallicities.  As these
characteristics are often attributed to galaxies with high ionizing
escape fractions, galaxies with high $W$ are likely candidates for the
population that dominates cosmic reionization.

\end{abstract}


\section{INTRODUCTION}
\label{sec:intro}

The rest-frame optical spectra ($\lambda_{\rm rest}\simeq
3700-6750$\,\AA) of star-forming galaxies contain a number of
important nebular emission and stellar photospheric absorption lines
that serve as diagnostics of the star-formation rate (SFR), dust
reddening, electron density, ionization parameter ($\mathscr{U}$),
gas-phase metallicity, and stellar population.  These basic properties
have been well-characterized for large samples of local galaxies by
surveys like the Sloan Digital Sky Survey (SDSS; \citealt{york00}).
Narrowband surveys have provided a view of subsets of the important
diagnostic emission lines for galaxies at $z\ga 1$ (e.g.,
\citealt{thompson96, malkan96, teplitz98, mannucci98, teplitz00,
  geach08, ly12a, sobral12, sobral13, ly15, suzuki15, ly16,
  khostovan16, matthee17}), while medium/broadband optical/near-IR and
        {\em Spitzer}/IRAC data have been used to estimate nebular
        line (e.g., \ha+\nii(+\sii), \oiii+\hb) equivalent widths and
        fluxes at $z\ga 2$ (e.g., \citealt{kriek11, shim11, vanderwel11,
          labbe13, stark13, smit14, shivaei15a, smit16, faisst16,
          rasappu16, forrest17, malkan17, caputi17, castellano17}).

Concurrent with these largely photometrically based efforts at high
redshift, advances in detector technology incorporated in the current
generation of moderate resolution multi-object near-IR spectrographs
have enabled direct and efficient spectroscopic measurements of
rest-frame optical nebular emission lines for large samples of
galaxies up to $z\sim 4$ (e.g., \citealt{forster09, kashino13, schenker13,
  steidel14, kriek15, wisnioski15, tran15, holden16, nanayakkara17}).
The {\em Hubble Space Telescope}'s grisms have provided an additional
low-resolution ($R\la 200$) probe of these features at $z\ga 0.7$
(e.g., \citealt{mccarthy99, atek10, atek11, trump11, straughn11, brammer12,
  fumagalli12, atek14a, maseda14}).  Taken together, these studies
point to an increase in the mean equivalent widths of galaxies at a
fixed stellar mass with redshift, reflecting an increase in the rate
at which galaxies are building up their stellar mass from $z\sim 2$ to
$z\sim 6$ (e.g., \citealt{reddy12a, stark13, gonzalez14, smit14,
  smit16, faisst16}).  The elevated mean equivalent widths of
rest-frame optical emission lines expected for high-redshift galaxies
has also prompted the community to re-evaluate the contribution of
such line emission to broadband photometry when modeling such
photometry to discern the stellar populations in high-redshift
galaxies, particularly those which may be faint and have low stellar
masses (e.g., \citealt{anders03, zackrisson08, schaerer09, trump11,
  stark13, debarros14, gonzalez14, smit14}).

Many of the advances in our understanding of the ISM conditions in
high-redshift galaxies have been based on data taken with the MOSFIRE
spectrograph \citep{mclean12} on the Keck~I telescope.  This
instrument operates from $0.97$ to $2.45$\,$\mu$m, and enables the
simultaneous spectroscopic observation of $\approx 30$ individual
galaxies while affording an increase in sensitivity of at least a
factor of five relative to the previous generation of near-IR
long-slit spectrographs on 8-10\,m class telescopes.  The resulting
combination of multiplexing and sensitivity results in an increase in
surveying efficiency of more than two orders of magnitude, a feature
that has been exploited by surveys including the Keck Baryonic
Structure Survey (KBSS; \citealt{steidel14}) and the MOSFIRE Deep
Evolution Field (MOSDEF) survey \citep{kriek15}.  Galaxies with
moderate resolution rest-frame optical emission line measurements at
$z\sim 1.4-3.8$ now number in the thousands.

Combining low-redshift galaxy survey data with near-IR spectroscopic
and narrowband imaging campaigns of galaxies at $z\simeq 1.5-3.8$ has
allowed the community to address how the physical conditions in
star-forming galaxies have evolved over the last $\simeq 12$\,Gyr of cosmic
time.  For example, several studies have suggested an increase in the
ionization parameter with redshift at a fixed stellar mass
\citep{hainline09, hayashi15, kewley15, shapley15, holden16,
  khostovan16, sanders16a, strom17}, an evolution that can be largely
explained by the decrease in metallicity with redshift at a fixed
stellar mass (i.e., the redshift evolution in the mass-metallicity
relationship; e.g., \citealt{sanders16a}).  Additionally, the mean
electron density shifts toward higher values with increasing
redshift, being an order of magnitude larger for typical star-forming
galaxies at $z\sim 2$ relative to local galaxies of similar stellar
masses \citep{sanders16a, strom17}.  The conclusions regarding the
high-excitation nature of high-redshift galaxies appear to be
supported by inferences of rest-frame optical emission line equivalent
widths in even higher redshift (e.g., $z\ga 4$) galaxies (e.g.,
\citealt{shim11, gonzalez12, stark13, labbe13, debarros14, smit14,
  smit16}), as well as by direct UV/optical spectroscopy of $z\ga 4$
galaxies that indicate on average a harder ionizing radiation field
relative to typical star-forming galaxies at $z\sim 2-3$ at a fixed
stellar mass \citep{stark15, nakajima16, mainali17,schmidt17, smit17,
  shapley17}.

The relative ease of detecting high equivalent width emission lines in
high-redshift galaxies leads us naturally to ask the question of how
such strong emission is connected to other characteristics of galaxies
and, from the standpoint of demographics, how the properties of those
strong line emitters compare with those of the general galaxy
population at similar redshifts.  In this vein, many of the
aforementioned investigations have emphasized the utility of
calibrations between rest-frame optical emission line equivalent
widths and other observable and derived quantities that characterize
galaxies, including their stellar population and ISM properties such
as stellar mass, age, reddening, gas-phase metallicity, and ionization
parameter.  These calibrations and their redshift evolution may be
used to estimate the physical conditions in galaxies from rather
straightforward measurements of the equivalent widths of rest-frame
optical emission lines.  Furthermore, these calibrations serve as a
reference against which rest-frame optical line measurements with {\em
  JWST} and future dark energy missions (e.g., {\em EUCLID}, {\em
  WFIRST}) may be interpreted.  Finally, such relationships may be
utilized to evaluate the impact of emission lines on broadband
photometry which, when uncorrected, may bias the characterization of
the stellar populations of high-redshift galaxies.

With these aims in mind, we use the extensive spectroscopic dataset of
the MOSDEF survey to characterize the rest-frame optical emission line
equivalent width distributions of star-forming galaxies as a function
of stellar population and ISM characteristics.  The sample analyzed in
this paper consists of $>1,000$ galaxies at redshifts $1.00\le z\le
3.80$, most with multiple emission line measurements, making it one of
the largest samples to date for which we can characterize the emission
line equivalent width distributions at these redshifts.  In this
paper, we consider equivalent width measurements of \oii, \hb, \oiii,
\oiii+\hb, \ha, and \ha+\nii, and how these equivalent widths vary
with certain galaxy and ISM properties.  These lines and their
rest-frame wavelengths are listed in Table~\ref{tab:lines}.

\begin{deluxetable}{lc}
\tabletypesize{\footnotesize}
\tablewidth{0pc}
\tablecaption{Lines Used in This Study}
\tablehead{
\colhead{Line} &
\colhead{Rest-frame Vacuum Wavelength (\AA)}}
\startdata
\oii & $3727,3730$ \\
\hb & $4863$ \\
\oiii & $4960, 5008$ \\
\ha & $6565$ \\
\nii & $6550, 6585$ 
\enddata
\label{tab:lines}
\end{deluxetable}

The outline of this paper is as follows.  In Section~\ref{sec:sample},
we discuss the criteria for selecting galaxies in our sample; the
methodology of fitting stellar population models to observed
photometry; and the calculation of H$\alpha$-based SFRs, metallicity-
and excitation-sensitive line ratios, and line equivalent widths.
Section~\ref{sec:calibrations} presents our main results on the
correlations between emission line equivalent widths and various
stellar population and ISM properties, both for the full sample and as
a function of redshift.  The physical context of the correlations
between equivalent widths and galaxy and ISM properties is discussed
in Section~\ref{sec:discussion}.  Here, we investigate the physical
causes of the trends we see and also discuss the redshift evolution of
the equivalent widths.  We also present a strategy for identifying
high-excitation and high-ionization galaxies at high redshift.  The
appendices present (a) a discussion of the line luminosity
completeness of the MOSDEF survey over the range of stellar masses
probed in this study; (b) the methodology used to calculate composite
spectra; (c) a comparison between the measured $\ha$ equivalent widths
and those predicted by models that assume different stellar
metallicities and dust attenuation curves; and (d) corrections that
can be used to estimate equivalent widths of single ionic species from
low-resolution spectral or narrowband data where emission line
features may be blended and/or confused.  AB magnitudes are assumed
throughout \citep{oke83}, and we adopt a \citet{chabrier03} initial
mass function (IMF) unless stated otherwise.  Wavelengths are
presented in the vacuum frame.  We adopt a cosmology with
$H_{0}=70$\,km\,s$^{-1}$\,Mpc$^{-1}$, $\Omega_{\Lambda}=0.7$, and
$\Omega_{\rm m}=0.3$.

\section{DATA, SAMPLE, AND MEASUREMENTS}
\label{sec:sample}

\subsection{MOSDEF Survey}

The MOSDEF survey \citep{kriek15} used the MOSFIRE instrument on the
Keck~I telescope to obtain {\em YJHK} spectroscopy of a large sample
of rest-frame optical selected galaxies in three redshift
windows---$z\simeq 1.37-1.70$, $2.09-2.70$, and $2.95-3.80$---in the CANDELS
fields (AEGIS, COSMOS, GOODS-N, GOODS-S, UDS; \citealt{grogin11,
  koekemoer11}).  These redshift windows ensure the coverage of as
many of the strong rest-frame optical emission lines (e.g., \oii,
H$\beta$, \oiii, H$\alpha$) that are accessible within the near-IR
atmospheric transmission windows.  The survey was conducted over 48.5
nights during the 2012B-2016A observing semesters.  

Objects were selected from the 3D-HST photometric catalogs
\citep{skelton14} to limiting rest-frame optical magnitudes of
$H=24.0$, $24.5$, and $25.0$\,mag, respectively, for the three
aforementioned redshift ranges, corresponding roughly to a limit in
stellar mass of $\sim 10^9$\,$M_\odot$.  For comparison, the
characteristic stellar mass derived from stellar mass functions at
redshifts $1.4\la z\la 3.8$ is $M^{\ast} \simeq (2.5-6.8)\times
10^{10}$\,$M_\odot$ (e.g., \citealt{davidzon17}).  Thus, the MOSDEF
sample probes galaxies that are anywhere from more than an order of
magnitude less massive, to a factor of $\approx 3\times$ more massive,
than typical galaxies at these redshifts.  For brevity, we refer to
the subsamples in the $z=1.0-1.8$, $z=1.8-2.7$, and $z=2.7-4.0$
redshift ranges as the $z\sim 1.5$ (low-redshift), $z\sim 2.3$
(middle-redshift), and $z\sim 3.4$ (high-redshift) samples,
respectively.  Galaxies in the two lower redshift ranges (e.g., where
some line ratios, such as $\nii/\ha$, are accessible) are hereafter
referred to as lying in the $z\sim 2$ sample.

Objects were targeted for MOSFIRE spectroscopy if they had external
spectroscopic redshifts, 3D-HST grism redshifts, and/or photometric
redshifts that placed them in the redshift ranges specified above.
Multi-object slit masks that primarily targeted objects in the two
higher redshift intervals had integration times of $\sim 2$\,hr in
each band ({\em JHK} for the $z\sim 2.3$ sample, and {\em HK} for the
$z\sim 3.4$ sample), while those targeting objects in the lowest
redshift range had integration times of $\sim 1$\,hr in each of the
{\em YJH} bands.  The adopted slit widths of $0\farcs7$ resulted in a
spectral resolutions of $R\approx 3000-3650$.  The very high
spectroscopic success rate ($\sim 80\%$) of our survey resulted in
redshifts for $\approx 1500$ objects.  Further details on the target
selection, observations, and spectroscopic data reduction are provided
in \citet{kriek15}.

\subsection{Slit Loss Corrections and Line Flux Measurements}

The continuum emission of the galaxies in the MOSDEF sample is
generally not detected in individual spectra, with the exception of
those objects with the brightest {\em H}-band magnitudes.  As such,
our method of computing the equivalent widths of lines relies on
combining spectroscopic line flux measurements with continuum flux
densities obtained from stellar population model fits to the
photometry of galaxies.  When computed in this way, the equivalent
widths rely on accurate slit loss corrections to the line fluxes, as
well as robust absolute flux calibrations based on slit stars observed
simultaneously with the primary targets.  The corrections for slit
loss are based on modeling the two-dimensional {\em H}-band light
distribution of galaxies and computing the amount of light passing
through the slit while taking into account the seeing at the time the
observations were obtained.  Details on the slit loss calculations are
provided in \citet{kriek15} and \citet{reddy15}.

Line fluxes from the MOSFIRE spectra were measured by fitting Gaussian
functions on top of a linear continuum \citep{kriek15}.  The $\ha$ and
$\hb$ emission line fluxes were corrected for underlying Balmer
absorption that arises from the stellar atmospheres of primarily A
stars (with weaker contributions from B and F stars).  These
absorption lines are typically wider than the emission lines due to
pressure broadening in stellar atmospheres.  As such, the degree to
which an emission line is corrected for Balmer absorption will depend
on the amount by which the emission line fills the absorption line, or
the ``emission filling fraction'' ($f_{\rm fill}$).  As the continuum,
and hence the Balmer absorption, is not detected in the individual
MOSDEF spectra, we computed an average filling fraction by
constructing composite spectra (see Appendix~\ref{sec:composite}) of
all galaxies where $\ha$ or $\hb$ were covered in the individual
spectra, regardless of whether the $\ha$ or $\hb$ emission lines were
detected.  The average continuum, and average Balmer absorption, are
significantly detected in the composite spectra.  We simultaneously
fit a Gaussian function with a negative amplitude to the $\hb$
absorption and a Gaussian function with a positive amplitude to the
$\hb$ emission line in the composite spectrum.  Due to its proximity
to the $\nii$ doublet, the $\ha$ emission line was fit simultaneously
with $\nii$, while the width of the underlying $\ha$ absorption line
was fixed to that measured for the $\hb$ absorption line.  The
two-component fit for $\hb$ and multi-component fit for $\ha$ yielded
a measure of the ``true'' average emission line flux, $f_{\rm em}^{\rm
  two}$, and the total absorbed flux, $f_{\rm abs}^{\rm two}$.

We then derived a second estimate of the emission line flux by
assuming no absorption ($f_{\rm em}^{\rm one}$).  To do this, we
fit a single Gaussian function to the $\hb$ emission line and a triple
Gaussian function to the $\ha+\nii$ complex in the composite spectrum,
and assumed a flat continuum with no absorption, effectively mimicking
the way in which line fluxes are measured in individual spectra.  In
order to compute $f_{\rm em}^{\rm one}$, we constrained the continuum
level using a linear fit to the flux density points on either side of
the Balmer absorption.  The emission filling fraction is then given by
\begin{eqnarray}
f_{\rm fill}=\frac{f_{\rm em}^{\rm two} - f_{\rm em}^{\rm one}}{f_{\rm abs}^{\rm two}}.
\end{eqnarray}
Performing these measurements on the composite spectra yielded values
of $f_{\rm fill}\simeq 0.36$ and $0.23$ for $\ha$ and $\hb$,
respectively.  For every galaxy, we computed the amount of absorbed
flux from the best-fit stellar population model, multiplied this flux
by $f_{\rm fill}$, and added the resulting number to the measured
$\ha$ and $\hb$ emission line fluxes.  The Balmer absorption
corrections serve to increase the \ha\, and \hb\, line fluxes by $\la
1\%$ and $\la 3\%$, respectively.

\subsection{Sample Selection}
\label{sec:sampleselection}

\begin{figure}
\epsscale{1.15}
\plotone{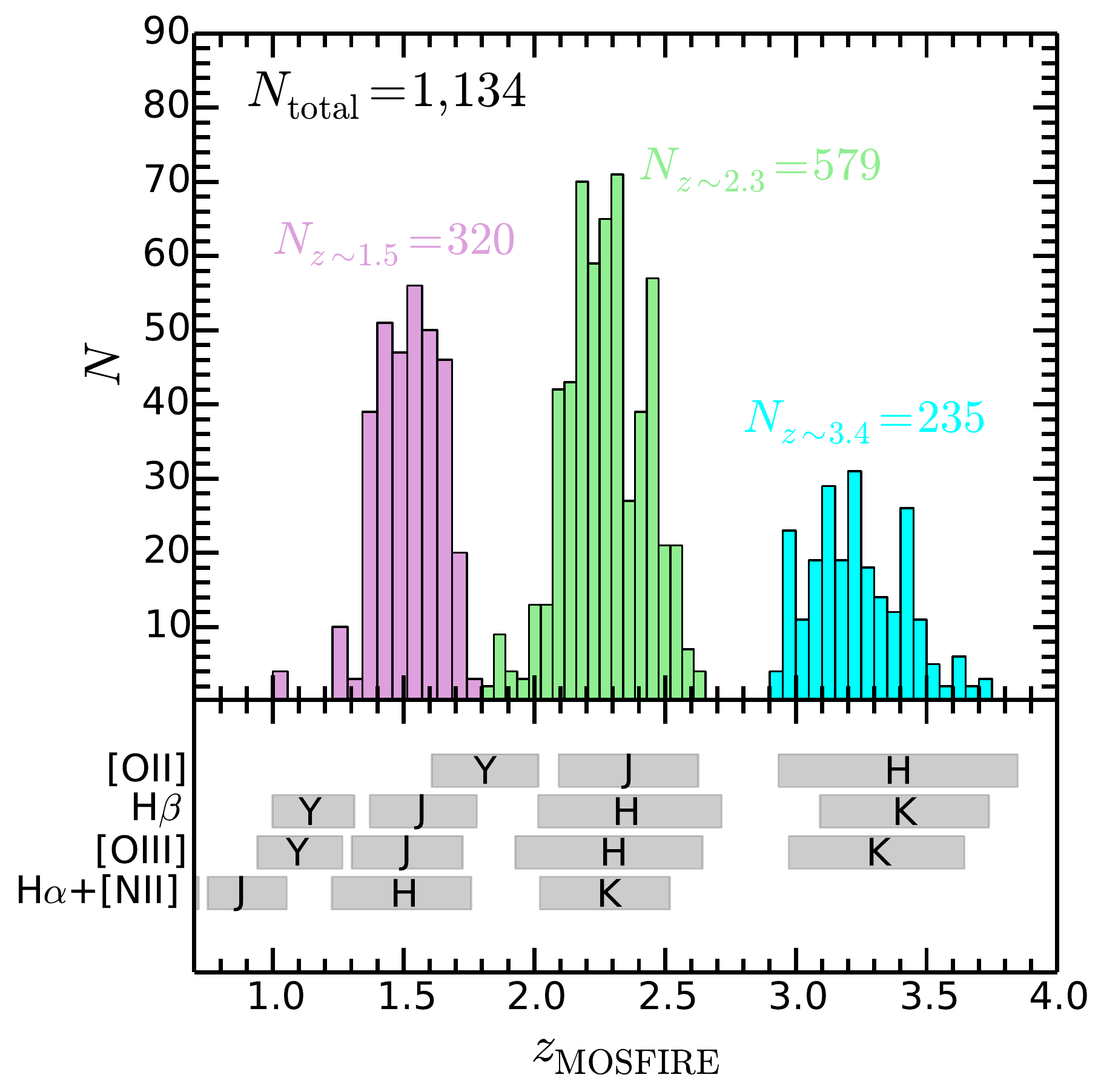}
\caption{{\em Top:} Redshift distribution of the $N=1,134$ galaxies in
  our sample, color-coded by the $z\sim 1.5$, $z\sim 2.3$, and $z\sim
  3.4$ subsamples to which the galaxies belong.  {\em Bottom:}
  Redshift ranges over which the strong rest-frame optical emission
  lines (\oii, \hb, \oiii, \ha, and \nii) are shifted into the MOSFIRE
  {\em YJHK} filters, as determined by the half-power points of the
  transmission profiles of these bands.}
\label{fig:zhist}
\end{figure}

To establish the sample analyzed in this paper, we considered only
those objects that had secure spectroscopic redshifts measured from
the MOSFIRE spectra.  Specifically, to be included in our sample, the
object must have either more than one emission line with an integrated
line flux with $S/N\ge 3$, or a single emission line with an
integrated line flux with $S/N\ge 3$ and a redshift from that single
line that agrees with a previously published spectroscopic redshift.
We excluded active galactic nuclei (AGNs) that were detected in the
X-ray, and/or detected in all four of the {\em Spitzer}/IRAC channels
and satisfied the \citet{donley12} criteria, and/or had an optical
emission line ratio $\log[\nii/\ha]>-0.4$ \citep{coil15, azadi17}.
Finally, objects lying outside of the redshift range $1.0\le z_{\rm
  MOSFIRE}\le 4.0$ were excluded.  These selection criteria resulted
in a final sample of 1,134 unique objects with the spectroscopic
redshift distribution shown in Figure~\ref{fig:zhist}.  There are 21,
20, and 12 objects in the $z\sim 1.5$, $z\sim 2.3$, and $z\sim 3.4$
subsamples, respectively, that are not in our parent sample and are
fainter than the {\em H} limits specified above, but for which we were
able to derive redshifts as they happened to fall on spectroscopic
slits assigned to other primary targets.

\begin{figure*}
\epsscale{1.15}
\plotone{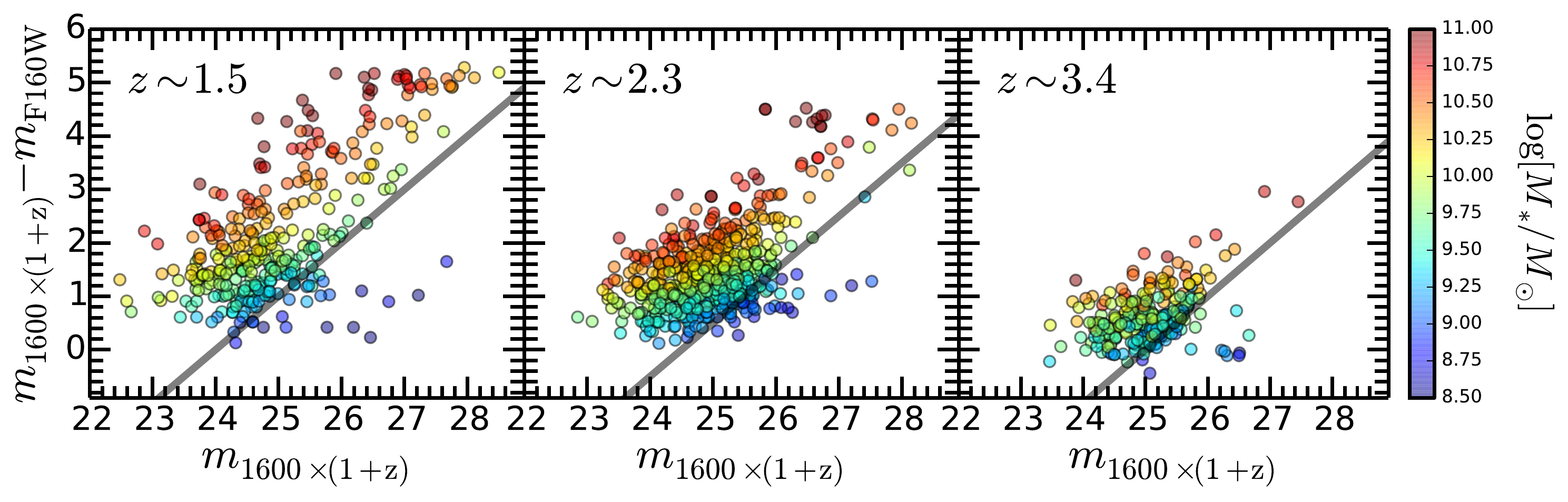}
\caption{Distribution of UV-optical colors as a function of UV
  magnitude for galaxies in three redshift subsamples.  The UV
  magnitude is taken to be the apparent magnitude at rest-frame
  1600\,\AA, while the optical magnitude is taken to be the {\em HST}
  F160W (i.e., {\em H}) magnitude.  The {\em H} limits of 24.0, 24.5,
  and 25.0 for the low, middle, and high redshift subsamples,
  respectively, are indicated by the solid lines in each panel.  The few
objects lying below these limits fell serendipitously on spectroscopic
slits assigned to other (primary) targets, and were ones where we were
able to derive redshifts that placed them in the redshift ranges of
interest.  All points are color-coded according to $\log[M_\ast/M_\odot]$.}
\label{fig:colors}
\end{figure*}

As discussed elsewhere (e.g., \citealt{shivaei15b}), the sample
includes galaxies that are largely representative of ``typical''
star-forming galaxies at redshifts $1.0\le z\le 4.0$, as judged by
their distributions of SFRs and stellar masses.  However, because the
sample consists of galaxies selected by their rest-frame optical
continuum (i.e., to the {\em H} limits specified above), it is highly
incomplete for UV-faint blue star-forming galaxies, i.e., those that
dominate the faint-end of the UV luminosity function.  This
incompleteness is illustrated in Figure~\ref{fig:colors}, where the
difference in apparent magnitude at rest-frame $1600$\,\AA\, and the
apparent magnitude at $1.6$\,$\mu$m (i.e., the UV-optical color) is
shown against the apparent magnitude at rest-frame $1600$\,\AA.  The
{\em H} limits adopted for the MOSDEF sample selection result in a
distribution of colors that favors very red galaxies ($m_{\rm
  1600\times (1+z)}-m_{\rm F160W} > 3$) at faint UV luminosities
($m_{\rm 1600\times(1+z)}>27$)---or at masses lower than $\simeq
10^{9}$\,$M_\odot$---over essentially the entire redshift range of
interest.  As such, in the following discussion we purposely avoid
presenting relationships between equivalent widths and UV magnitude or
UV luminosity, as these relationships will not be representative of
the general galaxy population.

As we are primarily interested in how line equivalent widths vary with
other galaxy properties, it is instructive to determine the degree to
which our sample is complete in line luminosity and equivalent width
for galaxies of different masses.  In Appendix~\ref{sec:completeness},
we present evidence that the MOSDEF sample is complete in line
luminosity over the range of stellar masses relevant for the current
analysis (i.e., $M_\ast \ga 10^9$\,$M_\odot$).

The local reference sample used to evaluate the redshift evolution in
line equivalent widths was drawn from Data Release 7 (DR7) of the SDSS
\citep{abazajian09}.  We combined the cataloged DR7 line fluxes and
continuum flux densities to compute equivalent widths.  SFRs and
stellar masses for the local sample were also included in our
analysis, where these values were computed using the techniques
discussed in \citet{brinchmann04} and \citet{salim07}.  The DR7 SDSS
sample was restricted to include only those objects with redshifts
$0.04\le z\le 0.1$ to minimize aperture effects and integrated line
fluxes in \oii, \hb, \oiii, \ha, and \nii\, with $S/N\ge 3$.  AGNs were
excluded based on the \citet{kauffmann03} criteria.  The resulting
local comparison sample consists of 49,865 galaxies.

\subsection{Stellar Population Modeling}
\label{sec:sedmodeling}

Stellar masses, ages, color excesses, and star-formation rates (SFRs)
were derived by modeling the photometry of galaxies in our sample---as
assembled in the 3D-HST photometric catalogs \citep{skelton14}.  The
number of photometric bands that were modeled varies from 18 bands in
UDS to 44 bands in COSMOS.  The fiducial modeling assumes the
\citet{bruzual03} (BC03) $Z=0.004$ model, corresponding to
$Z=0.28Z_\odot$ on the current abundance scale \citep{asplund09}, for
a constant SFR and reddened by an SMC extinction curve.  These choices
are based on previous studies that have suggested that typical
($L^{\ast}$) galaxies at $z\ga 1.5$ can be approximated on average
with a constant or rising star-formation history \citep{reddy12a}, and
that steeper (e.g., SMC-like) attenuation curves are required to
reproduce the dust obscurations (as constrained from far-infrared
data) of sub-solar metallicity galaxies at high redshift
\citep{reddy18}.  In Appendix~\ref{sec:sedpop}, we provide further
evidence that supports our default choice of the stellar metallicity
and attenuation curve.  The SMC curve assumed in our study is from
\citet{gordon03}, with a revised determination of the shape of the
curve at $\lambda \simeq 950 - 1250$\,\AA\, from \citet{reddy16a}.  If
applicable, the photometry was first corrected for the contribution
from the strongest emission lines in the MOSFIRE spectra, including
\oii, \hb, \oiii, and \ha.  The age was allowed to vary from 50\,Myr
(approximating the dynamical timescale; e.g., \citealt{reddy12b}) to
the age of the universe at the redshift of each galaxy.  We considered
a range of reddening from $0.0\le \ebmvstars\le 0.6$.  The best-fit
stellar mass, reddening, age, and SED-inferred SFR ($\sfrsed$) were
taken to be those from the model that gave the minimum value of
$\chi^2$ relative to the photometry.

As the MOSDEF sample includes galaxies over a range of stellar masses,
and hence stellar metallicities, we explore further in
Appendix~\ref{sec:sedpop} how adopting a different metallicity
stellar population model and attenuation curve for subsets of galaxies
in our sample affects the calibrations between line equivalent widths
and stellar population parameters.

\subsection{H$\alpha$ SFRs}
\label{sec:calcsfrha}

In addition to SED-derived SFRs, we also included H$\alpha$ SFRs
($\sfrha$, corrected for dust using the Balmer decrement) in our
analysis, where the latter are computed using the methodology
presented in \citet{reddy15} and \citet{shivaei15b}.  To maintain
consistency with the assumptions adopted for the SED fitting, the
\ha\, luminosities were converted to SFRs based on the assumed stellar
population models.  Specifically, the factor required to convert
$L(\ha)$ in units of erg\,s$^{-1}$ to $\sfrha$ in units of
$M_\odot$\,yr$^{-1}$ is $3.236\times 10^{-42}$ for our fiducial model
with a stellar metallicity of $0.28Z_\odot$.  The same factor for the
$1.4Z_\odot$ BC03 constant star formation models is $4.457\times
10^{-42}$.  For comparison, the \citet{kennicutt98} conversion factor
is $4.165\times 10^{-42}$ assuming a \citet{chabrier03} IMF, while the
\citet{hao11} conversion factor is $4.634\times 10^{-42}$ for the same
IMF.

\subsection{Excitation-sensitive Line Ratios, Gas-phase Metallicities, and $\xi_{\rm ion}$}
\label{sec:calcexcitation}

We present relationships between equivalent widths and several
excitation-sensitive line ratios.  These ratios are summarized in
Table~\ref{tab:diagnostics}, and include O32, which is sensitive to
the ionization parameter.  As the lines of this ratio are
well-separated in wavelength, the \oiii$\lambda 5008$ and \oii\, lines
were corrected for dust attenuation assuming the reddening derived
from the Balmer decrement (\ha/\hb)---i.e., $\ebmvgas$---and the
\citet{cardelli89} extinction curve (e.g., see \citealt{sanders16a}).
We also consider O3.  This ratio was also corrected for dust in a
manner similar to that for O32, though the dust corrections are small
relative to the flux measurement errors given the proximity of the
\oiii\, and \hb\, lines in wavelength.  The two strong-line
metallicity indicators examined here include the N2 and O3N2 indices.
These indices were converted to gas-phase oxygen abundances based on
the calibrations of \citet{pettini04}.

\begin{deluxetable*}{lcc}
\tabletypesize{\footnotesize}
\renewcommand{\arraystretch}{1.5}
\tablewidth{0pc}
\tablecaption{Diagnostics Used in this Study}
\tablehead{
\colhead{Diagnostic} &
\colhead{Definition} &
\colhead{Inferred Property}}
\startdata
$\beta$ (UV Slope) & $f_\lambda \propto \lambda^\beta$, $\lambda = 1268-2580$\,\AA & Reddening \\
\\
$\xi_{\rm ion}$ & $\frac{Q({\rm H})}{L_{\rm UV}}$ (Hz\,erg$^{-1}$)\tablenotemark{a} & Ionizing photon production efficiency \\
\\
O32 & $\frac{\oiii\lambda 5008}{\oii\lambda 3727, 3730}$ & Ionization parameter ($U$) \\
\\
O3 & $\frac{\oiii\lambda 5008}{\hb}$ & Excitation \\
\\
N2 & $\frac{\nii}{\ha}$ & Gas-phase Metallicity \\
\\
O3N2 & $\frac{{\rm O3}}{{\rm N2}}$ & Gas-phase Metallicity \\
\\
R23 & $\frac{\oii + \oiii}{\hb}$ & Gas-phase Metallicity 
\enddata
\tablenotetext{a}{$Q({\rm H})$ is the number of ionizing photons produced per second, and is commonly inferred from
the dust-corrected $\ha$ luminosity.  $L_{\rm UV}$ is the dust-corrected UV luminosity density in units of erg\,s$^{-1}$\,Hz$^{-1}$.}
\label{tab:diagnostics}
\end{deluxetable*}

All of the strong line ratios for SDSS galaxies were corrected for the
contribution of diffuse ionized gas (DIG) and the effects of
flux weighting from multiple emission regions using the prescriptions
of \citet{sanders17} and assuming a DIG fraction of $f_{\rm
  dig}=0.55$, the median value for the sample of SDSS galaxies
examined in \citet{sanders17}.  Aside from the DIG component,
flux weighting from multiple emission regions, each of which may have
different physical properties, can result in biases in
globally measured line ratios relative to the median \ion{H}{2}
region properties.  Some line ratios measured in local integrated
galaxy spectra, such as O3 and N2, are biased primarily by the effects
of flux weighting, while contributions from DIG emission dominate the
biases in certain low-ionization line ratios, such as O2 (e.g.,
\citealt{sanders17}).  The corrections applied to the local sample
result in O3, O32, N2, and O3N2 ratios that are on average 0.76, 1.21,
0.93, and 0.78$\times$ those of the uncorrected ratios, respectively.
Similarly, the corrected values of R23 are on average $0.76\times$
those of the uncorrected values.  Note that the equivalent widths of
the lines were not corrected for DIG emission, as we are interested in
establishing relations between the directly measured equivalent widths
and the aforementioned galaxy and ISM properties.

We have {\em not} applied similar corrections to the line ratios for
the high-redshift galaxies in the MOSDEF sample.  As discussed in
\citet{sanders17}, the DIG contribution to emission line ratios for
high-redshift galaxies is expected to be small ($\la 20\%$) given
their high SFR surface densities---a smaller DIG
contribution is also observed for local starburst galaxies.
Additionally, the high sSFRs and efficient feedback-driven metal
mixing expected for high-redshift galaxies would suggest that
flux-weighting effects play a minor role.  Having noted that DIG
contamination and flux-weighting effects are expected to be less
severe for typical star-forming galaxies at high redshift, there are
other possible factors that may complicate the interpretation of
globally measured line ratios at high redshift, such as the presence
of shocks or low-level AGN activity that may not reveal itself through
the methods typically used to identify AGNs at high redshift.

Finally, we include in our analysis the ionizing photon production
efficiency ($\xi_{\rm ion}$)---the ratio of the production rate of
ionizing photons to the non-ionizing UV continuum luminosity
density---as computed in \citet{shivaei18}.  To maintain consistency
with the attenuation curve adopted in modeling the SEDs of the
galaxies, we assumed the values of $\xi_{\rm ion}$ obtained with the
SMC extinction curve: these are generally $0.3$\,dex larger than those
obtained with the Calzetti attenuation curve (see \citealt{shivaei18}
for details).

\subsection{Equivalent Width ($W$) Measurements}
\label{sec:ewmeasurements}

Equivalent widths ($W$) for all but the Balmer recombination lines
were calculated by dividing the line flux by the continuum flux
density, where the latter was determined from the flux density at line
center from the best-fit stellar population model.  For $W(\ha)$ and
$W(\hb)$, the continuum flux density was computed as follows.  We fit
a linear function to the continuum flux density points of the best-fit
SED model in two wavelength windows bracketing the Balmer absorption:
$(\lambda_0-100)\times(1+z)$ to $(\lambda_0-30)\times(1+z)$ and
$(\lambda_0+30)\times(1+z)$ to $(\lambda_0+100)\times(1+z)$, where
$\lambda_0$ is the rest-frame wavelength of \ha\, or \hb\, in \AA.
The value of the linear function at $\lambda_0\times(1+z)$ was then
taken to be the continuum flux density at line center.

As the best-fit model is determined as that which gives the minimum
$\chi^2$ relative to the observed photometry, varying the modeling
assumptions (e.g., using a different stellar metallicity model or
star-formation history) typically results in a small variation in the
continuum flux density at line center.  For example, assuming the BC03
``solar'' metallicity models---corresponding to $1.4Z_\odot$ on the
current abundance scale \citep{asplund09}---results in continuum flux
densities that are on average $\simeq 5\%$ lower, and hence line
equivalent widths that are the same percentage higher, than those
obtained with our fiducial modeling.

The reported equivalent widths of \oii, \oiii, and \nii\, represent
the sum of the equivalent widths of the individual lines of the
doublets, in the case where both lines of the doublet are
significantly detected with an integrated $S/N\ge 3$.  If only one of
the lines of the \oiii\, or \nii\, doublet is undetected (typically
because it either falls out of the spectral coverage or falls on a sky
line), then the total equivalent width of the doublet is computed by
assuming a $2.97:1$ ratio of the flux of the stronger line of the
doublet to the flux of the weaker line.  For those galaxies where both
\oiii$\lambda 4960$ and \oiii$\lambda 5008$ are covered and at least
one line is detected, $\approx 96\%$ have a ratio, or limits on the
ratio, of \oiii$\lambda 5008$ to \oiii$\lambda 4960$ that is within
5$\sigma$ of the theoretical ratio of 2.97:1.  If both lines of the
doublet are undetected, then the upper limit in the total equivalent
width is taken to be the quadrature sum of the upper limits on the
equivalent widths of the individual lines of the doublet.  Galaxies
with upper limits in equivalent width are included in our analysis
through the construction of composite spectra and the measurement of
average equivalent widths, as described in
Appendex~\ref{sec:composite}.  The equivalent widths of emission lines
are taken to be positive and, unless stated otherwise, are divided by
the factor $1+z$ and hence refer to the {\em rest-frame} values.

\section{RELATIONS}
\label{sec:calibrations}

In this section, we present the relationships between rest-frame
optical emission line equivalent widths and several properties of the
SEDs of high-redshift galaxies, including stellar mass, rest-frame UV
slope (reddening), age, SFR, and specific SFR.  We also discuss how
the equivalent widths vary with several excitation-sensitive line
ratios including O32 and O3, gas-phase oxygen abundance indicators
including N2 and O3N2, and the ionizing photon production efficiency,
$\xi_{\rm ion}$.  Readers who wish to proceed directly to a discussion
of the physical context behind the redshift evolution in line
equivalent widths may skip to Section~\ref{sec:discussion}.

\subsection{Stellar Mass}
\label{sec:ewsedmass}

\begin{figure*}
\epsscale{1.15}
\plotone{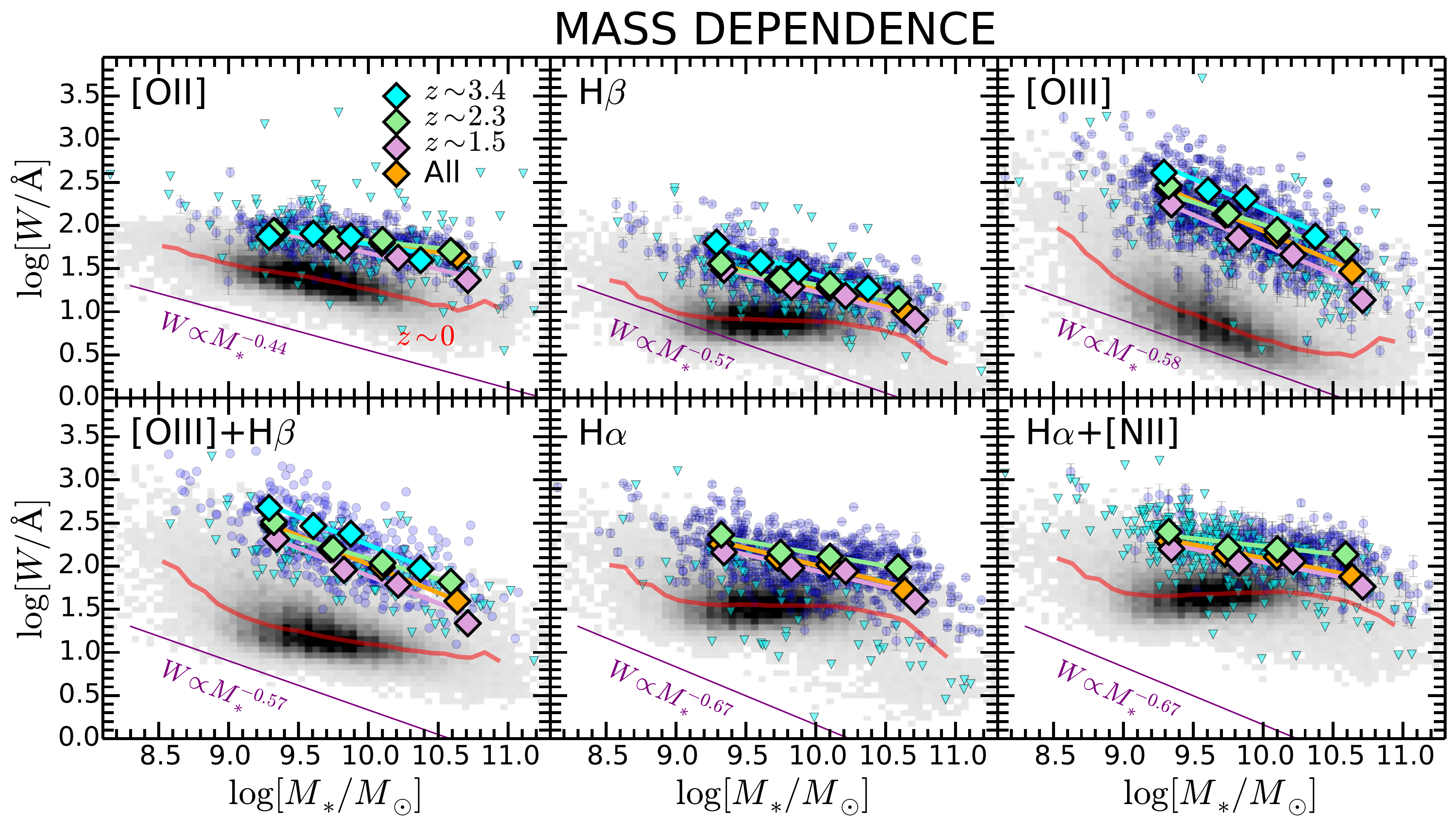}
\caption{Rest-frame equivalent widths ($W$) of \oii, \hb, \oiii,
  \oiii+\hb, \ha, and \ha+\nii, as a function of stellar mass for the
  full sample.  Objects where the lines have been detected are
  indicated by the small light blue points, while objects with
  undetected lines have $3\sigma$ upper limits on the equivalent
  widths indicated by the small downward-pointing triangles.  The
  large diamonds denote the average equivalent widths obtained from
  composite spectra of objects in four bins of stellar mass, and the
  thick solid line indicates the best-fit linear relation to the
  average values (parameters are given in
  Table~\ref{tab:relations_sedparms}).  Average equivalent widths and
  best-fit linear relations are also shown for galaxies in three
  redshift bins ($z\sim 1.5$, $z\sim 2.3$, and $z\sim 3.4$).  The
  distribution of equivalent width versus stellar mass for the local
  SDSS sample is shown in grayscale, where the red lines indicate the running
  median equivalent width as a function of stellar mass for the local
  sample.  For reference, the thin solid purple lines indicate the
  arbitrarily normalized relationships between equivalent width and
  stellar mass for a constant line luminosity for galaxies in the MOSDEF
  sample.}
\label{fig:ewsedmassz}
\end{figure*}

 The equivalent widths ($W$) of \oii, \hb, \oiii, \oiii+\hb, \ha, and
 \ha+\nii\, as a function of stellar mass are shown in
 Figure~\ref{fig:ewsedmassz}.  The equivalent width of the
 \ha+\nii+\sii\, complex is roughly $0.09$\,dex larger on
 average---going from 0.07\,dex larger to 0.11\,dex larger from the
 lowest to highest mass galaxies in our sample---than the \ha+\nii\,
 complex, and thus we only present results for the latter.  The
 Spearman rank correlation coefficient, significance of correlation,
 and slope and intercept for the best-fit linear correlation between
 $W$ and $M_{\ast}$---as is the case for all the other SED parameters
 considered here---are indicated in
 Table~\ref{tab:relations_sedparms}.\footnote{We note that massive
   galaxies with $M_\ast \ga 10^{11}$\,$M_\odot$ may be
   under-represented in our sample given their rarity and the
   relatively small volumes probed with spectroscopy.  However, the
   rarity of these massive galaxies suggests that their inclusion will
   not significantly affect the relationships between $W$ and $M_\ast$
   presented here, particularly if they follow the same relationship
   between $W$ and $M_\ast$ as galaxies with more moderate stellar
   masses.}  These linear correlations are based on fitting the
 average equivalent widths measured from composite spectra of galaxies
 in bins of stellar mass (Appendix~\ref{sec:composite}), and thus
 include both detections and upper limits in equivalent width.  Also
 listed in Table~\ref{tab:relations_sedparms} are the total numbers of
 galaxies with detected and undetected lines, the total number of
 galaxies, and the redshift range and mean redshift of the galaxies.
 The Spearman tests indicate a high degree of correlation between $W$
 and $M_{\ast}$ ($\sigma_{\rm P}>3$), such that galaxies with lower
 stellar masses have higher equivalent widths.

The trends between equivalent width and stellar mass are not
surprising given that $W$ is inversely proportional to the continuum
flux density, and the continuum flux density is tightly correlated
with the derived stellar mass.  To guide the eye, the purple lines in
Figure~\ref{fig:ewsedmassz} indicate the arbitrarily normalized
relationships between $W$ and $M_{\ast}$ under the assumption of a
constant line luminosity.  Because the equivalent widths of the
different lines are correlated with stellar mass to different degrees,
and with different slopes, we can conclude that the variation in $W$
with $M_{\ast}$ cannot solely be driven by changes in continuum flux
density.  In particular, it is clear from the comparison of the
constant-line-luminosity (purple) line and the best-fit relation to
$W(\oiii)$ versus $M_{\ast}$ that the \oiii\, luminosity is roughly
constant with mass above a stellar mass of $\simeq 10^9$\,$M_\odot$,
while the line luminosity decreases with decreasing mass for other
lines such as \oii, \hb, and \ha.  More generally, we find that the
equivalent widths of \oiii\, (and, hence, \oiii+\hb) present the
greatest variation with stellar mass.  Galaxies with stellar masses of
$M_{\ast}\approx 10^9$\,$M_\odot$ have \oiii\, and \oiii+\hb\,
equivalent widths that are factor of $\simeq 7-9$ times higher than
those of galaxies with $M_\ast \approx 10^{10.5}$\,$M_\odot$.  As we
will see shortly, the \oiii\, and \oiii+\hb\, equivalent widths are
more sensitive than any other line equivalent widths to essentially
all of the observables and derived quantities examined here.

Figure~\ref{fig:ewsedmassz} shows in more detail how the relationship
between $W$ and stellar mass varies with redshift,
when dividing the MOSDEF sample into low, middle, and high-redshift
bins.  For each of these redshift bins, we further subdivided the
galaxies into bins of stellar mass, constructed the
composite spectrum in each of these redshift/stellar mass bins, and
computed the average equivalent width from these composite spectra.

The relationship between equivalent width and stellar mass evolves
with redshift (Figure~\ref{fig:ewsedmassz}).  Among galaxies in the
MOSDEF sample, we find that the equivalent width increases with
redshift at a fixed stellar mass, a trend that is most pronounced when
considering $W(\oiii)$.  This trend has been noted before in the
context of $W(\ha)$ (e.g., \citealt{fumagalli12, sobral14}) and
$W(\oiii+\hb)$ (e.g., \citealt{khostovan16}).  More generally, the
equivalent widths of $z\sim 2.3$ galaxies in the MOSDEF sample are a
factor of $\simeq 30\times$ larger than those of local galaxies at a
fixed mass of $10^{9.7}$\,$M_\odot$, corresponding roughly to the
average stellar mass of galaxies in our sample.  In
Section~\ref{sec:residuals}, we discuss how the redshift evolution in
$W$ versus $M_{\ast}$ can be explained in terms of corresponding redshift
evolutions in the SFR versus $M_{\ast}$ and gas-phase metallicity versus
$M_{\ast}$ (i.e., the mass-metallicity or MZR) relations.

\subsection{UV Slope}

\begin{figure}
\epsscale{1.15}
\plotone{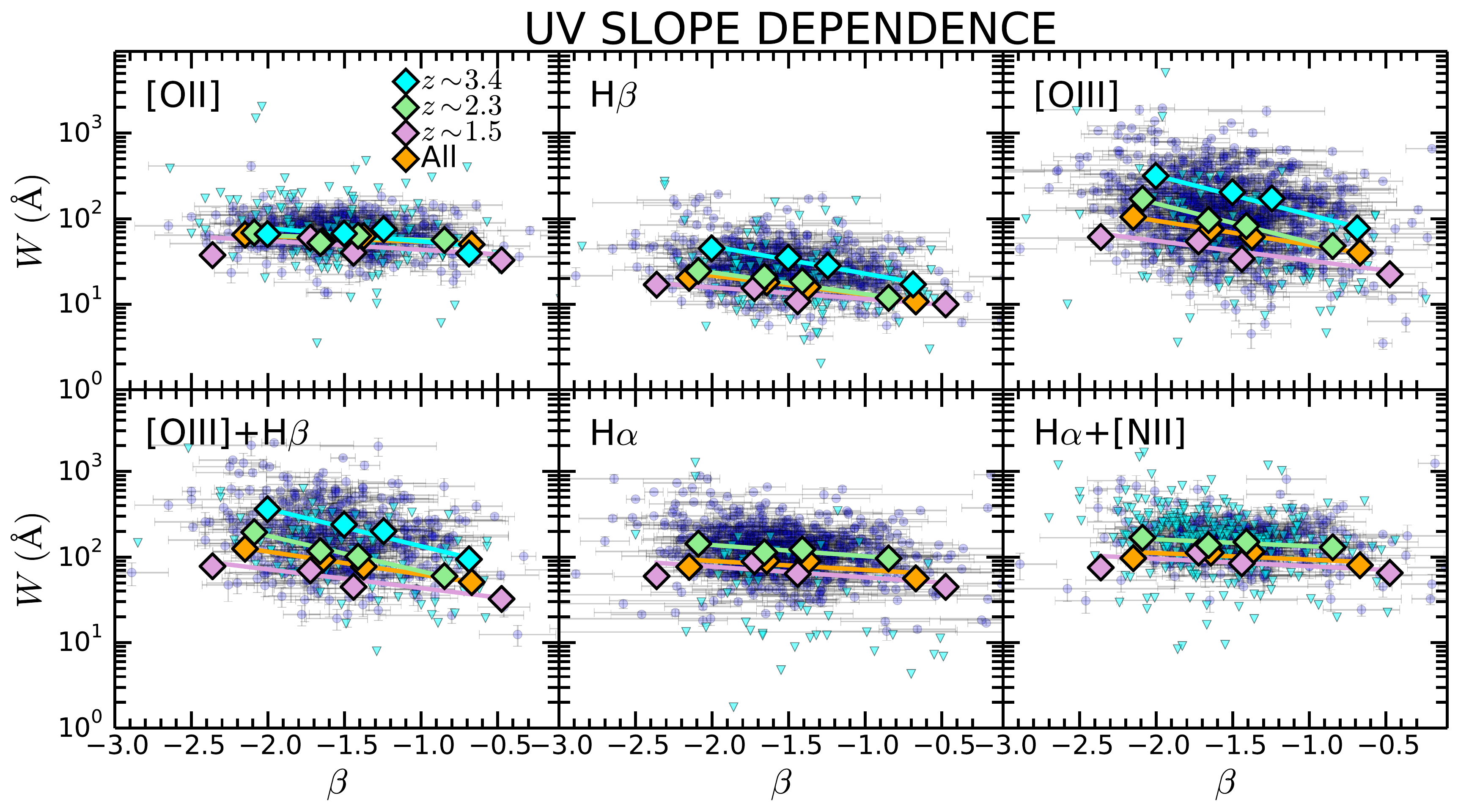}
\caption{Rest-frame equivalent widths ($W$) of \oii, \hb, \oiii,
  \oiii+\hb, \ha, and \ha+\nii, as a function of UV slope, $\beta$,
  for objects in the full sample.  Symbols are the same as in
  Figure~\ref{fig:ewsedmassz}.}
\label{fig:ewbetaphotz}
\end{figure}

Another fundamental property of galaxy SEDs is the UV slope, an
observable quantity that can be directly related to the reddening of
the stellar population for star-forming galaxies given a dust
attenuation curve (Table~\ref{tab:diagnostics}).
Figure~\ref{fig:ewbetaphotz} shows the line equivalent widths as a
function of the UV spectral slope.  The UV slope was computed by
fitting a power law through the observed photometry spanning the
rest-frame wavelength range $\lambda = 1268-2580$\,\AA\,
\citep{calzetti94}.  Spearman tests indicate that there are
significant correlations between $W$ and $\beta$, such that galaxies
with bluer UV spectral slopes have larger equivalent widths
(Table~\ref{tab:relations_sedparms}).  In particular, while $W(\oii)$
changes by only $\approx 30\%$ from galaxies with $\beta = -2.2$ to
$-0.7$, $W(\oiii)$ and $W(\oiii+\hb)$ change by a factor of $\simeq
2.5$ over the same range in UV slope.  Given the monotonic
relationship between UV slope and $\ebmvstars$ for simple
star-formation histories, the trends between $W$ and $\ebmvstars$ are
similar to those obtained between $W$ and $\beta$, such that galaxies
with bluer stellar color excesses have higher equivalent widths on
average than galaxies with redder stellar color excesses.  Likewise,
the trends observed between $W$ and $\beta$ are similar to those
between $W$ and the nebular, or gas, color excess, $\ebmvgas$.  For
brevity, the trends between $W$ and $\ebmvstars$/$\ebmvgas$ are not
shown here.

The redshift evolution in the relationship between $W$ and $\beta$ is
such that $W$ increases with redshift at a fixed $\beta$.  The
mildest evolution occurs with $W(\oii)$, while the strongest evolution
occurs with $W(\oiii)$ and $W(\oiii+\hb)$.  In the latter case, the
equivalent widths increase on average by a factor of $\simeq 2$ and
$\simeq 4$ at $z\sim 2.3$ and $z\sim 3.4$, respectively, relative to
the values at $z\sim 1.5$.  As the UV slope can be directly related to
the reddening of the UV continuum, the redshift dependence of $W$
versus $\beta$ can be largely explained in terms of the relationship
between UV slope/reddening and stellar mass, as discussed in
Section~\ref{sec:discussion}.

\subsection{Age}
\label{sec:ewsedage}

\begin{figure}
\epsscale{1.15}
\plotone{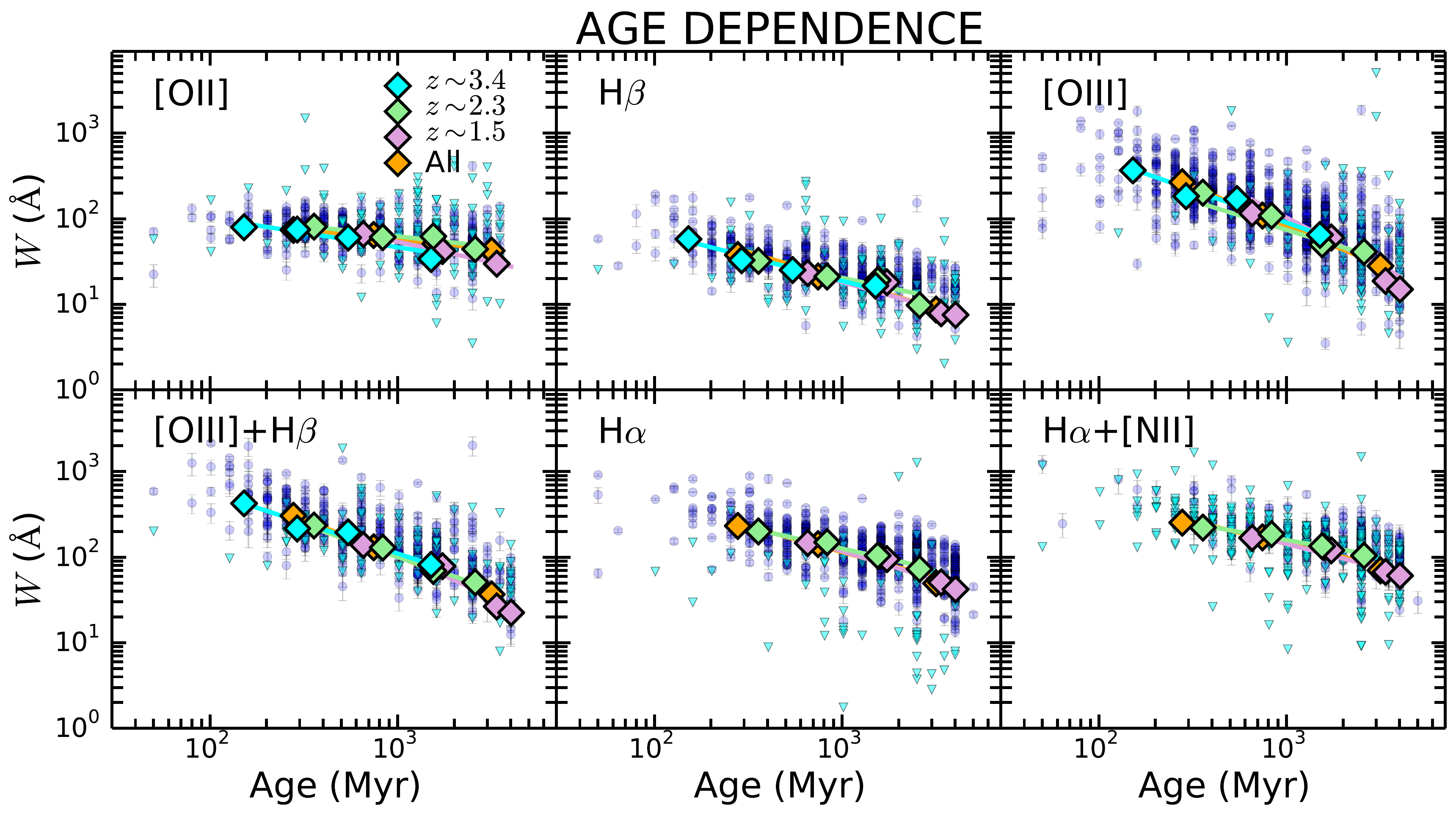}
\caption{Rest-frame equivalent widths ($W$) of \oii, \hb, \oiii,
  \oiii+\hb, \ha, and \ha+\nii, as a function of the
  luminosity-weighted stellar age for the full sample.  Symbols are
  the same as in Figure~\ref{fig:ewsedmassz}.}
\label{fig:ewsedagez}
\end{figure}

In the broadband stellar population modeling considered here, the
luminosity-weighted stellar population age is constrained primarily by
the shape or strength of the Balmer and $4000$\,\AA\, breaks.  The
equivalent widths of all the strong rest-frame optical emission lines
as a function of age are shown in Figure~\ref{fig:ewsedagez}.  The
equivalent widths of recombination lines have long been used as
proxies for stellar population age in star-forming galaxies for a
simple reason.  The recombination line equivalent widths are the
ratios of the recombination line luminosities, which are sensitive to
the current rate of star formation, to the continuum luminosity
densities, which are sensitive to the past average SFR.  Thus, for
simple star-formation histories, there is a direct correspondence
between equivalent width and age \citep{stasinska96}.  While we find
significant correlations between equivalent width and age for both
\ha\, and \hb, the most significant correlation is found between
$W(\oiii)$ (and $W(\oiii+\hb)$) versus age.  While the \oiii\, line
luminosity is partly dependent on SFR, it is also sensitive to other
factors including the hardness of the ionizing radiation field and
metallicity.  The fact that $W(\oiii)$ correlates more strongly with
age than the Balmer recombination lines then suggests that these
additional factors must also depend on age (e.g., see also
\citealt{stasinska96, gallazzi05}).  In terms of the MOSDEF sample,
the youngest galaxies have \oiii\, equivalent widths that are an order
of magnitude higher than those of the oldest galaxies.

Figure~\ref{fig:ewsedagez} shows that while the average age of
galaxies in our sample decreases toward higher redshifts, the
relationship between average equivalent width and age does not evolve
strongly with redshift.  Given that the age is sensitive to the shape
of the SED (see above), we conclude that the equivalent width can
serve as a rough proxy for spectral shape, independent of redshift.
As we discuss below, this conclusion is supported by another probe of
the spectral shape, namely the specific SFR.  From a practical
standpoint, the redshift invariance of the $\langle W \rangle$
versus $\langle {\rm age}\rangle$ relationship would nominally suggest a
promising avenue for estimating the {\em average} age of stellar
populations based on simple measurements of line equivalent widths for
high-redshift galaxies (e.g., \citealt{stasinska96}).  However, we
must keep in mind that there is typically at least an order of
magnitude spread in ages at a given $W$ (Figure~\ref{fig:ewsedagez}),
not even accounting for an increase the spread of ages if one were to
allow the star-formation history to vary between individual galaxies.
Thus, the aforementioned relations may be useful for predicting
average ages for large samples of high-redshift galaxies, but age
estimates for individual galaxies should be treated with an abundance
of caution.

\subsection{SFR and sSFR}
\label{sec:ewsedsfr}

\begin{figure}
\epsscale{1.15}
\plotone{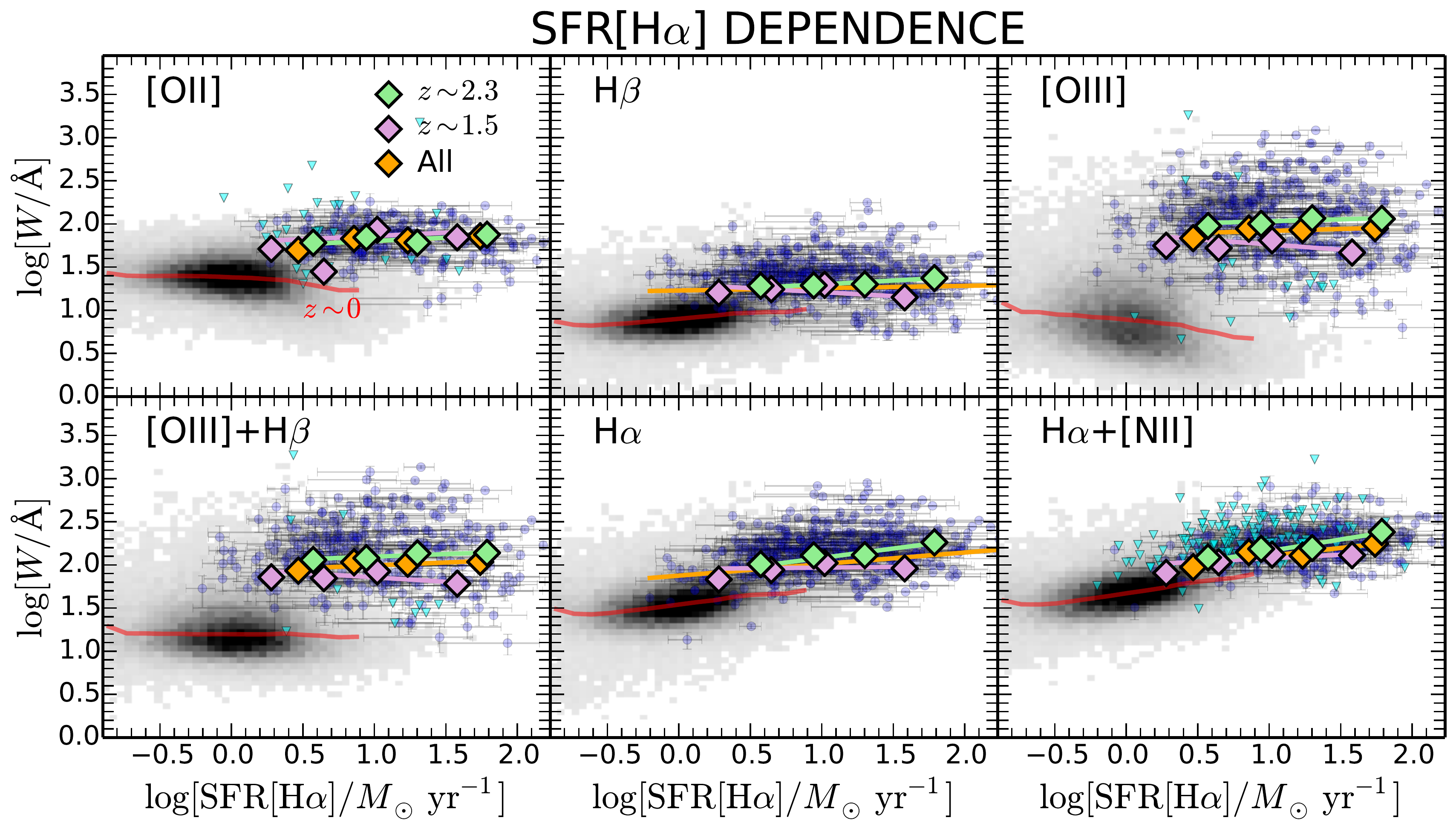}
\caption{Rest-frame equivalent widths ($W$) of \oii, \hb, \oiii,
  \oiii+\hb, \ha, and \ha+\nii, as a function of the SFR inferred from
  the dust-corrected H$\alpha$ luminosity.  Only those galaxies with significantly detected \ha\,
  and \hb\, lines are shown.  Symbols are the same as in
  Figure~\ref{fig:ewsedmassz}.}
\label{fig:ewsfrhaz}
\end{figure}

Because $\sfrsed$ is tightly correlated with stellar mass (i.e., both
quantities are sensitive to the normalization of the best-fit SED), we
have chosen to focus instead on SFR measurements made independently of
the SED fitting, namely SFRs derived from $\ha$.
Figure~\ref{fig:ewsfrhaz} shows how the equivalent widths vary with
H$\alpha$-based SFRs ($\sfrha$) for the two lower redshift bins where
$\ha$ is accessible.  In general, the correlations between
$W$ and $\sfrha$ are less significant than those found for other
trends (e.g., between $W$ and $M_\ast$, or $W$ and age).
Specifically, there is only a weak dependence of the equivalent width
as a function of $\sfrha$ at a fixed redshift, though we note that
galaxies at a fixed SFR have equivalent widths that increase with
redshift---the only exception being $W(\oii))$ versus $\sfrha$, trends
that do not change significantly between $z\sim 1.5$ and
$z\sim 2.3$.

\begin{figure}
\epsscale{1.15}
\plotone{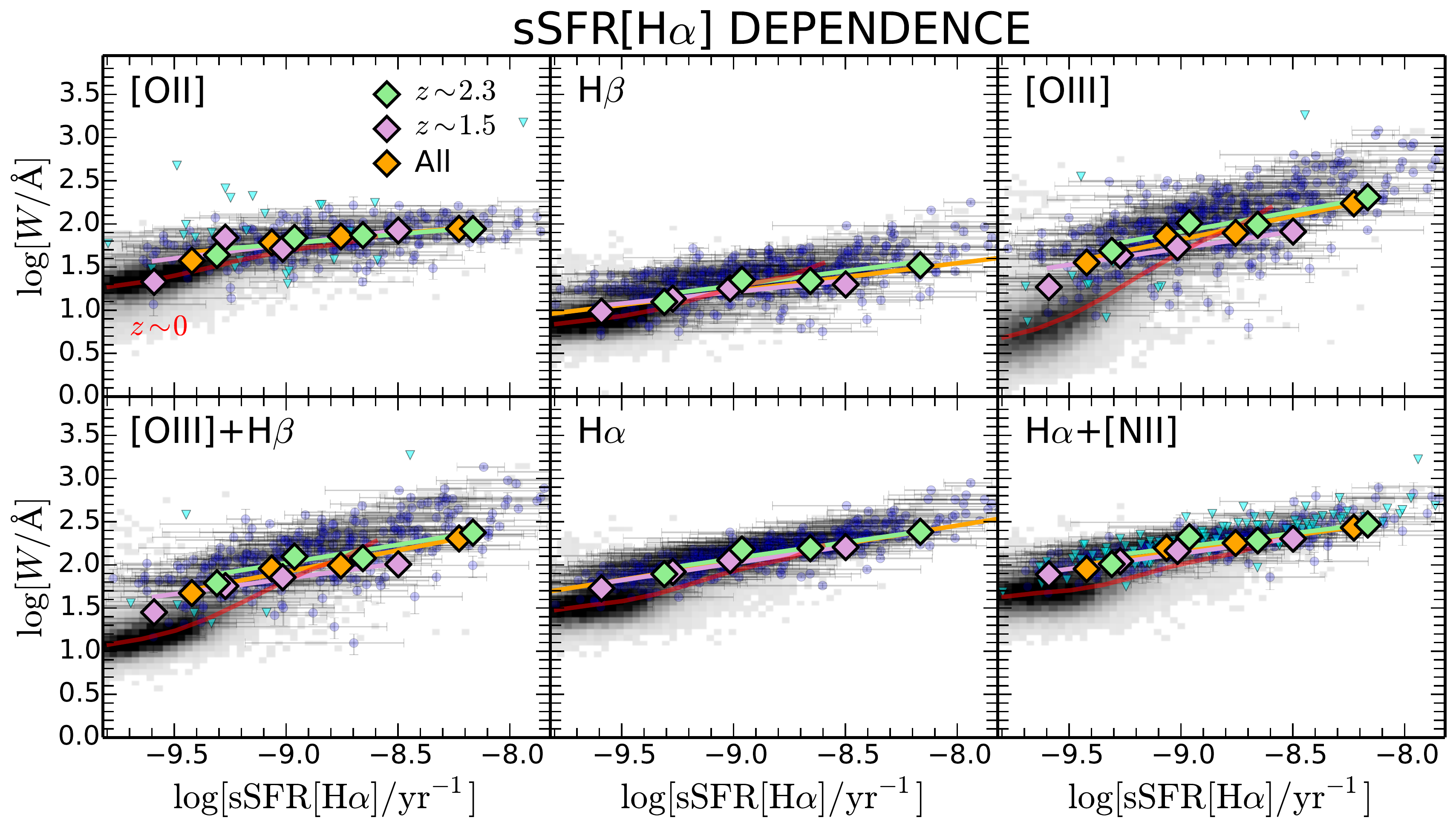}
\caption{Rest-frame equivalent widths ($W$) of \oii, \hb, \oiii,
  \oiii+\hb, \ha, and \ha+\nii, as a function of the specific SFR,
  where the SFR is inferred from the dust-corrected H$\alpha$
  luminosity.  Only those galaxies with significantly detected \ha\,
  and \hb\, lines are shown.  Symbols are the same as in
  Figure~\ref{fig:ewsedmassz}.}
\label{fig:ewssfrhaz}
\end{figure}

We find that $W$ is very tightly correlated with the specific SFR,
$\ssfrha$, such that galaxies with larger sSFRs have higher equivalent
widths (Figure~\ref{fig:ewssfrhaz}).  These tight relationships are
directly related to the fact that equivalent width is a ratio of the
line luminosity to continuum luminosity density, where the line
luminosity is sensitive to SFR and where the continuum luminosity
density is sensitive to stellar mass.  Figure~\ref{fig:ewsedmassz}
shows that the equivalent width rises with redshift at a fixed mass.
The redshift evolution of the SFR-$M_\ast$ relation implies that the
sSFR rises with redshift at a fixed stellar mass.  Hence, $W$ is
correlated with sSFR.  It is also evident that the relationships
between $W$ and sSFR are largely redshift-invariant, a result that is
also supported by the trends between $W$ and $\ssfrsed$ for galaxies
in the $z\sim 1.5$, $z\sim 2.3$, and $z\sim 3.4$ redshift subsamples.
This issue is discussed further in Section~\ref{sec:ewssfrmet}.

\subsection{Excitation and Gas-phase Metallicity}

The response of the ISM to the radiation field in high-redshift
galaxies is reflected in the excitation-sensitive line ratios
including O32 and O3.  The ionizing properties of the massive stars
that give rise to that radiation field can be ascertained from
calculations of $\xi_{\rm ion}$ (e.g., \citealt{robertson13,
  bouwens15b, shivaei18}).  Another significant quantity associated
with the state of the ISM is the gas-phase oxygen abundance.  Here we
examine how line equivalent widths vary with these properties of the
ISM.

\subsubsection{O32 (Ionization Parameter)}

\begin{figure}
\epsscale{1.15}
\plotone{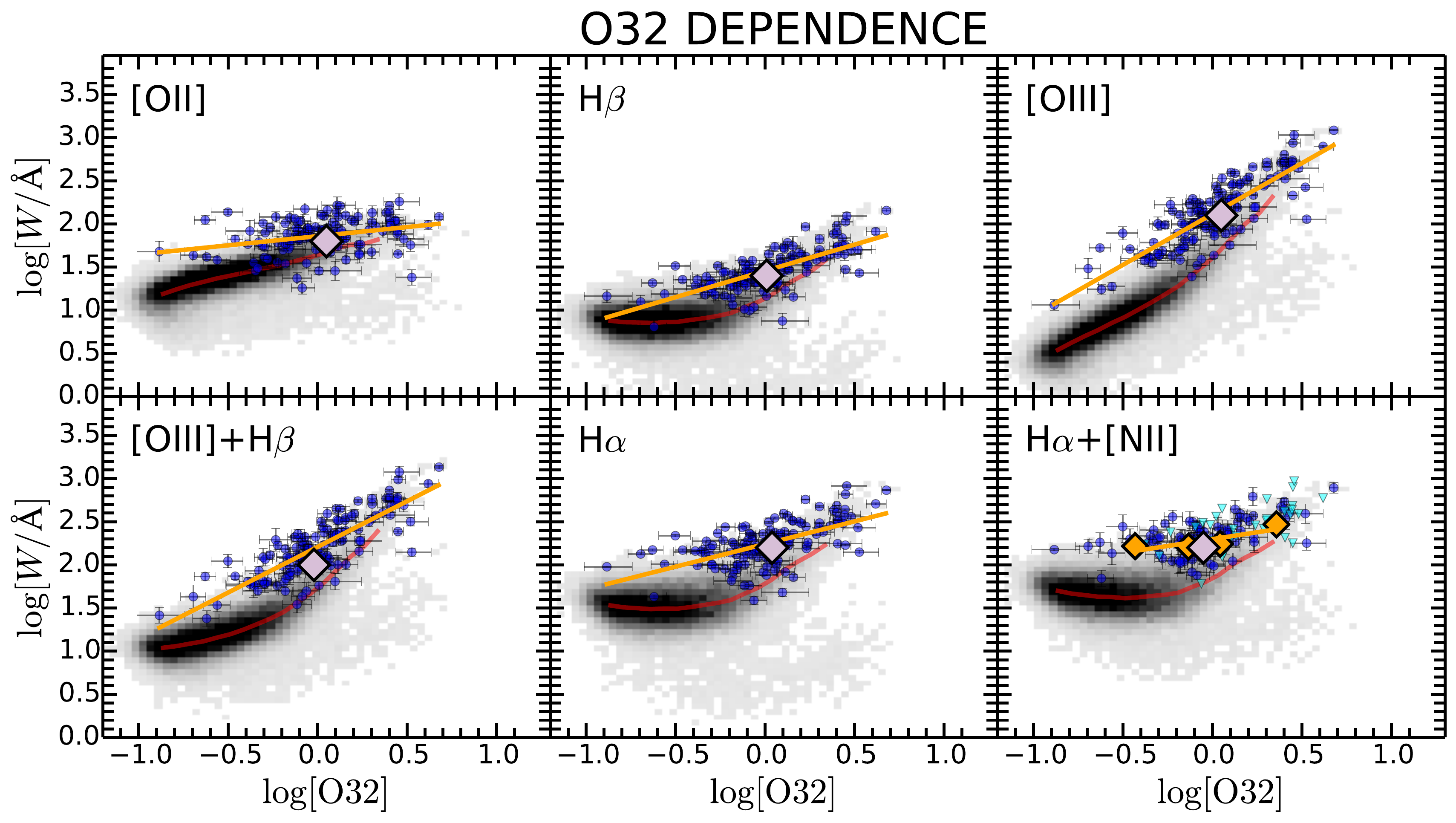}
\caption{Rest-frame equivalent widths ($W$) of \oii, \hb, \oiii,
  \oiii+\hb, \ha, and \ha+\nii, as a function of $\log[{\rm O32}]$ for
  galaxies in the MOSDEF sample ({\em blue} circles and {\em cyan}
  triangles).  The O32 values have been corrected for dust based on
  the Balmer decrement (\ha/\hb).  Thus, the figure shows only those
  objects where \oii, \oiii, \hb, and \ha\, were all significantly
  detected, and only shows objects in the two lower redshift bins of
  our sample that have coverage of \ha.  Orange diamonds in the lower
  rightmost panel indicate the mean equivalent width of \ha+\nii\, in
  bins of $\log[{\rm O32}]$, while the orange lines in all panels
  indicate the mean trend in $W$ versus $\log[{\rm O32}]$.  Purple
  diamonds denote the average values obtained from composite spectra
  of objects with spectral coverage of \oii, \oiii, \hb, and \ha,
  irrespective of whether the lines were detected.  The distribution
  of local SDSS galaxies is shown in grayscale, with the running
  median trend indicated by the red line.}
\label{fig:ewo32}
\end{figure}

The dependence of the equivalent widths on O32, a proxy for the
ionization parameter (${\sc U}$; e.g., \citealt{sanders16a}), is shown in
Figure~\ref{fig:ewo32}, with details of the sample redshift ranges,
number of galaxies, significance of correlation, and best-fit
parameters given in Table~\ref{tab:relations_ism}.  As noted earlier,
the O32 values were corrected for dust attenuation based on the Balmer
decrement, and thus the figure only shows those galaxies that had
significant (i.e., $\ge 3\sigma$) detections of \oii, \oiii, \hb, and
\ha. For reference, also shown are $\langle W\rangle$ and $\langle
\log[{\rm O32}]\rangle$ obtained from composite spectra that include
galaxies where all four aforementioned lines were covered in the
spectra, but not necessarily detected ({\em purple} diamonds in
Figure~\ref{fig:ewo32}).  These average values lie close to the mean
trend of $W$ versus $\log[{\rm O32}]$ obtained for objects with
significant detections of all four lines, suggesting that galaxies
where one or more of these lines may be undetected have a similar
distribution in $W$ and $\log[{\rm O32}]$.

\begin{deluxetable*}{llllrrcrc}
\tabletypesize{\footnotesize}
\tablewidth{0pc}
\tablecaption{Dependence of Equivalent Widths on O32, Oxygen Abundance, and $\xi_{\rm ion}$}
\tablehead{
\colhead{Attribute\tablenotemark{a}} &
\colhead{Line\tablenotemark{a}} &
\colhead{$z$-Range ($\langle z\rangle$)\tablenotemark{b}} &
\colhead{$N$ (det/undet)\tablenotemark{c}} &
\colhead{$\rho$\tablenotemark{d}} &
\colhead{$\sigma_{\rm P}$\tablenotemark{d}} &
\colhead{Intercept\tablenotemark{e}} &
\colhead{Slope\tablenotemark{e}} & 
\colhead{RMS\tablenotemark{e}}}
\startdata
$\log[{\rm O32}]$ & \oii & $1.604 - 2.545$ (2.229) & 125 (125/0) & 0.34 & 4.7 & $1.858\pm 0.017$ & $0.212\pm 0.059$ & 0.18 \\
                  & \hb & $1.604 - 2.545$ (2.229) & 125 (125/0) & 0.75 & 10.0 & $1.456\pm 0.015$ & $0.609\pm 0.053$ & 0.17 \\
                  & \oiii & $1.604-2.545$ (2.229) & 125 (125/0) & 0.81 & 11.8 & $2.114\pm 0.019$ & $1.178\pm 0.068$ & 0.21 \\
                  & \oiii+\hb & $1.604 - 2.545$ (2.229) & 125 (125/0) & 0.80 & 10.2 & $2.208\pm0.018$ & $1.058 \pm 0.064$ & 0.20 \\
                  & \ha & $1.604- 2.545$ (2.229) & 125 (125/0) & 0.51 & 7.8 & $2.241\pm 0.018$ & $0.525\pm 0.064$ & 0.20 \\
                  & \ha+\nii & $1.688 - 2.545$ (2.213) & 118 (82/36) & 0.48 & 5.9 & $2.298 \pm 0.003$ & $0.318\pm 0.007$ & 0.18 \\
\hline
$12+\log[{\rm O/H}]_{\rm N2}$ & \oii & $1.604 - 2.580$ (2.246) & 169 (158/11) & -0.53 & 6.9 & $7.806\pm0.727$ & $-0.719\pm 0.086$ & 0.18 \\
                             & \hb & $1.358 - 2.529$ (2.053) & 201 (189/12) & -0.62 & 8.9 & $11.775\pm 0.352$ & $-1.245\pm 0.046$ & 0.22 \\
                             & \oiii & $1.358 - 2.580$ (2.044) & 277 (257/20) & -0.61 & 10.4 & $19.111\pm 1.406$ & $-2.049\pm 0.165$ & 0.32 \\
                             & \oiii+\hb & $1.358 - 2.527$ (2.024) & 170 (152/18) & -0.62 & 8.0 & $18.887\pm 1.927$ & $-2.012\pm 0.305$ & 0.30 \\
                             & \ha & $1.005 - 2.580$ (2.027) & 360 (360/0) & -0.64 & 12.6 & $13.631\pm 0.482$ & $-1.367\pm 0.069$ & 0.21 \\
                             & \ha+\nii & $1.005 - 2.580$ (2.027) & 360 (360/0) & -0.55 & 10.4 & $10.960\pm 0.672$ & $-1.039\pm 0.074$ & 0.20 \\
\hline
$12+\log[{\rm O/H}]_{\rm O3N2}$ & \oii & $1.604-2.483$ (2.219) & 81 (80/1) & -0.62 & 7.0 & $10.607\pm 1.246$ & $-1.054\pm 0.150$ & 0.16 \\
                              & \hb & $1.390 - 2.527$ (2.026) & 151 (151/0) & -0.69 & 9.5 & $12.121\pm 0.846$ & $-1.286 \pm 0.101$ & 0.19 \\
                              & \oiii & $1.390 - 2.527$ (2.026) & 151 (151/0) & -0.88 & 13.0 & $25.074\pm 1.047$ & $-2.771\pm 0.125$ & 0.19 \\
                              & \oiii+\hb & $1.390 - 2.527$ (2.026) & 151 (151/0) & -0.86 & 11.1 & $21.965\pm 0.981$ & $-2.385\pm 0.117$ & 0.19 \\
                              & \ha & $1.390 - 2.527$ (2.026) & 151 (151/0) & -0.70 & 10.8 & $12.987\pm 0.851$ & $-1.297\pm 0.102$ & 0.16 \\
                              & \ha+\nii & $1.390- 2.527$ (2.026) & 151 (151/0) & -0.63 & 9.5 & $10.898\pm 0.862$ & $-1.035\pm 0.103$ & 0.16 \\
\hline
$\log[\xi_{\rm ion}/{\rm Hz\, erg^{-1}}]$ & \oii & $1.581 - 2.654$ (2.251) & 170 (150/20) & 0.29 & 4.1 & $-6.013\pm 1.025$ & $0.307\pm 0.040$ & 0.19 \\
              & \hb & $1.368 - 2.654$ (2.080) & 332 (332/0) & 0.30 & 5.5 & $-4.423\pm 1.053$ & $0.230\pm 0.042$ & 0.26 \\
              & \oiii & $1.369 - 2.586$ (2.064) & 281 (270/11) & 0.33 & 6.2 & $-10.969\pm 1.562$ & $0.506\pm 0.061$ & 0.41 \\
              & \oiii+\hb & $1.369 - 2.586$ (2.064) & 281 (270/11) & 0.35 & 5.9 & $-10.116\pm 1.587$ & $0.476\pm 0.063$ & 0.37 \\
              & \ha & $1.369 - 2.654$ (2.080) & 332 (332/0) & 0.69 & 13.6 & $-10.343\pm 0.699$ & $0.494\pm 0.028$ & 0.19 \\
              & \ha+\nii & $1.390 - 2.654$ (2.050) & 304 (201/103) & 0.66 & 10.6 & $-11.324\pm 0.819$ & $0.531\pm 0.032$ & 0.21 
\enddata
\tablenotetext{a}{Statistics are presented for the relationship between $\log[W/{\rm \AA}]$ for the line (or lines)
listed under column heading ``Line'' and the property listed under column heading ``Attribute.''}
\tablenotetext{b}{Redshift range and mean redshift of objects in this subsample.}
\tablenotetext{c}{Total number of objects and the number of detections and non-detections of the line (or lines)
listed under column heading ``Line.''}
\tablenotetext{d}{Spearman rank correlation coefficient and the number of standard deviations by which the correlation
deviates from the null hypothesis of no correlation.}
\tablenotetext{e}{Intercept and slope of the best-fit linear function to the composite averages, and the rms of the data points about this best-fit linear function.}
\label{tab:relations_ism}
\end{deluxetable*}

In general, we find that $W$ is strongly correlated with O32.  Of
course, this is by construction for lines like \oii\, and \oiii, which
enter into the calculation of O32.  On the other hand, while the
dust-corrected O32 values are dependent on the \ha/\hb\, ratio, they
are less sensitive to the individual \ha\, and \hb\, line
luminosities.  Even in this case, we find that the equivalent widths
of \ha\, and \hb\, increase with O32 (see also \citealt{kewley15}).
We also computed the dust-corrected O32 for the SDSS sample of
galaxies in the same manner as was done for the MOSDEF sample, with
values indicated by the grayscale distribution in
Figure~\ref{fig:ewo32}.  We note that correcting the \ha\, and \hb\,
emission lines for Balmer absorption in the local galaxies using the
empirically derived prescription of \citet{groves12} does not
significantly alter the dust corrections derived for O32 and, as such,
the trends between $W$ and O32 for local galaxies are affected only
minimally.  We remind the reader that the O32 values for the SDSS
sample shown in this figure were corrected for DIG emission
(Section~\ref{sec:calcexcitation}).

A comparison between the MOSDEF and SDSS samples suggests that there
may be some mild redshift evolution in the $W$ versus $\log[{\rm
    O32}]$ relation, though the amount of evolution depends
sensitively on whether O32 for the local sample is corrected for DIG
emission.  We return to this point in Section~\ref{sec:discmets}.  At
any rate, while the $z\sim 2$ MOSDEF sample is on average offset to
higher O32 and $W$ relative to the local sample, there are still
substantial numbers of local ``analog'' galaxies that occupy the same
region of $W$-O32 parameter space as the higher redshift galaxies.

\subsubsection{O3}

\begin{figure}
\epsscale{1.15}
\plotone{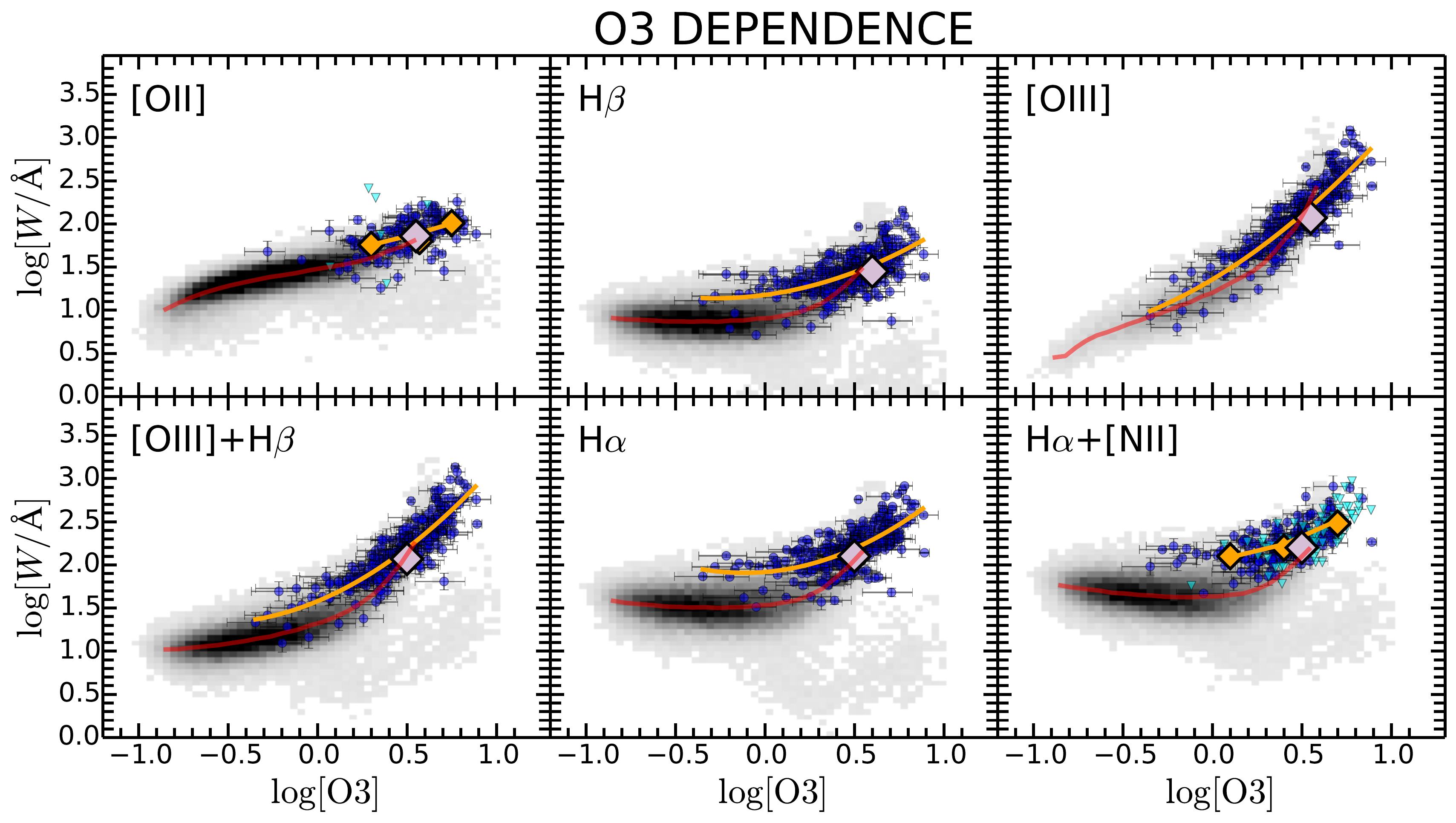}
\caption{Rest-frame equivalent widths ($W$) of \oii, \hb, \oiii,
  \oiii+\hb, \ha, and \ha+\nii, as a function of $\log[{\rm O3}]$ for
  galaxies in the MOSDEF sample ({\em blue} circles and {\em cyan}
  triangles ).  The O3 values have been corrected for dust based on
  the Balmer decrement (\ha/\hb).  Thus, the figure shows only those
  objects where \oiii, \hb, and \ha\, were all significantly detected,
  and only shows objects in the two lower redshift bins of our sample
  which have coverage of \ha.  The orange lines indicate the best-fit
  quadratic trend between $W$ and $\log[{\rm O3}]$.  Other symbols are
  the same as in Figure~\ref{fig:ewo32}.}
\label{fig:ewo3hb}
\end{figure}

The O3 ratio is sensitive to both the ionization parameter and
metallicity of the ISM, and thus may be expected to exhibit the same
behavior with respect to equivalent widths as the O32 ratio.
Figure~\ref{fig:ewo3hb} shows that this is indeed the case.  Once
again, we find a high significance of (positive) correlation between
the equivalent widths and O3, particularly as it concerns $W(\oiii)$.
As the $z\sim 2$ data indicate some curvature in the response of $W$
to $\log[{\rm O3}]$, we chose to fit second-order polynomials to the
trends of $W$ versus $\log[{\rm O3}]$, with the best-fit parameters
given in Table~\ref{tab:relations_o3hb}.  Comparison with the local
SDSS sample suggests that there is little, if any, redshift evolution
in the trends between $W$ and $\log[{\rm O3}]$ as far as those
galaxies with $\log[{\rm O3}]\ga 0.3$ are concerned.  On the other
hand, there may be some mild evolution of higher $W$ at a given
$\log[{\rm O3}]$ below this threshold.  However, as with O32, there
are local galaxies that span the same parameter space in $W$ versus O3
as the $z\sim 2$ galaxies.

\begin{deluxetable*}{lllrrrc}
\tabletypesize{\footnotesize}
\tablewidth{0pc}
\tablecaption{Dependence of Equivalent Widths on $\log[{\rm O3}]$ }
\tablehead{
\colhead{Line\tablenotemark{a}} &
\colhead{$z$-Range ($\langle z\rangle$)\tablenotemark{b}} &
\colhead{$N$ (det/undet)\tablenotemark{c}} &
\colhead{$\rho$\tablenotemark{d}} &
\colhead{$\sigma_{\rm P}$\tablenotemark{d}} &
\colhead{Coefficients\tablenotemark{e}} & 
\colhead{RMS\tablenotemark{e}}}
\startdata
\oii & $1.604 - 2.545$ (2.220) & 124 (116/8) & 0.53 & 5.6 & $c_0 = 1.635\pm 0.053$; $c_1 = 0.346\pm 0.224$; $c_2 = 0.197\pm 0.249$ & 0.19  \\
\hb & $1.357-2.586$ (2.069) & 241 (241/0) & 0.61 & 9.4 & $c_0 = 1.178\pm 0.029$; $c_1 = 0.278\pm 0.124$; $c_2 = 0.498\pm 0.155$ & 0.20 \\
\oiii & $1.357-2.586$ (2.069) & 241 (241/0) & 0.88 & 13.6 & $c_0 = 1.357\pm 0.028$; $c_1 = 1.241\pm 0.121$; $c_2 = 0.530\pm 0.151$ & 0.20 \\
\oiii+\hb & $1.357-2.586$ (2.069) & 241 (241/0) & 0.85 & 13.2 & $c_0 = 1.580\pm0.028$; $c_1 = 0.863\pm0.121$; $c_2 = 0.727\pm 0.152$ & 0.20 \\
\ha & $1.357-2.586$ (2.069) & 241 (241/0) & 0.67 & 10.4 & $c_0 = 1.917\pm 0.029$; $c_1 = 0.171\pm 0.127$; $c_2 = 0.750\pm 0.159$ & 0.20 \\
\ha+\nii & $1.390 - 2.586$ (2.040) & 217 (140/77) & 0.64 & 7.6 & $c_0 = 2.067\pm 0.028$; $c_1 = 0.150\pm 0.126$; $c_2 = 0.699\pm 0.181$ & 0.18 
\enddata
\tablenotetext{a}{Statistics are presented for the relationship between $\log[W/{\rm \AA}]$ for the line (or lines)
listed in this column and $\log[{\rm O3}]$.}
\tablenotetext{b}{Redshift range and mean redshift of objects in this subsample.}
\tablenotetext{c}{Total number of objects and the number of detections and non-detections of the line (or lines)
listed under column heading ``Line.''}
\tablenotetext{d}{Spearman rank correlation coefficient and the number of standard deviations by which the correlation
deviates from the null hypothesis of no correlation.}
\tablenotetext{e}{Coefficients of the best-fit quadratic function of the form $c_0 + c_1 x + c_2 x^2$, and the rms of the data 
points about this best-fit quadratic function.}
\label{tab:relations_o3hb}
\end{deluxetable*}

\subsubsection{Metallicity-sensitive Line Indices}

\begin{figure}
\epsscale{1.15}
\plotone{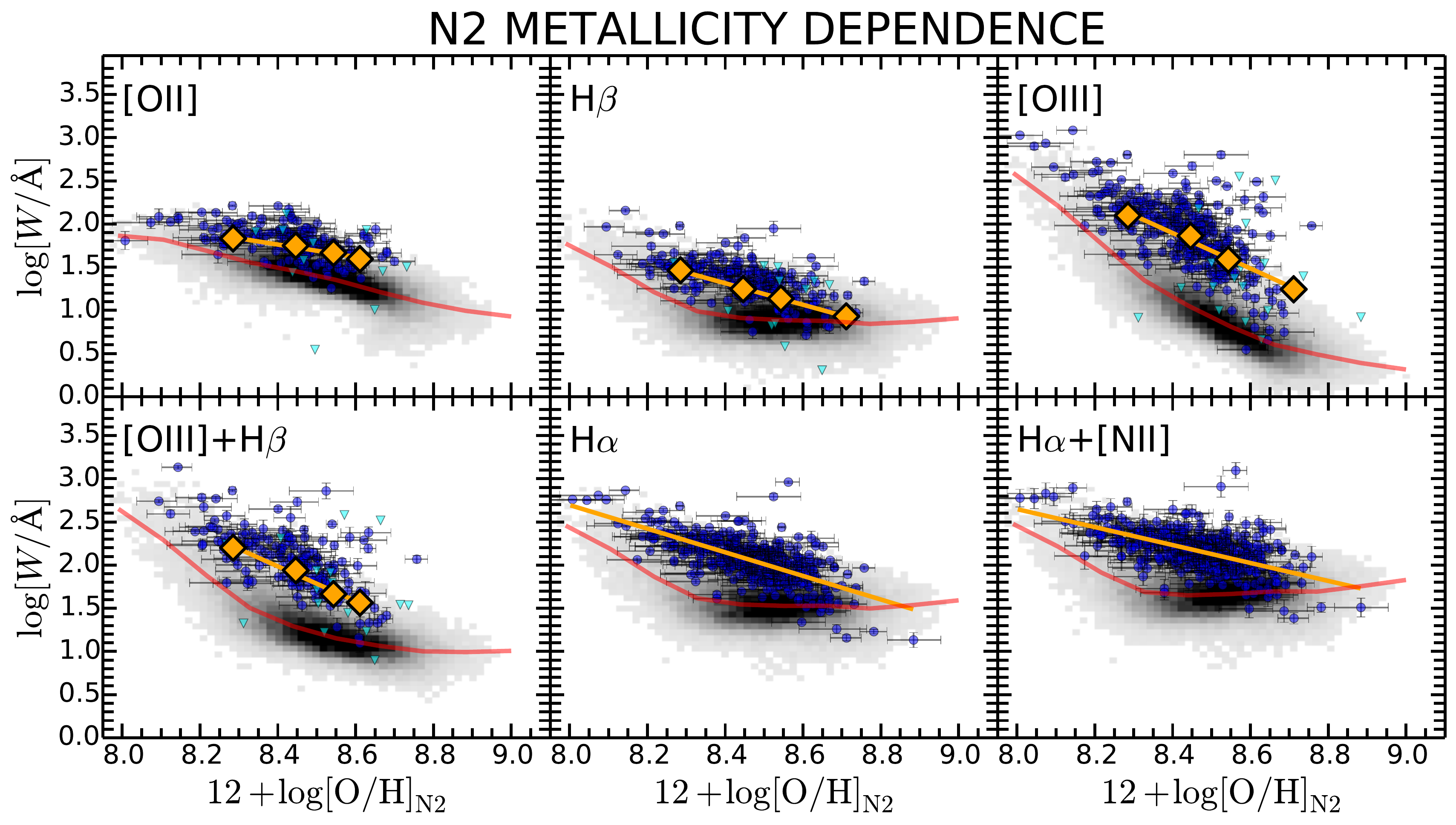}
\caption{Rest-frame equivalent widths ($W$) of \oii, \hb, \oiii,
  \oiii+\hb, \ha, and \ha+\nii, as a function of the gas-phase oxygen
  abundance inferred from the N2 index, and assuming the
  \citet{pettini04} calibration, for galaxies in the MOSDEF sample
  ({\em blue} circles and {\em cyan} triangles).  The same is shown
  for the local SDSS sample in grayscale.  Symbols are the same as in
  Figure~\ref{fig:ewo32}.  Stacked points are shown only in those
  panels where for some galaxies the corresponding line equivalent
  width have upper limits.}
\label{fig:ewn2met}
\end{figure}

\begin{figure}
\epsscale{1.15}
\plotone{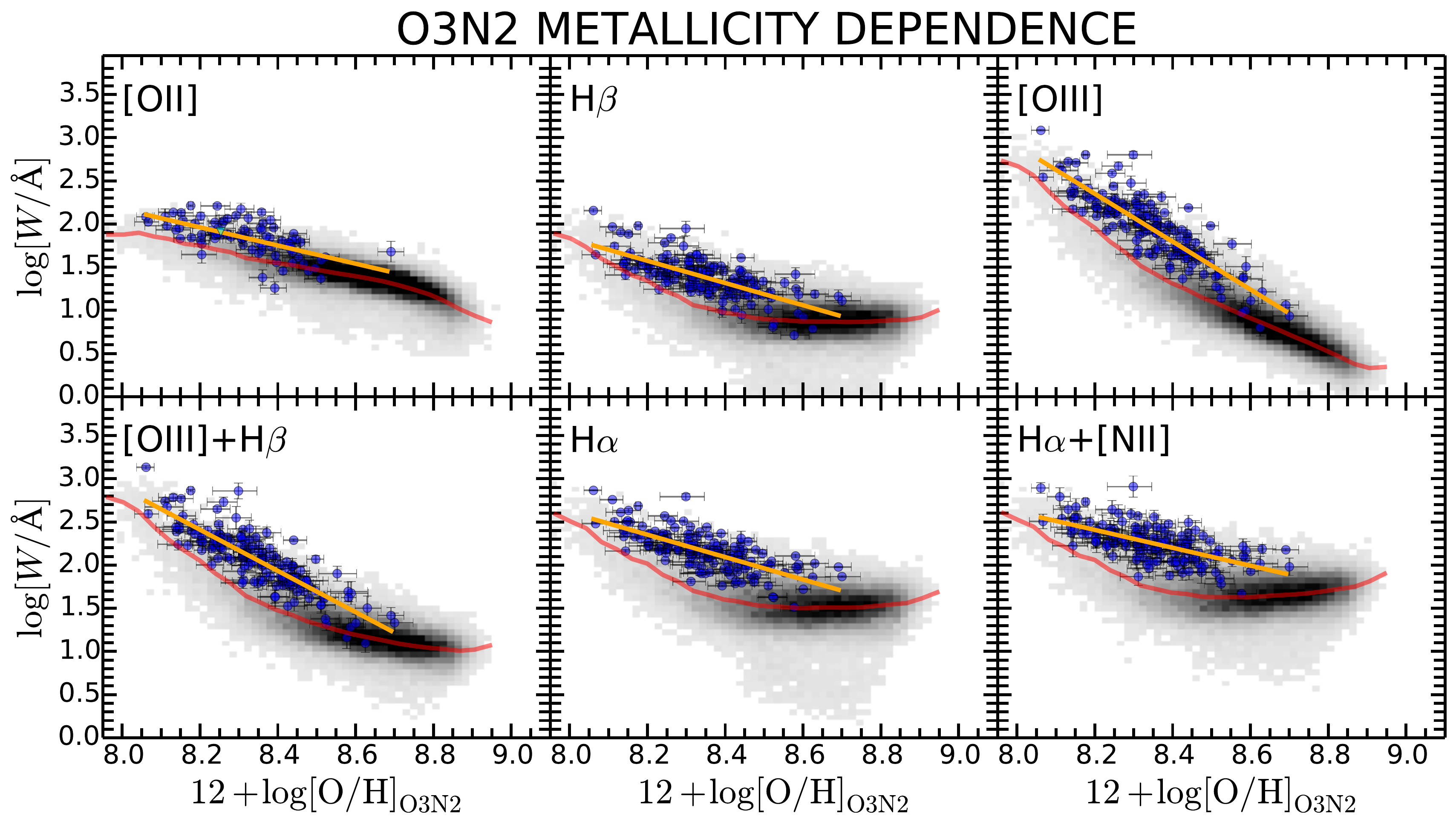}
\caption{Rest-frame equivalent widths ($W$) of \oii, \hb, \oiii,
  \oiii+\hb, \ha, and \ha+\nii, as a function of the gas-phase oxygen
  abundance inferred from the O3N2 index and assuming the
  \citet{pettini04} calibration.  Symbols are the same as in
  Figure~\ref{fig:ewn2met}.}
\label{fig:ewo3n2met}
\end{figure}

There are several indices based on strong rest-frame optical emission
lines that are used to estimate gas-phase oxygen abundances
\citep{kewley08}.  Abundances estimated from the N2 and O3N2 indices
are advantageous as they rely on lines that are typically detected in
$\sim L^{\ast}$ star-forming galaxies at high redshift, dust
corrections are negligible owing to the close proximity in wavelength
of the lines used to compute the O3 and N2 indices, and the indices
vary monotonically with abundance, unlike the R23 index.  On the other
hand, the R23 index may be preferable given apparent anomalies in
oxygen abundances derived based on nitrogen lines for high-redshift
galaxies.  We discuss these issues further in
Section~\ref{sec:discmets}.

Oxygen abundances estimated from the N2 and O3N2 indices are shown in
Figures~\ref{fig:ewn2met} and \ref{fig:ewo3n2met}, and the redshift
ranges, significance of correlations, and best-fit parameters for the
different subsamples are listed in Table~\ref{tab:relations_ism}.  Not
shown are galaxies with either upper or lower limits in oxygen
abundance: most of these galaxies have upper limits in N2, and
including them will make the best-fit slopes of the relations shown in
Figure~\ref{fig:ewn2met} slightly less negative.  In general,
inclusion of galaxies with either upper or lower limits in oxygen
abundance will not shift the relations shown in
Figures~\ref{fig:ewn2met} and \ref{fig:ewo3n2met} enough for them to
completely overlap with the locus of where most $z\sim 0$ galaxies
lie.  The data indicate significant correlations between equivalent
widths and oxygen abundance, such that galaxies with higher $W$ have
lower oxygen abundances.  These correlations are undoubtedly a
by-product of the relationships between metallicity and stellar mass,
and between $W$ and stellar mass (Figure~\ref{fig:ewsedmassz}).  As
with many of the other parameters we have examined, $W(\oiii)$ shows
the most significant variation with metallicity given the steeper
(i.e., more negative) slope of $W(\oiii)$ versus $12+\log[{\rm O/H}]$
relative to those derived for the other lines.

Abundances for the local SDSS sample were computed in the same way as
for the MOSDEF sample, and are denoted by the grayscale distributions
in Figures~\ref{fig:ewn2met} and \ref{fig:ewo3n2met}.  
Comparison with the local sample suggests that 
galaxies at $z\sim 2$ have $\simeq 0.2-0.5$\,dex larger $\log[W/{\rm \AA}]$
at a given oxygen abundance.  Despite these average offsets, the
$z\sim 2$ galaxies lie within the distribution of $W$ versus
$12+\log[{\rm O/H}]$ found for local galaxies.

\subsection{Ionizing Photon Production Efficiency ($\xi_{\rm ion}$)}

\begin{figure*}
\epsscale{1.15}
\plotone{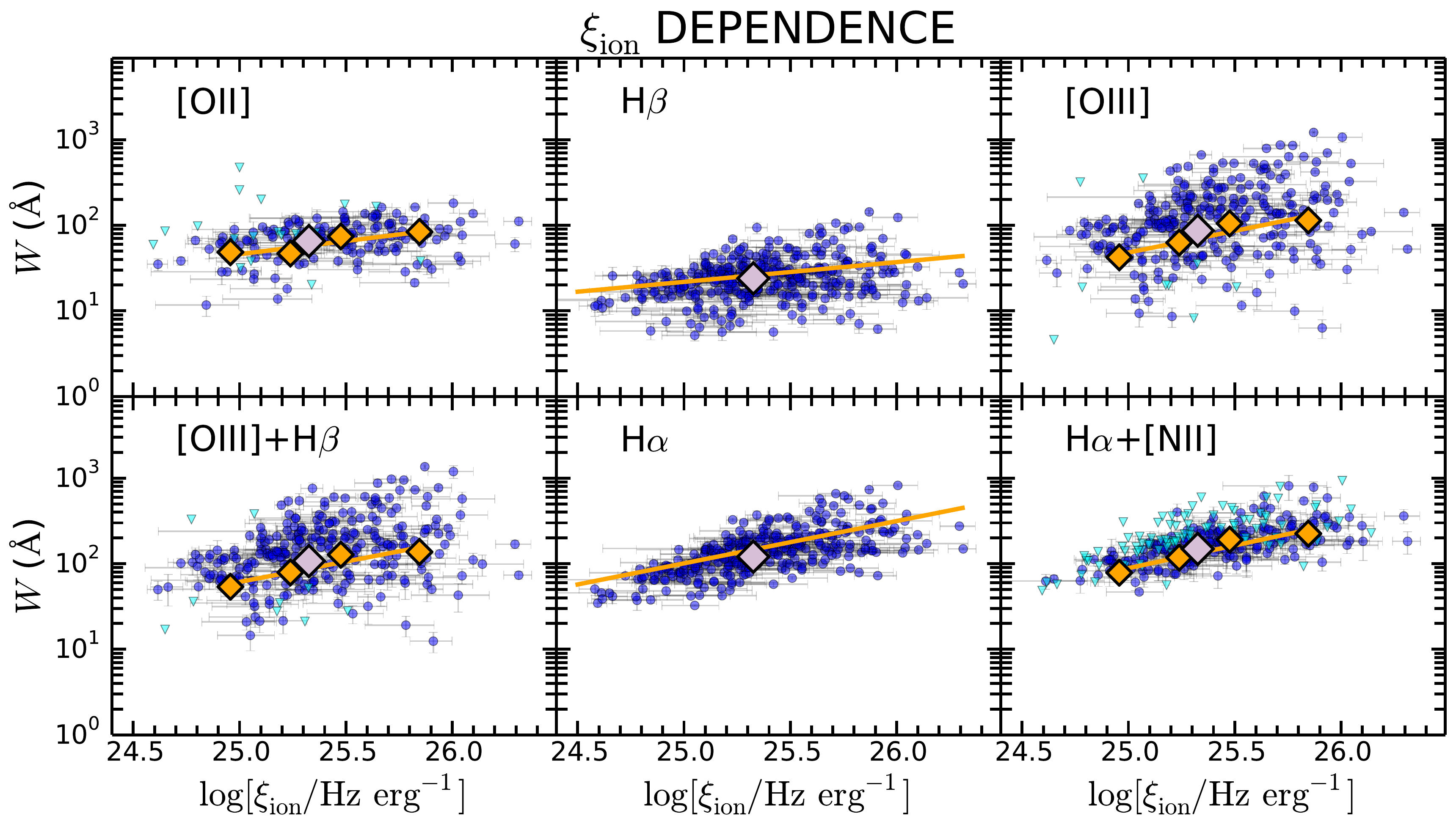}
\caption{Rest-frame equivalent widths ($W$) of \oii, \hb, \oiii,
  \oiii+\hb, \ha, and \ha+\nii, as a function of the ionizing photon
  production efficiency, $\xi_{\rm ion}$.  Symbols are the same as in
  previous figures.  As detailed in the text, the values of $\xi_{\rm
    ion}$ shown here assume the SMC curve when dust-correcting the UV
  continuum.}
\label{fig:ewxiion}
\end{figure*}

As noted earlier, $\xi_{\rm ion}$ is a convenient metric for
characterizing the hardness of the far-UV spectra of galaxies, in
particular the production rate of ionizing photons relative to the
SFR.  \citet{shivaei18} present the most comprehensive analysis of
how $\xi_{\rm ion}$ varies with a number of other galaxy/ISM
properties, including stellar mass, ionization parameter, metallicity,
and age.  Here, we focus on how the equivalent widths of the strong
rest-frame optical emission lines vary with $\xi_{\rm ion}$.

Figure~\ref{fig:ewxiion} shows the equivalent widths of \oii, \hb,
\oiii, \oiii+\hb, \ha, and \ha+\nii\, as a function of $\xi_{\rm
  ion}$.  Spearman tests indicate a high degree of significance in the
correlations between $W$ and $\xi_{\rm ion}$ (e.g., see
Table~\ref{tab:relations_ism}), such that galaxies with larger
$\xi_{\rm ion}$ have higher line equivalent widths.  Note that the
scatter in the trends between the Balmer lines (\ha\, in particular)
and $\xi_{\rm ion}$ is artificially tight as \ha\, and the Balmer
decrement are used, in part, to compute $\xi_{\rm ion}$.
Nevertheless, if we consider \oiii---a line that is not used in the
computation of $\xi_{\rm ion}$---we still find a significant positive
correlation between line equivalent width and $\xi_{\rm ion}$, albeit
with a large scatter (e.g., see also \citealt{chevallard18} and
\citealt{tang18}).  Specifically, $W(\oiii)$ varies by an order of
magnitude (or more) at a given $\xi_{\rm ion}$.  However, it is clear
from the upper rightmost panel of Figure~\ref{fig:ewxiion} that
galaxies with the highest $W(\oiii)\ga 500$\,\AA\, exclusively have
$\xi_{\rm ion}$ that exceed the value typically assumed for
high-redshift galaxies ($\xi_{\rm ion}= 25.2$; \citealt{robertson13}).
The relations between $W$ and $\xi_{\rm ion}$ provided in
Table~\ref{tab:relations_ism}, can be inverted to infer $\xi_{\rm ion}$
from a measurement of $W$.  For convenience, we have refit the average
(binned) values of $\xi_{\rm ion}$ as a function of $W(\oiii)$ to obtain
the following relation:
\begin{eqnarray}
\log\left[\frac{\xi_{\rm ion}}{\rm Hz\, erg^{-1}}\right] \nonumber & = & (1.83\pm 0.31) \times\log\left[\frac{W(\oiii)}{\rm \AA}\right] \nonumber \\
& & + (21.96 \pm 0.57).
\end{eqnarray}
We remind the reader that the relation above is valid for $\xi_{\rm ion}$ values
computed using the SMC extinction curve.

\section{DISCUSSION}
\label{sec:discussion}

Having presented the relations between equivalent widths and various
SED-derived parameters and ISM properties in
Section~\ref{sec:calibrations}, we now turn to a discussion of how the
various aforementioned trends arise, and how the equivalent widths
evolve with redshift when placing our measurements in context with
those of the literature.  We also consider practical applications of
our results in terms of selecting high-ionization and high-excitation
galaxies, and those that may dominate cosmic reionization.

\subsection{The Sensitivity of $\oiii$ Equivalent Width to Mass, Metallicity, and SFR}
\label{sec:residuals}

In Section~\ref{sec:ewsedmass}, we presented evidence that the
relationships between equivalent width and stellar mass evolve with
redshift, such that galaxies of a fixed $M_{\ast}$ have larger $W$
with redshift.  This evolution in $W$ at a fixed mass may be
associated with the increase in SFR with redshift at a given mass
(e.g., \citealt{noeske07,reddy06a,daddi07a,rodighiero11, wuyts11,
  reddy12b,whitaker14, schreiber15}), and/or the decrease in
metallicity with redshift at a given mass (e.g., \citealt{erb06a,
  maiolino08, troncoso14, sanders15, onodera16}).  The former is
indicated by Figure~\ref{fig:ewsfrhaz}, where the higher redshift
galaxies are found to have higher average $\sfrha$ and $W$ relative
to the local sample.  The
latter is indicated by Figures~\ref{fig:ewn2met} and
\ref{fig:ewo3n2met}, showing that galaxies with lower metallicities
have higher equivalent widths.

\subsubsection{Calculation of Residuals}

To further explore the cause of this evolution, we focused on \oiii,
and we computed the amount by which each galaxy's $W(\oiii)$,
$\sfrha$, oxygen abundance, and O32 deviates from the mean values of
these quantities for galaxies of the same stellar mass.  We considered
oxgyen abundances estimates from N2, as there is a larger number of
galaxies with N2-based metallicities than O3N2-based metallicities.
O32 is included in this analysis as it is inversely correlated with
metallicity, and thus serves as a second proxy for metallicity.  We
first determined the best-fit linear functions describing the behavior
of $\log[M_\ast/M_\odot]$ with $\log[W/{\rm \AA}]$ (e.g., as
exemplified in Figure~\ref{fig:ewsedmassz}),
$\log[\sfrha/M_\odot\,{\rm yr}^{-1}]$ (i.e., the SFR-$M_{\ast}$
relation), $12+\log({\rm O/H})_{\rm N2}$ (i.e., the mass-metallicity
relation), and $\log[{\rm O32}]$, where all these quantities could be
directly measured---i.e., we required direct detections of \oii,
\oiii, \hb, \ha, and \nii, leaving a sample of 141 objects.  The
functional forms of these linear relations are:
\begin{eqnarray}
\log[W(\oiii)/{\rm \AA}] & = & 8.41 - 0.64\log[M_\ast/M_\odot] \\
\log[\sfrha/M_\odot\,{\rm yr}^{-1}] & = & -4.11 +0.54\log[M_\ast/M_\odot] \\
12+\log({\rm O/H}) & = & 6.70 + 0.17\log[M_\ast/M_\odot] \\
\log({\rm O32}) & = & 3.98- 0.40\log[M_\ast/M_\odot]. 
\end{eqnarray}
We then computed the amount by which each galaxy deviates at a
fixed stellar mass from the best-fit relations listed above (i.e., the
residuals) in $\log[W(\oiii)/{\rm \AA}]$, $\log[\sfrha/M_\odot\,{\rm
    yr}^{-1}]$, $12+\log({\rm O/H})$, and $\log({\rm O32})$.  These
residuals are referred to as $\delw$, $\delsfr$, $\delz$, and
$\delion$, respectively.  Below, we consider how the relationship
between the residuals in equivalent width versus residuals in SFR
varies in bins of residual metallicity, and how the relationships
between the residuals in equivalent width versus residuals in
metallicity and O32 vary in bins of residual SFR (e.g.,
Figure~\ref{fig:residuals}).  The residuals allow us to examine how
the quantities of interest vary at a given stellar mass in a way that
is analogous to the methods employed by \citet{salim15},
\citet{kashino17}, and \citet{sanders18}.

\begin{deluxetable*}{lccrrr}
\tabletypesize{\footnotesize}
\tablewidth{0pc}
\tablecaption{$\Delta\log[W(\oiii)/{\rm \AA}]$ Residual Spearman Correlation Tests and Best-fit Parameters}
\tablehead{
\colhead{Attribute\tablenotemark{a}} &
\colhead{Criteria\tablenotemark{b}} &
\colhead{$\rho$\tablenotemark{c}} &
\colhead{$\sigma_{\rm P}$\tablenotemark{c}} &
\colhead{Intercept\tablenotemark{d}} &
\colhead{Slope\tablenotemark{d}}}
\startdata
$\Delta\log[{\rm SFR}(\ha)/M_\odot\,{\rm yr}^{-1}]$ & All & 0.55 & 6.5 & $0.00\pm 0.02$ & $0.64\pm 0.06$ \\
& $\Delta[12+\log({\rm O/H})_{\rm N2}] < -0.1$ & 0.56 & 2.8 & $0.10\pm 0.06$ & $0.62\pm 0.15$ \\
& $-0.1\le \Delta[12+\log({\rm O/H})_{\rm N2}] < 0.05$ & 0.26 & 1.9 & $0.02\pm 0.03$ & $0.40\pm 0.11$ \\
& $\Delta[12+\log({\rm O/H})_{\rm N2}] \ge 0.05$ & 0.50 & 3.4 & $-0.09\pm 0.03$ & $0.47\pm 0.08$ \\
\hline
$\Delta[12+\log({\rm O/H})_{\rm N2}]$ & All & -0.58 & 6.8 & $0.00\pm 0.03$ & $-2.95\pm 0.42$ \\
& $\Delta[{\rm SFR(\ha)}/M_\odot\,{\rm yr}^{-1}] < -0.4$ & -0.57 & 2.4 & $-0.07\pm 0.16$ & $-1.98\pm 0.10$ \\
& $-0.4 \le \Delta[{\rm SFR(\ha)}/M_\odot\,{\rm yr}^{-1}] < 0.3$ & -0.46 & 4.4 & $-0.01\pm 0.05$ & $-2.39\pm 0.35$ \\
& $\Delta[{\rm SFR(\ha)}/M_\odot\,{\rm yr}^{-1}] \ge 0.3$ & -0.45 & 2.4 & $0.09\pm 0.12$ & $-1.47\pm 0.82$ \\
\hline
$\Delta[\log({\rm O32})]$ & All & 0.71 & 8.5 & $0.00\pm 0.02$ & $1.11\pm 0.07$ \\
& $\Delta[{\rm SFR(\ha)}/M_\odot\,{\rm yr}^{-1}] < -0.4$ & 0.61 & 2.6 & $-0.22\pm 0.03$ & $1.09\pm 0.20$ \\
& $-0.4 \le \Delta[{\rm SFR(\ha)}/M_\odot\,{\rm yr}^{-1}] < 0.3$ & 0.76 & 7.2 & $-0.01\pm 0.02$ & $0.88\pm 0.06$ \\
& $\Delta[{\rm SFR(\ha)}/M_\odot\,{\rm yr}^{-1}] \ge 0.3$ & 0.91 & 4.9 & $0.18\pm 0.02$ & $1.09\pm 0.09$ 
\enddata
\tablenotetext{a}{Statistics are presented for the relationship between $\Delta\log[W(\oiii)/{\rm \AA}]$ and the attribute
listed in this column.}
\tablenotetext{b}{Criteria used to construct the subsamples.}
\tablenotetext{c}{Spearman rank correlation coefficient and the number of standard deviations by which the correlation
deviates from the null hypothesis of no correlation.}
\tablenotetext{d}{Intercept and slope of the best-fit linear function.}
\label{tab:residuals}
\end{deluxetable*}

In the context of the present analysis, our goal is to use the
residuals as a diagnostic tool to determine whether, at a given mass,
$W(\oiii)$ is sensitive to SFR, metallicity, or both.  For instance,
if $W(\oiii)$ is primarily sensitive to SFR (and not metallicity) at a
given mass, then $\delw$ and $\delsfr$ should be correlated and
exhibit no systematic offsets when determined in different bins of
metallicity.  On the other hand, if $W(\oiii)$ is primarily sensitive
to metallicity (and not SFR) at a given mass, then $\delw$ and
$\delsfr$ should be uncorrelated and the locus of points should shift
systematically for galaxies in different bins of metallicity.  Similar
behaviors will manifest themselves in the relations between $\delw$
and $\delz$: if $W(\oiii)$ is only sensitive to SFR at a given mass,
then $\delw$ and $\delz$ should be uncorrelated in different bins of
SFR, with the locus of points shifting systematically with SFR; if
$W(\oiii)$ is only sensitive to metallicity at a given stellar mass,
then $\delw$ and $\delz$ should be correlated and exhibit no
systematic offsets when determined in bins of SFR.

\subsubsection{A Physical Context for the Sensitivity of \oiii\, to 
SFR and Metallicity}

\begin{figure}
\epsscale{1.15}
\plotone{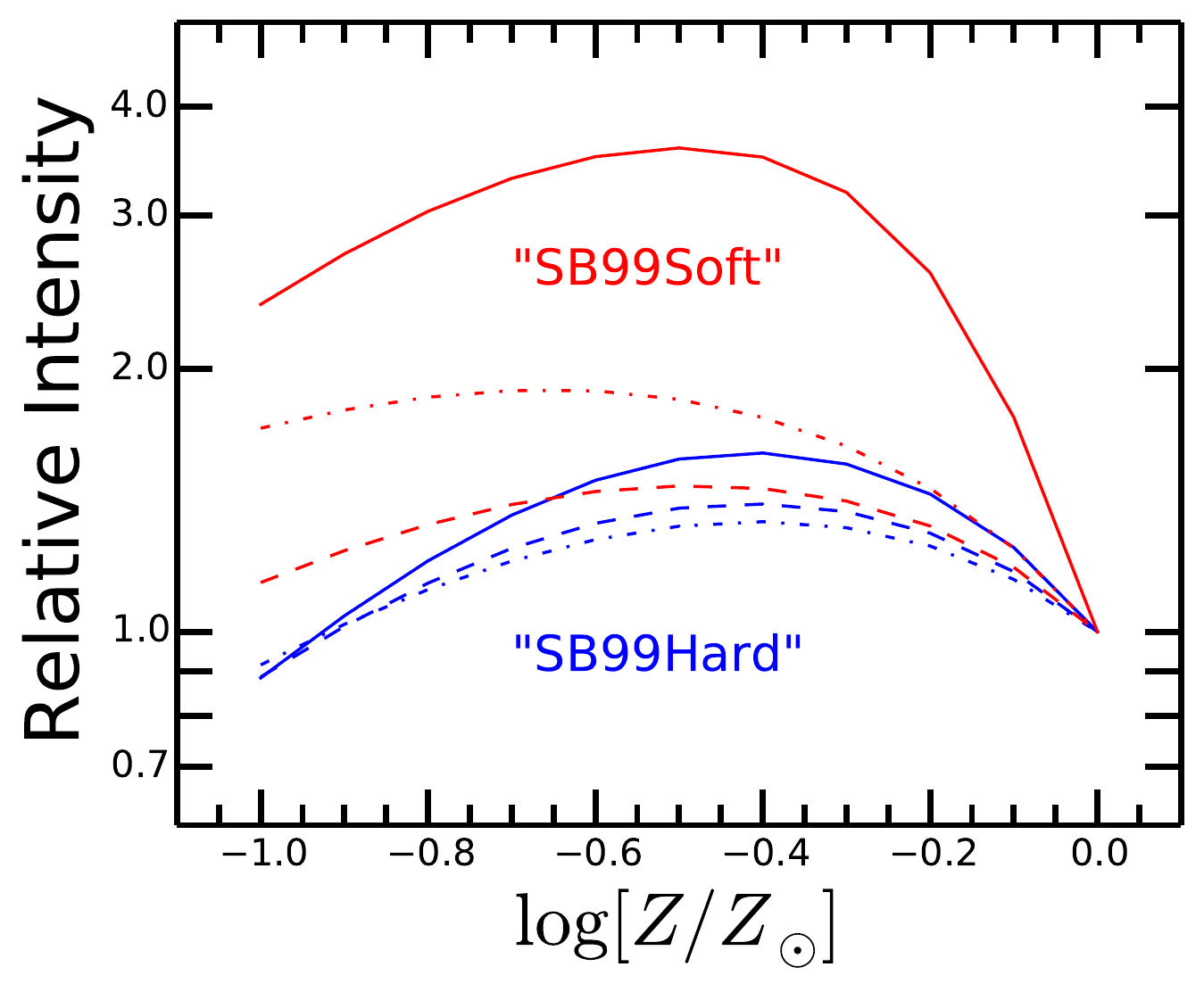}
\caption{Intensity of the $\oiii\lambda\lambda 4960,5008$ doublet
  relative to its solar value as a function of gas-phase metallicity
  for two stellar population models with different ionizing spectral
  shapes and $\log(\mathscr{U})=-3.5,-2.5,-1.5$ (solid, dashed, and
  dotted-dashed lines, respectively).  Curves corresponding to the
  SBsoft and SBhard models are indicated in {\em red} and {\em
    blue}, respectively.}
\label{fig:cloudy}
\end{figure}

In reality, the equivalent width is sensitive to both SFR and
metallicity at a fixed mass.  For example, for galaxies of a fixed
mass and metallicity, an increase in SFR translates to a larger
ionizing photon rate and hence higher \oiii\, luminosities.  For
galaxies of a fixed mass and SFR, a lower metallicity implies stellar
populations with a harder ionizing spectrum and hence higher \oiii\,
luminosities.  The sensitivity of \oiii\, to gas-phase metallicity, as
predicted from Cloudy photoionization modeling \citep{ferland17}, is
shown in Figure~\ref{fig:cloudy}.  For the modeling, we assumed an
open geometry (i.e., constant density slab of gas) with $n_{\rm
  e}=250$\,cm$^{-3}$; a range of ionization parameters
$\log(\mathscr{U})=-3.5$ to $-1.5$; and two Starburst99
\citep{leitherer99} stellar population model spectra taken from
\citet{sanders16b}, called ``SB99hard'' and ``SB99soft,'' which
encompass the range of ionizing spectral shapes relevant for galaxies
in our sample.  Specifically, the ``SB99hard'' model assumes a
$1/7$\,$Z_\odot$ stellar population formed in a single burst of star
formation that occurred $0.5$\,Myr ago.  The ``SB99soft'' model
assumes a solar metallicity stellar population formed with a
continuous SFR of 1\,$M_\odot$\,yr$^{-1}$.  This figure indicates that
for metallicities $\ga \frac{1}{3} Z_\odot$, and over the range of
metallicities spanned by most of the galaxies in the MOSDEF sample,
the \oiii\, line luminosity at a fixed SFR and stellar mass will
increase with decreasing metallicity.  Correspondingly, $W(\oiii)$
will also increase with decreasing metallicity in this regime at a
fixed SFR and stellar mass.

\subsubsection{Results for the High-redshift Sample}

\begin{figure}
\epsscale{1.10}
\plotone{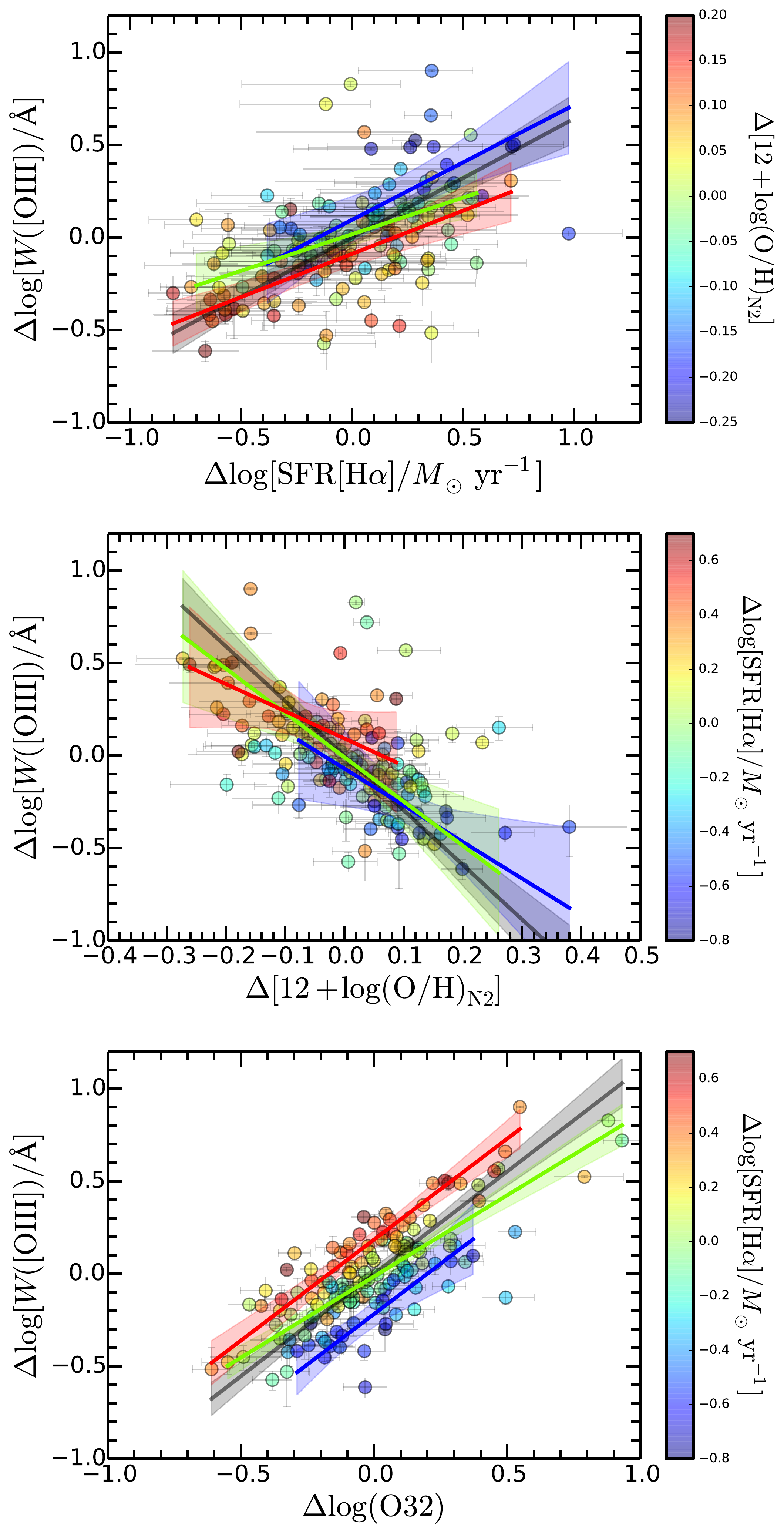}
\caption{Residuals of equivalent width versus residuals in $\sfrha$
  ({\em top}), oxygen abundance inferred from N2 ({\em middle}), and
  O32 ({\em bottom}).  The residual is defined as the amount by which
  a quantity for a galaxy deviates from its mean value expected for
  galaxies of the same stellar mass.  Only the 141 galaxies with significantly
  detected \oii, \oiii, \hb, \ha, and \nii\, lines are shown, such
  that the same set of galaxies appears in every panel.  The gray
  lines and gray-shaded regions indicate the best-fit linear function
  and $95\%$ confidence intervals of the fit, respectively, for all
  galaxies shown.  Similarly, the blue, green, and red lines and
  similarly colored shaded regions show the best-fit linear functions
  and $95\%$ confidence intervals of the fits when dividing galaxies
  into different bins of residuals in oxygen abundance and $\sfrha$.
  Details of the binning, Spearman correlation tests, and best-fit
  intercepts and slopes are provided in Table~\ref{tab:residuals}.}
\label{fig:residuals}
\end{figure}

The expected sensitive of $W(\oiii)$ to both SFR and metallicity (at a
fixed mass) is borne out by the results of Figure~\ref{fig:residuals},
where we show how $\delw$ depends on the other residuals: $\delsfr$,
$\delz$, and $\delion$.  Focusing on the top panel of this figure (and
Table~\ref{tab:residuals}), we find that the correlations between
$\delw$ and $\delsfr$ in different bins of $\delz$ are generally
significant at the $\ga 2\sigma$ level.  Moreover, the intercept of
$\delw$ versus $\delsfr$ increases monotonically from the bin of
highest to lowest metallicity.  Thus, at a given stellar mass and SFR,
the equivalent width increases with decreasing metallicity.

The correlations between $\delw$ and $\delion$ are also significant,
though we note that the scatter between these quantities is tightened
artificially as the \oiii\, line luminosity factors into both
$W(\oiii)$ and O32.  Nevertheless, it is clear from the bottom panel
of Figure~\ref{fig:residuals} that the normalization of the relation
between $\delw$ and $\delion$ increases systematically with $\delsfr$.
In the context of the anti-correlation between ionization parameter
and metallicity (e.g., \citealt{dopita86, dopita06, perez14,
  sanders16a}; c.f., \citealt{strom17}), this increase in
normalization implies that the equivalent width changes with SFR at a
fixed stellar mass and metallicity.  This behavior is marginally
supported by the $\delw$ versus $\delz$ relations, which also increase
in normalization with $\delsfr$.

\subsubsection{Results for the Local Sample}

\begin{figure}
\epsscale{1.15}
\plotone{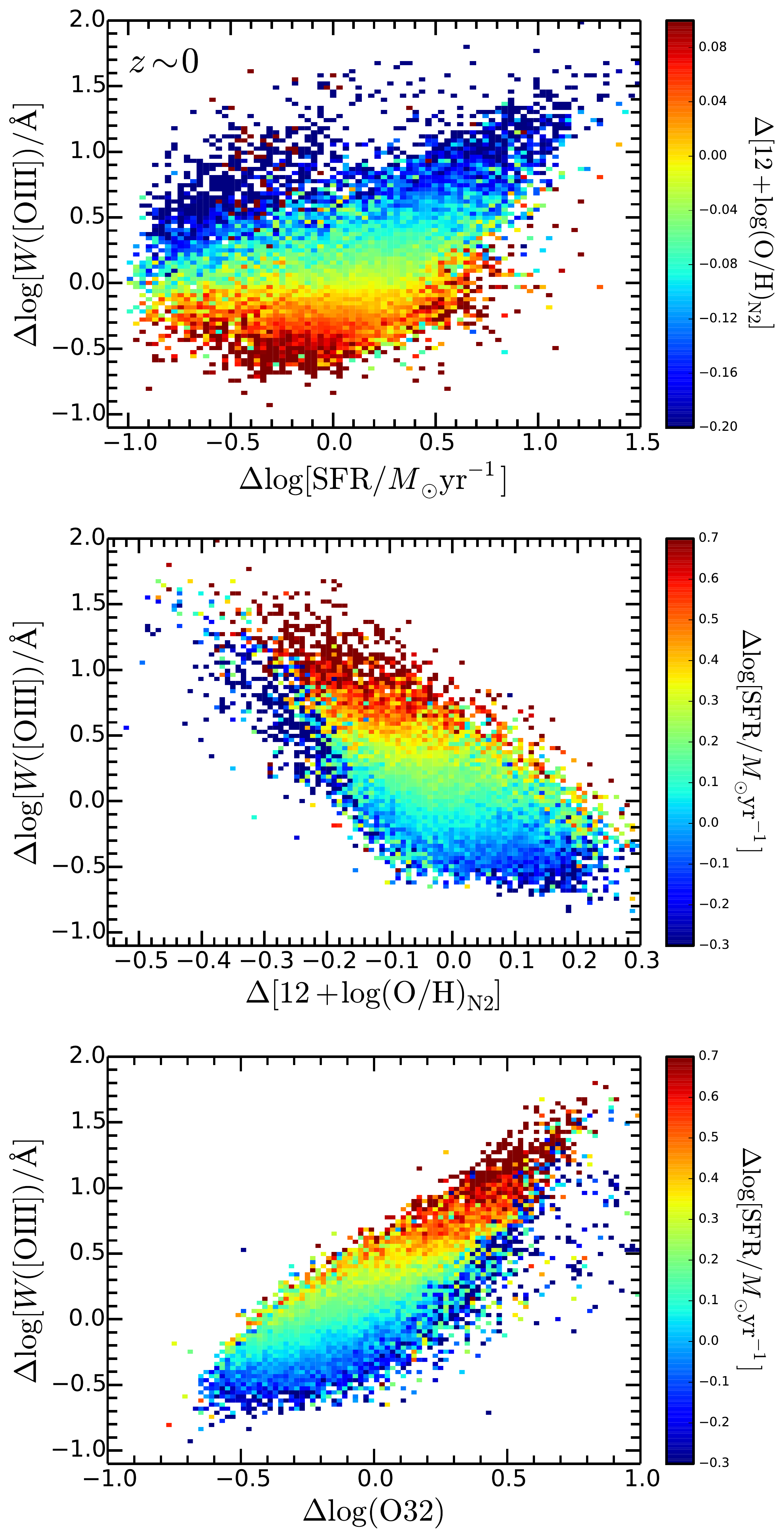}
\caption{Same as Figure~\ref{fig:residuals} for the $z\sim 0$ SDSS
  sample.  Shown are the residuals in \oiii\, equivalent width versus
  the residuals in SFR (top), $12+\log({\rm O/H})_{\rm N2}$ (middle),
  and $\log({\rm O32})$ (bottom).  The residuals are obtained in a
  manner similar to that of the MOSDEF sample, where the mean
  relations between stellar mass, metallicity, and SFR are taken from
  \citet{andrews13}, and the mean relation between O32 and stellar
  mass is taken from \citet{sanders16a}.  The same trends observed for
  the $z\sim 2$ MOSDEF sample are also observed locally: galaxies of a
  fixed sSFR have equivalent widths that increase with decreasing
  metallicity, galaxies of a fixed metallicity have equivalent widths
  that increase with SFR, and galaxies of a fixed ionization parameter
  have equivalent widths that increase with SFR.}
\label{fig:residuals_sdss}
\end{figure}

It is instructive to determine whether the high-redshift residual
trends are reflected in the local sample.  For the SDSS galaxies, the
residuals were computed in a manner similar to those for the
high-redshift sample, where we assumed the median relations between
stellar mass, SFR, and metallicity in \citet{andrews13}, and the
median relationship between O32 and stellar mass in
\citet{sanders16a}.  For the local galaxies, we find significant
correlations ($\sigma_{\rm P} \ga 4$; Figure~\ref{fig:residuals_sdss})
between $\delw$ and $\delsfr$, $\delz$, and $\delion$.  Moreover, the
local sample segregates in the same way as the high-redshift sample,
in the sense that the median value of $\delw$ increases with $\delz$
at a fixed $\delsfr$, while $\delw$ increases with $\delsfr$ at a
fixed $\delz$ or $\delion$.

\subsubsection{Summary}

The residual plots for the high-redshift and local
galaxies indicate that the equivalent width of \oiii\, is sensitive to
both the metallicity and SFR at a given stellar mass.  A $z\sim 2$
galaxy with $M_\ast = 10^{10}$\,$M_\odot$ has an average $\log[{\rm
    SFR}/M_\odot\,{\rm yr}^{-1}]$ that is $\simeq 1.0$\,dex larger,
and a $12+\log({\rm O/H})$ that is $\simeq 0.2$\,dex lower, than those
of a $z\sim 0$ galaxy with the same stellar mass, according to the
redshift evolution of the SFR-$M_\ast$ and MZR relations,
respectively.  The relations shown in Figure~\ref{fig:residuals} and
listed in Table~\ref{tab:residuals} indicate that these changes in SFR
and metallicity translate to a $\simeq 0.64$\,dex and a $\simeq
0.59$\,dex increase, respectively, in $\log[W(\oiii)/{\rm \AA}]$, for
a total increase of $\simeq 0.64 + 0.59 \approx 1.23$\,dex.  This
overall change in equivalent width is entirely consistent with the
$\delta\log[W(\oiii)/{\rm \AA}] \approx 1.3$\,dex offset indicated in
the evolution of $W(\oiii)$ versus $M_\ast$ with redshift shown the upper
rightmost panel of Figure~\ref{fig:ewsedmassz}.  Thus, the redshift
evolution in the SFR-$M_\ast$ and metallicity versus $M_\ast$
relations is sufficient to explain the increase in $W(\oiii)$ with
redshift at a fixed stellar mass.

\subsection{Trends for Emission Lines Other Than \oiii}

The other emission lines of interest are sensitive to SFR and
metallicity to varying degrees.  For example, not surprisingly,
$W(\ha)$ and $W(\hb)$ are more sensitive to SFR and less sensitive to
metallicity than $W(\oiii)$ is at a given stellar mass.  As a result,
the evolution in the $W(\ha)$ and $W(\hb)$ versus $M_\ast$ relations
is not as pronounced as it is for $W(\oiii)$ versus $M_\ast$, a result
noted in Section~\ref{sec:ewsedmass}.  For \oii, we find no
significant correlations between the residuals discussed above, a
finding that at face value suggests that $W(\oii)$ is not as sensitive
as $W(\oiii)$ is to either SFR or metallicity for the high-redshift
sample.  As a consequence, the $W(\oii)$ versus $M_\ast$ relations
among MOSDEF galaxies show markedly less evolution with redshift
relative to the $W(\oiii)$ versus $M_\ast$ relations
(Figure~\ref{fig:ewsedmassz}).  Nonetheless, when examined over a
large enough baseline in redshift or lookback time, $W(\oii)$ is
systematically larger for the MOSDEF galaxies relative to local
galaxies at a fixed stellar mass
(Figure~\ref{fig:ewsedmassz}).\footnote{Some of this evolution may be
  due to the evolving contribution of DIG emission to the \oii\, line
  (e.g., \citealt{sanders17}).}

\subsection{Other Redshift-dependent Trends}
\label{sec:betairxtrend}

There are several other trends noted in Section~\ref{sec:calibrations}
that can be explained by the dependence of equivalent width on SFR and
metallicity at a given stellar mass.  The rest-frame UV slope
($\beta$) is sensitive to the reddening, or $\ebmvstars$, and there is
a monotonic relation between the two for a given intrinsic stellar
population and dust attenuation curve.  Figure~\ref{fig:ewbetaphotz}
shows that for all of the lines of interest except \oii, the
equivalent width increases with redshift at a fixed $\beta$, or
reddening.

\begin{figure}
\epsscale{1.15}
\plotone{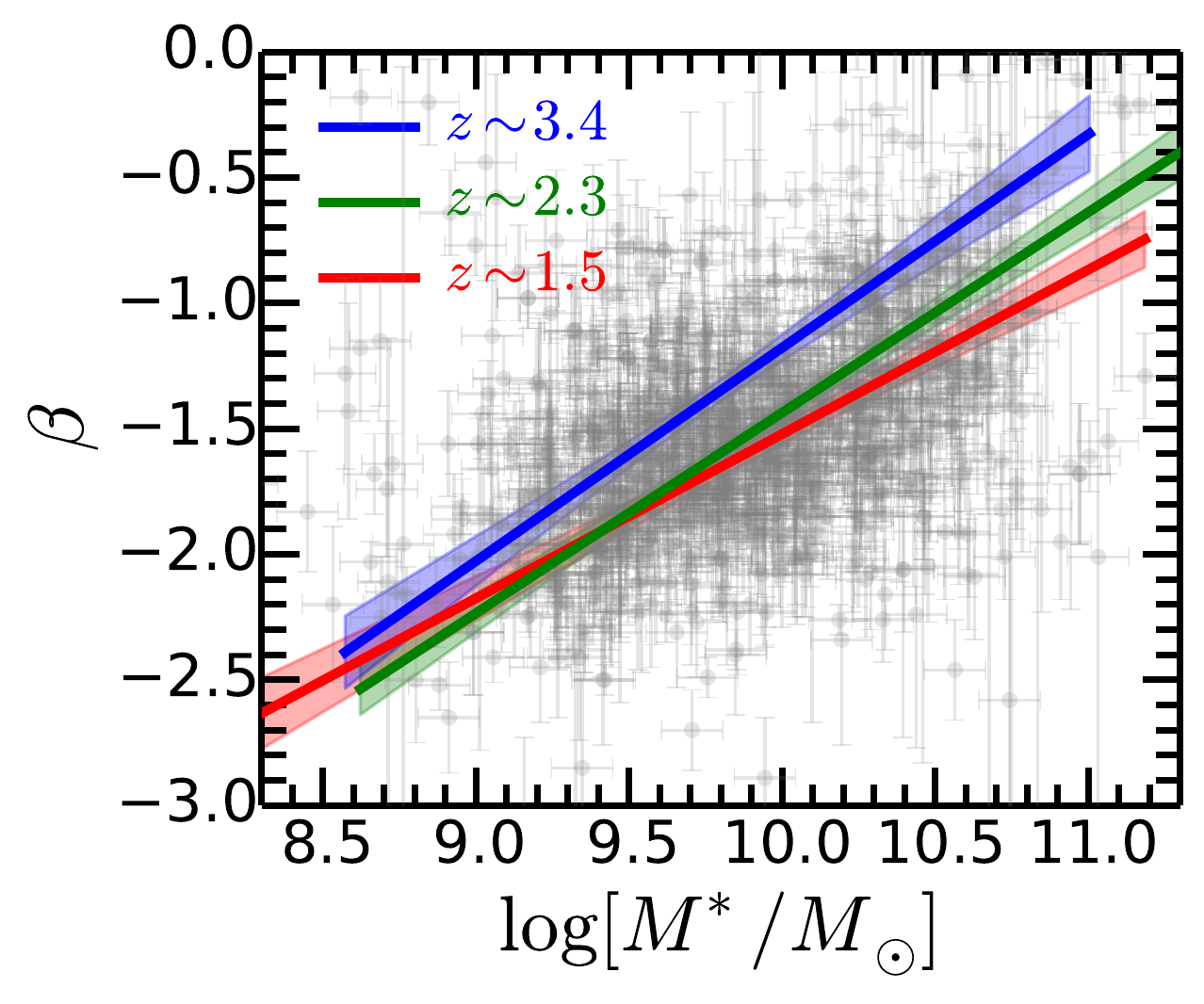}
\caption{Variation of UV slope, $\beta$, with stellar mass for the
  MOSDEF sample.  Spearman tests indicate a high degree of
  significance ($\sigma_{\rm P}\ga 10$) in the correlations between
  $\beta$ and $M_\ast$ for the low, middle, and high redshift
  subsamples.  The best-fit linear correlations their $95\%$
  confidence intervals are indicated by the red, green, and blue lines
  and shaded regions for the $z\sim 1.5$, $z\sim 2.3$, and $z\sim 3.4$
  subsamples, respectively.  The best-fit parameters are listed in
  Table~\ref{tab:betavsmass}.}
\label{fig:betavsmass}
\end{figure}

\begin{deluxetable}{lccrr}
\tabletypesize{\footnotesize}
\tablewidth{0pc}
\tablecaption{UV Slope versus Stellar Mass}
\tablehead{
\colhead{Redshift Subsample} &
\colhead{$\rho$\tablenotemark{a}} &
\colhead{$\sigma_{\rm P}$\tablenotemark{a}} &
\colhead{Intercept\tablenotemark{b}} &
\colhead{Slope\tablenotemark{b}}}
\startdata
$z\sim 1.5$ & 0.51 & 9.6 & $-8.06\pm 0.42$ & $0.65\pm 0.04$ \\
$z\sim 2.3$ & 0.44 & 10.7 & $-9.42\pm 0.39$ & $0.80\pm 0.04$ \\
$z\sim 3.4$ & 0.55 & 8.7 & $-9.67\pm 0.57$ & $0.85\pm 0.06$ 
\enddata
\tablenotetext{a}{Spearman rank correlation coefficient and the number of standard deviations 
by which the correlation
deviates from the null hypothesis of no correlation.}
\tablenotetext{b}{Intercept and slope of the best-fit linear function between 
$\beta$ and $\log[M_\ast/M_\odot]$.}
\label{tab:betavsmass}
\end{deluxetable}

The observed redshift evolution of the trends between $W$ and $\beta$
is a logical outcome of the dependence of UV slope (or reddening) on
stellar mass which, for the MOSDEF sample, is shown in
Figure~\ref{fig:betavsmass}.  A formal fit to the data indicates that
$\beta$ is correlated with $M_\ast$ with $\ga 8\sigma$ significance
for the $z\sim 1.5$, $z\sim 2.3$, and $z\sim 3.4$ subsamples
(Table~\ref{tab:betavsmass}).  Between the two lower redshift
subsamples, galaxies of a given $\beta$ will on average have roughly
the same stellar mass, while there is some indication that galaxies in
the highest redshift subsample have stellar masses that are a factor
of $2-3\times$ lower than those of $z\sim 2$ galaxies at a given
$\beta$.  Or, alternatively, the $z\sim 3.4$ galaxies have $\beta$
that are on average $\delta\beta \simeq 0.2$ redder than $z\sim 2$
galaxies of the same stellar mass.

At face value, the apparent redshift evolution in the $\beta$ versus
$M_\ast$ relation runs counter to the conclusions of a number of
studies that have suggested that the correlation between stellar mass
and dust attenuation---which is directly related to $\beta$ and is
typically parameterized by the ratio of the infrared-to-UV luminosity
(IRX; \citealt{meurer99})---does not evolve with
redshift \citep{pannella09, reddy10, bouwens16a, reddy18}.  However,
the relationship between IRX and $\beta$ may evolve with redshift
(e.g., \citealt{bouwens16a, reddy18}).  For example, if higher
redshift galaxies follow steeper attenuation curves than lower
redshift galaxies of a fixed stellar mass, then the higher redshift
galaxies will exhibit redder $\beta$ for a given IRX (e.g., see
\citealt{reddy18} for further discussion).  Thus, an evolution of
$\beta$ with redshift at a fixed stellar mass does not necessarily 
imply that IRX evolves in the same manner, and in fact, the relation
between IRX and stellar mass does not appear to evolve strongly
with redshift as noted above.

Viewed in this context of the strong correlation between $\beta$ and
$M_\ast$ and the fact that the correlation does not evolve strongly
with redshift (or, that it evolves toward lower stellar masses at a
given $\beta$), it follows that the increase in $W$ with redshift at a
given stellar mass will translate to an increase in $W$ with redshift
at a given $\beta$, for essentially the same reasons delineated in the
previous section.

Finally, a consequence of the redshift evolution in the SFR-$M_\ast$ relation
is that high-redshift galaxies have on average lower stellar masses
than local galaxies at a fixed SFR.  Therefore, the increasing
normalization of the $W$ versus SFR relation with redshift (e.g.,
Figure~\ref{fig:ewsfrhaz}) reflects the
trend of higher equivalent widths at lower stellar masses and higher
redshifts, precisely the behavior shown in
Figure~\ref{fig:ewsedmassz}.

\subsection{Redshift Evolution in the Trends between Equivalent Width and 
ISM Physical Conditions}
\label{sec:discmets}

\subsubsection{Equivalent Width versus Metallicity}
\label{sec:ewversusmetallicity}

We can use similar lines of reasoning as those invoked above to
explain the redshift evolution in the relationship between equivalent
width and oxygen abundance, at least as determined through nitrogen
lines (Figures~\ref{fig:ewn2met} and \ref{fig:ewo3n2met}).  An outcome
of the redshift evolution in the MZR \citep{sanders18} is that
galaxies of a fixed N2-inferred (or O3N2-inferred) oxygen abundance of
$12+\log({\rm O/H}) = 8.45$---roughly the mean abundance for the
$z\sim 2$ sample---have an average $\log[M_\ast/M_\odot]$ that is
$\approx 1$\,dex larger than that of $z\sim 0$ galaxies.  On the other
hand, such galaxies at $z\sim 2$ have SFRs that are {\em two} orders
of magnitude larger than those of $z\sim 0$ galaxies at the same
metallicity.  Thus, the stellar mass increases with redshift by a
modest amount relative to the SFR at a fixed (N2-based) metallicity,
resulting in an overall increase in sSFR of $\approx 1$\,dex with
redshift.  According to the relations listed in
Table~\ref{tab:relations_sedparms} for $\log[{\rm sSFR[H\alpha]}/{\rm
    yr}^{-1}]$, this increase in sSFR results in an $\approx 0.5$\,dex
increase in $\log[W(\oiii)/{\rm \AA}]$, similar to the difference in
$\log[W(\oiii)/{\rm \AA}]$ between $z\sim 2$ and $z\sim 0$ galaxies
for a fixed oxygen abundance of $12+\log({\rm O/H})_{\rm N2} = 8.45$
(Figure~\ref{fig:ewn2met}).

\begin{figure}
\epsscale{1.15}
\plotone{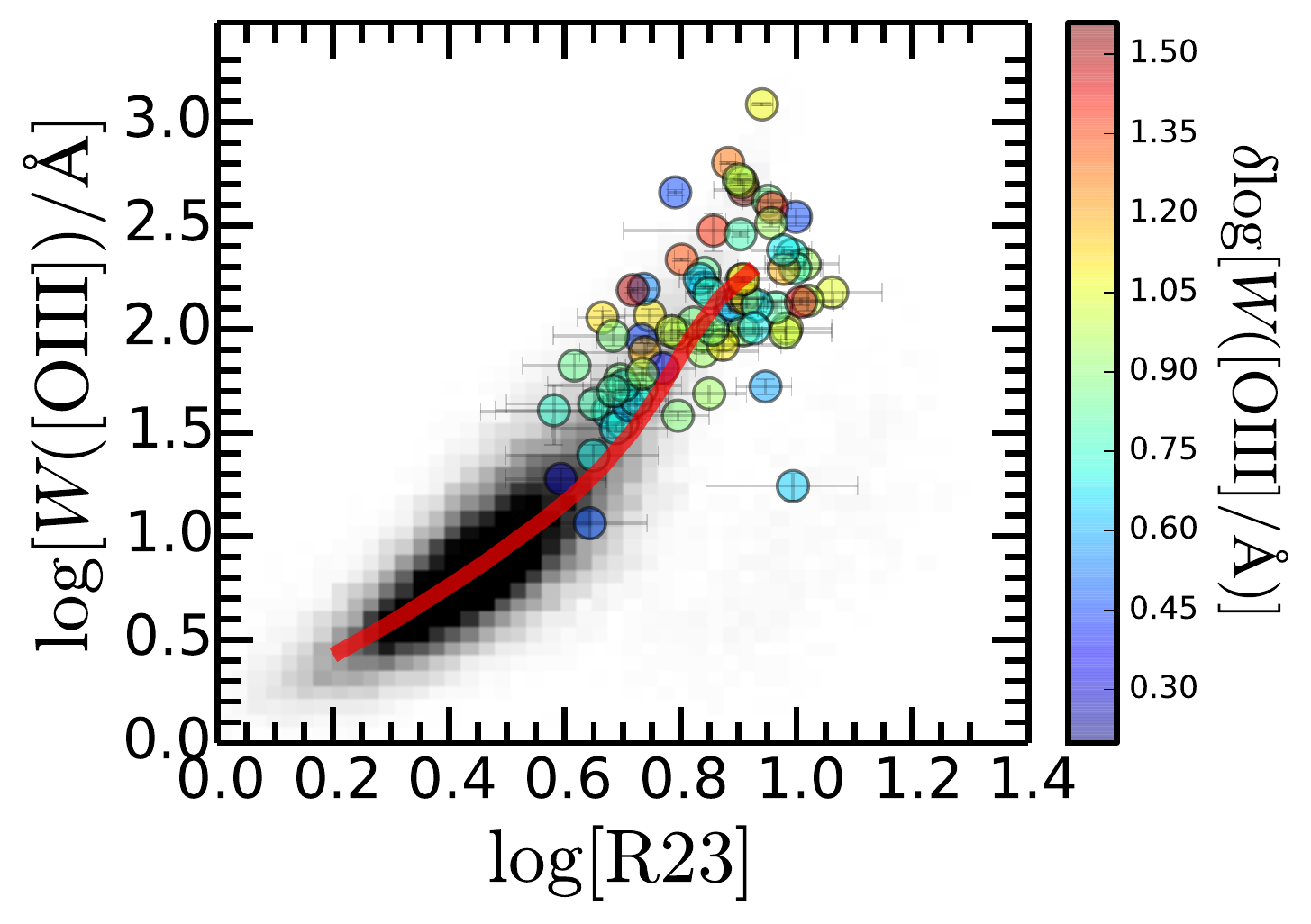}
\caption{Equivalent width of the \oiii\, line versus the R23
  metallicity index for the 114 galaxies where \oii, \hb, \oiii, \ha,
  and \nii\, are significantly detected, color-coded by their offset
  in $\log[W(\oiii)/{\rm \AA}]$ from the median trend of
  $\log[W(\oiii)/{\rm \AA}]$ versus $12+\log({\rm O/H})_{\rm N2}$ for
  local SDSS galaxies (red line in the upper rightmost panel of
  Figure~\ref{fig:ewn2met}).  Also shown is the distribution of R23
  for the SDSS sample, corrected for DIG emission and flux-weighting
  effects \citep{sanders17}, with the running median for the local
  galaxies indicated by the red line.}
\label{fig:r23}
\end{figure}

Several studies have found that high-redshift galaxies exhibit larger
N2 ratios at a given oxygen abundance and/or are offset toward higher
N2 in the BPT plane \citep{kewley13,masters14,steidel14, shapley15,
  sanders16a}.  Various possible explanations have been given for this
offset, including elevated nitrogen abundances at a fixed oxygen
abundance \citep{masters14, shapley15}, or differences in ISM
conditions, including gas pressure \citep{kewley13} or ionization
parameter \citep{steidel14}, and/or differences in stellar populations
\citep{masters14, steidel14}.  Thus, metallicity estimates based on N2
may be biased to higher values than metallicities computed using other
strong lines for galaxies of a fixed ionizing spectrum (e.g.,
\citealt{steidel16}).  We find some indication of this metallicity
bias in our analysis.  For example, we find that the trend between
$W(\oiii)$ and $12+\log({\rm O/H})_{\rm O3N2}$ evolves less strongly
with redshift than the trend between $W(\oiii)$ and $12+\log({\rm
  O/H})_{\rm N2}$, a difference that may be related to the fact that
the O3N2 index is less sensitive to the nitrogen lines than N2.  In
view of the N2-based metallicity offsets, it is useful to examine
whether the offsets in $W$ at a given N2- or O3N2-inferred oxygen
abundance between high- and low-redshift galaxies persist when using
abundance indicators that do not involve nitrogen lines.

To this end, we calculated the R23 values for the local SDSS galaxies
as well as the 114 MOSDEF galaxies with direct detections of the lines
used in calculating this index---i.e., \oii,\oiii,\hb, and \ha\,
(\ha\, is required in addition to \hb\, in order to correct the lines
for dust attenuation based on the Balmer decrement)---as well as
detections of \nii\, (i.e., to ensure that the galaxies have measured
N2 values).  As noted previously, the line ratios (including R23) were
corrected for the effects of DIG emission and flux weighting using the
prescriptions of \citet{sanders17}.  As discussed in \citet{shapley15}
(see also \citealt{sanders16a}), galaxies follow a monotonic increase
in metallicity proceeding from high O32 and R23 values to low O32 and
R23 values (e.g., \citealt{sanders16a}).  From Figure~\ref{fig:ewo32},
we see that $W(\oiii)$ correlates tightly with O32
(Figure~\ref{fig:ewo32}).  It then follows that there will be a
similar monotonic increase in metallicity as $W(\oiii)$ and R23
decrease in tandem.

Figure~\ref{fig:r23} shows that the $z\sim 2$ galaxies mostly follow
the local relationship between $\log[W(\oiii)]$ and R23.  In
particular, we find that those $z\sim 2$ galaxies that exhibit the
largest offsets in their \oiii\, equivalent widths relative to local
SDSS galaxies at the same $12+\log({\rm O/H})_{\rm N2}$ scatter around
the median trend of $\log[W(\oiii)/{\rm \AA}]$ versus $\log[{\rm R23}]$
for $z\sim 0$ galaxies.\footnote{If the line ratios of the SDSS
  galaxies are not corrected for DIG emission and flux weighting, the
  relationship between $\log[W(\oiii)]$ and $\log[R23]$ for the local
  sample would be shifted by $\simeq 0.1$\,dex to larger $\log[R23]$.
  In this case of no correction, we find that at a fixed $\log[{\rm
      R23}] = 0.8$, corresponding roughly to $12+\log({\rm O/H})_{\rm
    R23} \simeq 8.5$, galaxies at $z\sim 2$ have an average
  $\log[W(\oiii)]$ that is $\approx 0.3$\,dex larger than that of
  local galaxies.  This offset is smaller than the $\approx 0.6$\,dex
  offset between $W(\oiii)$ versus $12+\log({\rm O/H})_{\rm N2}$ at a
  similar oxygen abundance, implying a smaller degree of redshift
  evolution in $W(\oiii)$ versus $12+\log({\rm O/H})_{\rm R23}$
  relative to the redshift evolution in $W(\oiii)$ versus
  $12+\log({\rm O/H})_{\rm N2}$.}  In summary, our results suggest
that, when using metallicity indicators that do not depend on nitrogen
lines, the relationship between equivalent width and metallicity at
$z\sim 2$ is similar to that seen at $z\sim 0$, or at the least, does
not evolve with redshift as strongly as that observed when using
N2-based metallicities.  Further evidence supporting this conclusion
is provided by an analysis of the relationship between $W$ and O3, as
the latter is also sensitive to metallicity (see next section).

On a note of caution, the sensitivity of the R23 index to
metallicity is limited over the range of metallicities represented in
the MOSDEF sample, as it is for this range that the R23 index
effectively saturates.  Going forward, robust calibrations of
strong-line metallicity indicators---such as those obtained through
[\ion{O}{3}]$\lambda 4364$ auroral line measurements of electron
temperatures---will be crucial for robustly quantifying some of the
redshift offsets in $W$ versus metallicity discussed here.  We note that
the aforementioned bias in N2-based metallicities does not affect our
analysis of the correlations between residuals in equivalent width,
metallicity, and SFR presented in Section~\ref{sec:residuals}, since
these correlations are examined in a relative sense at a fixed
redshift (e.g., \citealt{sanders18}).

\subsubsection{Equivalent Width versus Excitation Conditions}

\begin{figure}
\epsscale{1.15}
\plotone{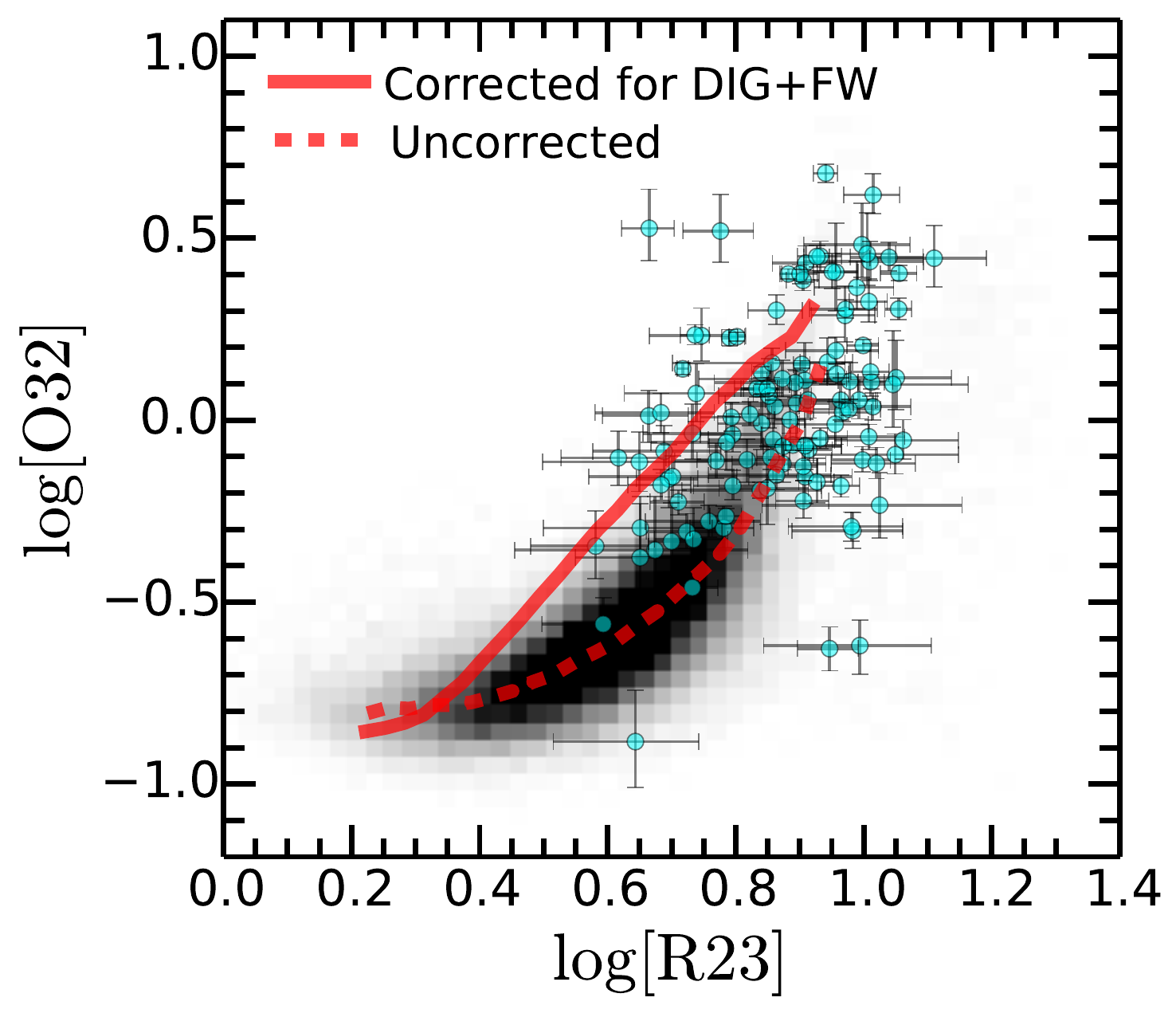}
\caption{Relationship between $\log[{\rm O32}]$ and $\log[{\rm R23}]$
  for the $z\sim 2$ MOSDEF galaxies, shown by the cyan circles.  For
  comparison, we also show the median relationship for SDSS galaxies
  when correcting both $\log[{\rm O32}]$ and $\log[{\rm R23}]$ for DIG
  emission and flux-weighting effects using the prescriptions of
  \citet{sanders17} (solid line).  The grayscale distribution shows
  $\log[{\rm O32}]$ versus $\log[{\rm R23}]$, with the median relation
  indicated by the dashed line, when leaving these quantities
  uncorrected for DIG emission and flux-weighting effects.  A full
  discussion of how the inferred redshift evolution in O32 versus R23
  depends on corrections made to these ratios for DIG emission and
  flux-weighting effects will be discussed elsewhere (Shapley et~al. 2018,
  in preparation).}
\label{fig:o32vsr23}
\end{figure}

A primary conclusion of Section~\ref{sec:residuals} is that the
equivalent width of $\oiii$ is sensitive to both SFR and metallicity
at a given stellar mass.  Here, we focus on the extent to which the
sensitivity of $W$ to SFR and metallicity translates to the observed
behavior of $W$ versus O32 and O3.  Specifically, the anti-correlation
between ionization parameter and metallicity implies that the redshift
invariance of the $W$ versus R23-based-metallicity relation
(Figure~\ref{fig:r23}) should translate to a similar redshift
invariance of the $W$ versus O32 relation if we assume that the O32
versus R23 relation does not evolve with redshift.  However, this
expected behavior is not observed in Figure~\ref{fig:ewo32}.  Rather,
we observe some mild redshift evolution in the $W$ versus O32
relation.  The redshift evolution in $W(\oiii)$ versus O32 and the
apparent redshift invariance of $W(\oiii)$ versus R23 implies that the
$O32$ versus R23 relation evolves with redshift, as shown in
Figure~\ref{fig:o32vsr23}.  Interestingly, if we do not correct the
SDSS galaxies for DIG emission and flux-weighting effects, the
resulting relation between $\log[W(\oiii)]$ and $\log[{\rm O32}]$ for
the local sample provides an adequate description of this same
relation for the $z\sim 2$ galaxies.

On the other hand, the O3 ratio---sensitive to both ionization
parameter and metallicity---is not as significantly affected by DIG
emission and flux-weighting effects as O32.  Indeed, for local
galaxies, the O3 ratio is biased $<0.1$\,dex over the range of
excitation and metallicity displayed by the majority of $z\sim 0$ SDSS
galaxies \citep{sanders17}.  We find that the $\log[W(\oiii)]$ versus
$\log[{\rm O3}]$ relation does not evolve significantly, on average,
between the low- and high-redshift samples (Figure~\ref{fig:ewo3hb}).
This fact, combined with the observation that the $\log[W(\oiii)]$
versus $12+\log({\rm O/H})_{\rm O3N2}$ does not evolve with redshift
as strongly as the $\log[W(\oiii)]$ versus $12+\log({\rm O/H})_{\rm
  N2}$ relation, suggests that the observed redshift evolution in the
latter (i.e., $\log[W(\oiii)]$ versus $12+\log({\rm O/H})_{\rm N2}$)
results from a systematic bias in the nitrogen-based metallicity
indicators at high redshift.  A full analysis of the relationships
between O32, O3, and metallicity, and how these relations are affected
by DIG, flux-weighting effects, and shock emission, is beyond the
scope of this paper and will be presented elsewhere (Shapley et
al. 2018, in preparation).  Nonetheless, at face value, our results
suggest a redshift invariance in both $\log[W(\oiii)]$ versus and
$\log[{\rm R23}]$ and $\log[W(\oiii)]$ versus $\log[{\rm O3}]$,
implying that the metallicity plays a fundamental role in modulating
the \oiii\, equivalent width.  This conclusion is an accordance with
expectations that lower stellar metallicities result in a harder
ionizing spectrum and hence larger \oiii\, equivalent widths when
keeping both stellar mass and SFR fixed (e.g.,
Figure~\ref{fig:cloudy}).  As noted in Section~\ref{sec:residuals},
SFR and stellar mass are additional variables that will affect the
equivalent widths, an issue that we discuss further in the next
section.

\subsection{Redshift Invariant Trends with sSFR and Age}
\label{sec:ewssfrmet}

\begin{figure}
\epsscale{1.15}
\plotone{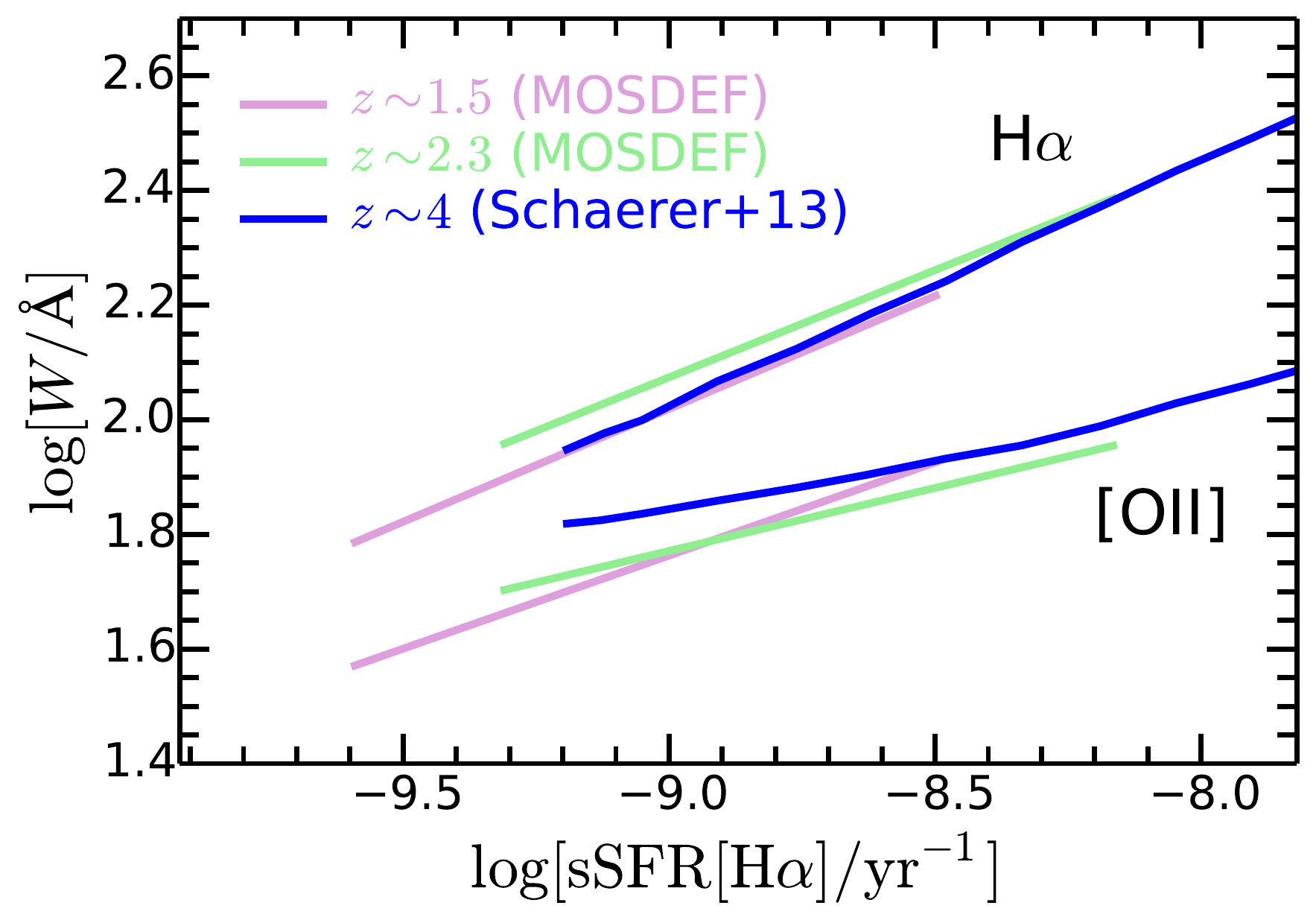}
\caption{Average trends of \ha\, and \oii\, equivalent widths versus $\ssfrha$
for galaxies in the two lower redshift bins of the MOSDEF
sample ($z\sim 1.5$ and $z\sim 2.3$) and the $z\sim 4$ {\em B}-dropout
sample analyzed by \citet{schaerer13}.  The dispersion of objects around
these average trends is typically of order $\simeq 0.2$\,dex.}
\label{fig:compbdrops}
\end{figure}

\begin{figure*}[!t]
\epsscale{1.15}
\plotone{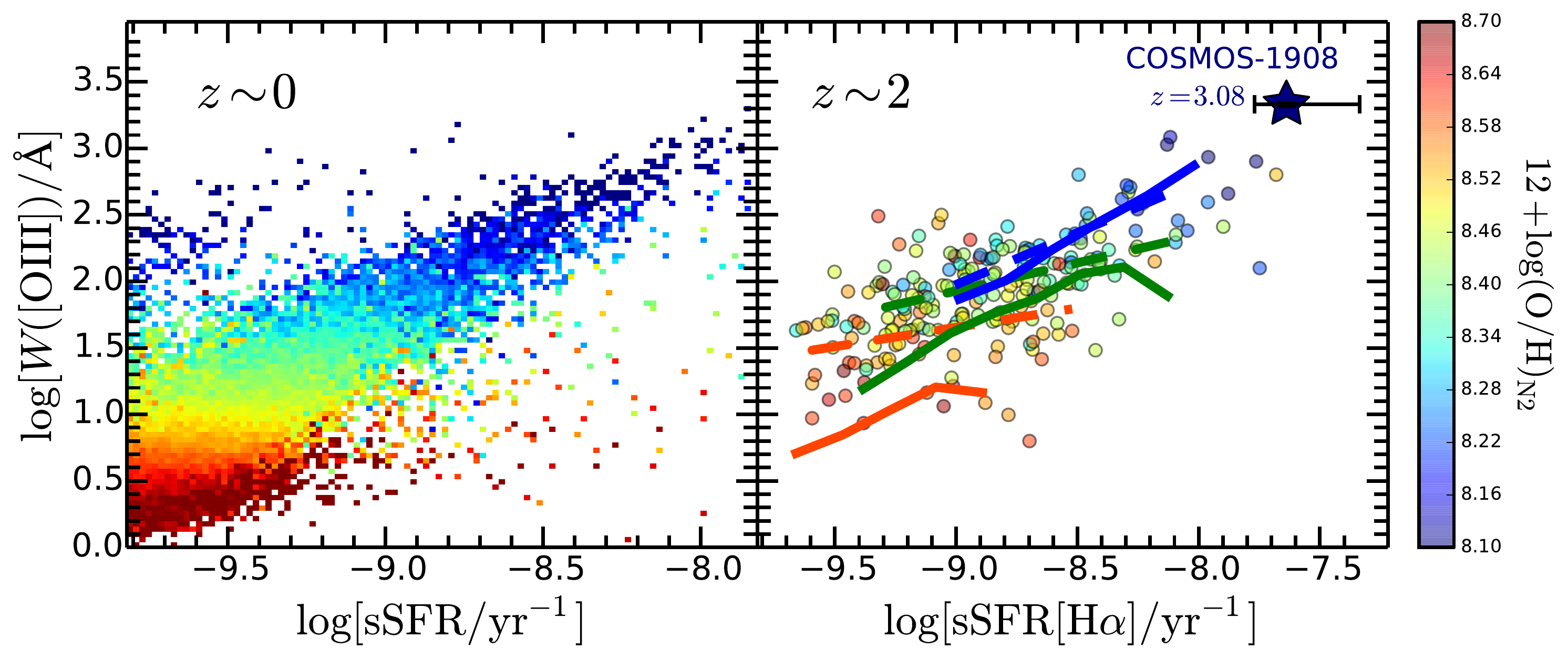}
\caption{{\em Left:} relationship between $\log[W(\oiii)/{\rm \AA}]$
  and $\log[{\rm sSFR}/{\rm yr}^{-1}]$ for $z\sim 0$ galaxies from the
  SDSS, color-coded by N2-inferred oxygen abundance.  {\em Right:}
  same as the {\em left} panel, for the $z\sim 2$ galaxies.  The solid
  lines indicate the median trend of $\log[W(\oiii)/{\rm \AA}]$ with
  $\log[{\rm sSFR}/{\rm yr}^{-1}]$ for SDSS galaxies in three bins of
  oxygen abundance: $12+\log[{\rm O/H}] < 8.26$, $8.26\le 12+\log[{\rm
      O/H}] < 8.49$, and $12+\log[{\rm O/H}] \ge 8.49$.  The dashed
  lines show the best-fit linear relations between $\log[W(\oiii)/{\rm
      \AA}]$ and $\log[{\rm sSFR}/{\rm yr}^{-1}]$ for the $z\sim 2$
  galaxies in the same bins of oxygen abundance specified above.  The large
star shows the position of the $z=3.08$ galaxy, COSMOS-1908, with a direct
electron temperature-based metallicity from \citet{sanders16b}.}
\label{fig:ewssfrmet}
\end{figure*}

\subsubsection{Redshift Invariance of $W$ versus sSFR at $1.5\la z\la 4$}

As noted in Section~\ref{sec:calibrations}, the equivalent width
depends on those properties that are sensitive to spectral shape, or
mass-to-light ratio, such as sSFR and age.  Furthermore, the
relationship between equivalent width and sSFR or age for
high-redshift galaxies is similar to that seen for local galaxies.
The redshift invariance of the $W$ versus sSFR relationship persists
at even higher redshifts.  Based on simultaneous fitting of the
stellar and nebular continuum and nebular line emission in a sample of
$z\sim 4$ galaxies, \citet{schaerer13} derived relationships between
sSFR and the equivalent widths of both \ha\, and \oii\, for a number
of different star formation histories.  As shown in
Figure~\ref{fig:compbdrops}, the relationships between equivalent
width and sSFR derived for a constant star-formation history for the
$z\sim 4$ galaxies are consistent with the mean trends found for
galaxies at $z\sim 1.5$ and $z\sim 2.3$ from the MOSDEF sample.

\subsubsection{$W$ versus sSFR at a Fixed Metallicity}

It is of interest to examine in more detail how the relationship
between equivalent width and sSFR segregates by metallicity.  In
particular, the equivalent width is sensitive to both line luminosity
and continuum luminosity density.  Hence, $W$ is, not surprisingly,
correlated with both SFR and stellar mass, as shown previously.
However, the equivalent width (e.g., of \oiii), and in particular the
line luminosity, is also sensitive to metallicity, as demonstrated in
Figure~\ref{fig:residuals}.  Figure~\ref{fig:ewssfrmet} shows how
$z\sim 0$ and $z\sim 2$ galaxies segregate by metallicity in \oiii\,
equivalent width versus sSFR space.

In both the low- and high-redshift samples, there is a monotonic
decrease in metallicity with both increasing sSFR and equivalent
width.  Also shown in the right panel of Figure~\ref{fig:ewssfrmet}
are the median trends of equivalent width versus sSFR for galaxies in
three bins of oxygen abundance for the high redshift sample (dashed
lines) and the SDSS sample (solid lines).\footnote{Note that the
  slopes of the median trends of equivalent width versus sSFR for
  galaxies in different bins of metallicity depend on the widths of
  the bins: the slopes decrease in magnitude with decreasing bin
  width.  The sensitivity of these slopes to bin width applies to both
  the $z\sim 0$ and $z\sim 2$ trends, such that the relative
  comparison of the low- and high-redshift slopes remains valid.}  In
comparing the local and high-redshift samples, we find that the median
trends of equivalent width with sSFR for the two lower bins of oxygen
abundance for the $z\sim 0$ galaxies are roughly within a factor of
two of the best-fit linear trends between these quantities in the same
bins of oxygen abundance for the $z\sim 2$ galaxies.  There is a more
significant ($\simeq 0.3-0.5$\,dex) offset in $\log[W(\oiii)]$ versus
$\log[\ssfrha]$ for the bin of highest metallicity.  This offset could
plausibly be explained by an enhanced nitrogen abundance in
high-redshift galaxies relative to local ones at a fixed oxygen
abundance.  Specifically, Figure~\ref{fig:ewn2met} shows that the
largest offset between the high-redshift and local trends between
$\log[W(\oiii)]$ and $12+\log({\rm O/H})_{\rm N2}$ occurs for galaxies
at higher metallicities.  However, taking into consideration the
apparent redshift invariance of the relationship between
$\log[W(\oiii)]$ and $12+\log({\rm O/H})$, where the latter is
determined from lines other than nitrogen
(Section~\ref{sec:ewversusmetallicity}), we find that the
high-redshift galaxies segregate with metallicity in $\log[W(\oiii)]$
versus $12+\log({\rm O/H})$ space in the same way as seen for local
galaxies.  In other words, at a fixed sSFR and metallicity, $z\sim 2$
galaxies have \oiii\, equivalent widths that are within a factor of
two of those of $z\sim 0$ galaxies.

\begin{figure}
\epsscale{1.15}
\plotone{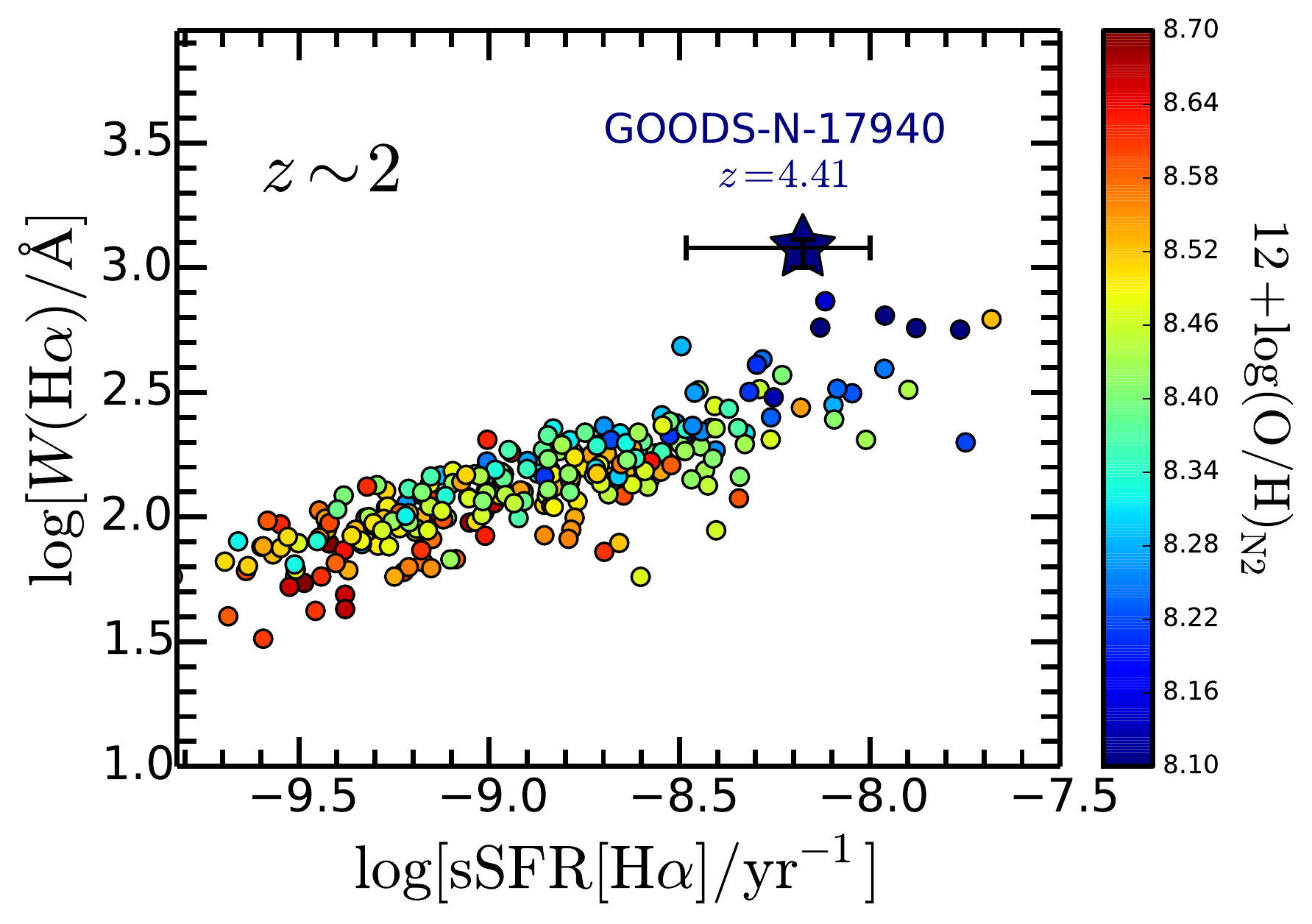}
\caption{Same as the right panel of Figure~\ref{fig:ewssfrmet} for $\log[W(\ha)]$ versus
$\log[\ssfrha]$, again color-coded by N2-inferred oxygen abundance.  For clarity,
the median trends for the $z\sim 0$ and $z\sim 2$ samples are not shown.  
The single galaxy for which a nebular oxygen abundance has been measured at $z\ga 4$ 
\citep{shapley17} is shown by the large star.}
\label{fig:haewssfrmet}
\end{figure}

We also note that the $W$ versus sSFR relationship appears to
separate with metallicity at $z\ga 3$ in the same way as seen for
$z\sim 2$ galaxies.  For instance, COSMOS-1908---one of the galaxies
in the MOSDEF survey for which the auroral line $\oiii\lambda 4364$ is
detected---has a redshift of $z=3.08$, a direct electron temperature
based metallicity of $12+\log({\rm O/H}) = 8.00$, and $W(\oiii)\simeq
2140$\,\AA\, \citep{sanders16b}.  This object's sSFR ($\simeq
2.3\times 10^{-8}$\,yr$^{-1}$) places it in the same region of the
$W(\oiii)$ versus sSFR relation as lower redshift ($z\sim 2$) galaxies of
the same metallicity (Figure~\ref{fig:ewssfrmet}).
Figure~\ref{fig:haewssfrmet} shows the relationship between $W(\ha)$
and $\ssfrha$ for the MOSDEF sample, again color-coded by N2-based
metallicities.  There is a single galaxy for which a nebular oxygen
abundance has been derived at $z\ga 4$, namely, GOODS-N-17940, also
from the MOSDEF survey.  Using the excess emission in the {\em
  Spitzer}/IRAC 3.6\,$\mu$m band relative to the surrounding
photometry for this object, \citet{shapley17} derived an H$\alpha$
(rest-frame) equivalent width of $W(\ha) \simeq 1200$\,\AA\, and a
$\ssfrha \simeq 9.67\times 10^{-9}$\,yr$^{-1}$.\footnote{The
  H$\alpha$-based sSFR was derived by assuming a conversion from
  $L(\ha)$ to $\sfrha$ consistent with that of our fiducial model, and
  assuming the Balmer decrement measured from the H$\gamma$ and
  H$\delta$ lines to correct for dust.  Similarly, the stellar mass
  was computed assuming our fiducial model.}  Based on spectroscopic
detections of \oii\, and \neiii, \citet{shapley17} derived an oxygen abundance
of $12+\log({\rm O/H})\simeq 8.00$, again placing this object in approximately
the same region of the $W(\ha)$-$\ssfrha$ parameter space occupied by lower
redshift objects of the same metallicity.

\subsubsection{Summary}

Based on the relationships between equivalent width and sSFR for local
galaxies, $z\sim 2$ galaxies from the MOSDEF sample, two galaxies with
metallicity determinations at $z\ga 3$, and the $z\sim 4$ galaxies
analyzed by \citet{schaerer13}, we conclude that the $W$ versus sSFR
relation is largely invariant with redshift, particularly when
considering galaxies of the same metallicity.  This conclusion is
supported by the residual analysis presented previously, which
indicates that the line equivalent width is approximately fixed for a
given sSFR and metallicity.

Finally, the tight relationship between equivalent width and sSFR at a
given metallicity implies a similarly tight relationship between
equivalent width and mass-to-light ($M/L$) ratio.  For the constant
star-formation history assumed in our analysis, the age of the stellar
population is directly related to $M/L$, modulo the effects of dust.
As a consequence, the equivalent width correlates with age in a manner
that is independent of the redshifts of the galaxies, as discussed in
Section~\ref{sec:ewsedage}.

\subsection{Redshift Evolution of Equivalent Widths}

\subsubsection{$W(\oiii)$ versus $M_{\ast}$}

\begin{figure}
\epsscale{1.20}
\plotone{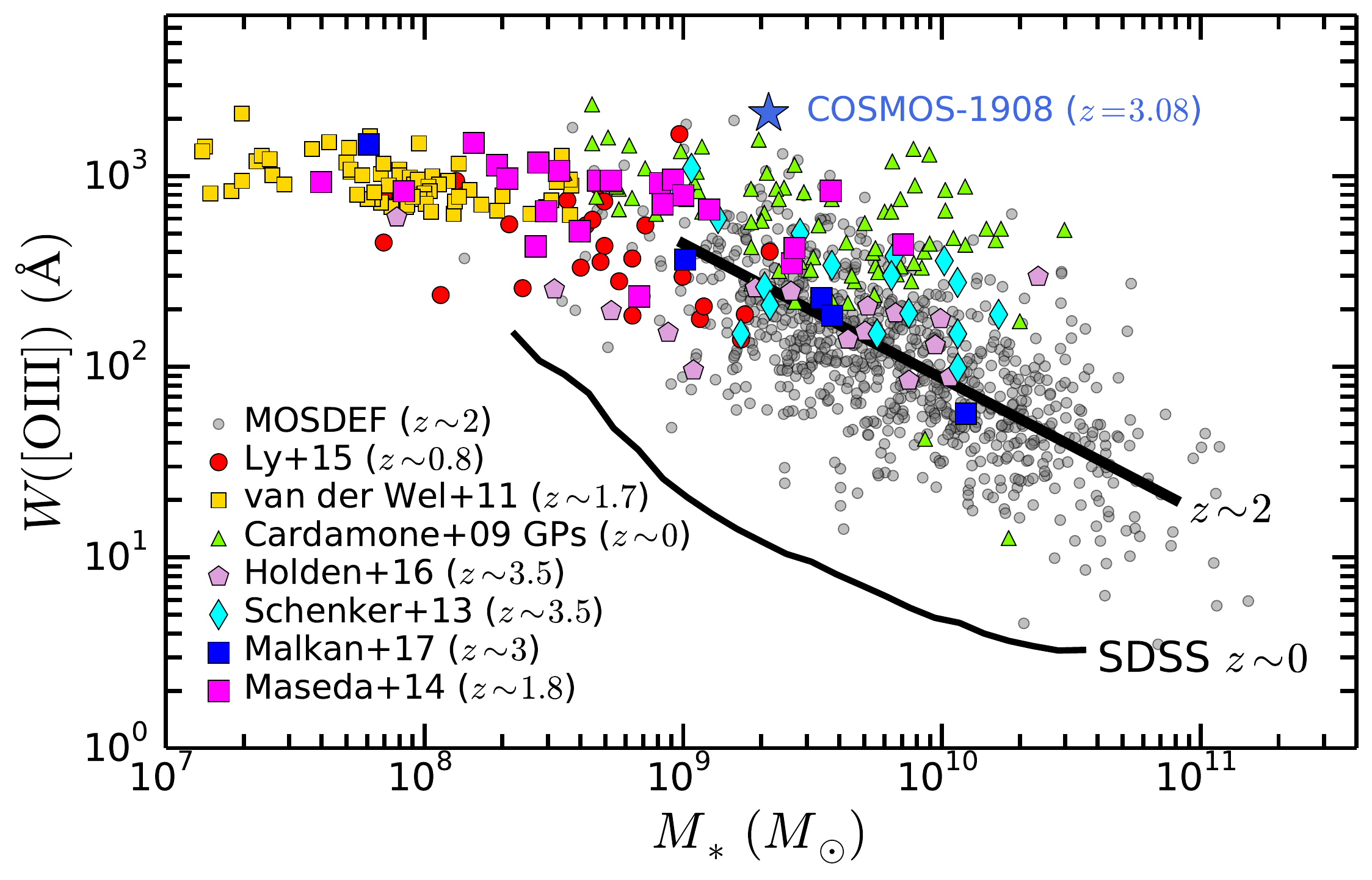}
\caption{Rest-frame equivalent width of the $\oiii\lambda\lambda
  4960,5008$ doublet versus stellar mass from MOSDEF (excluding upper
  limits in $W(\oiii)$, for clarity), compared with values from the
  literature at lower and similar redshifts.  These include the
  $\oiii\lambda 4364$-detected sample from \citet{ly15}, extreme
  emission line galaxy samples of \citet{vanderwel11} and
  \citet{maseda14}, the Green Peas sample of \citet{cardamone09}, and
  the LBG-selected samples of \citet{schenker13}, \citet{holden16},
  and \citet{malkan17}.  Also shown is COSMOS-1908 from
  \citet{sanders16b}.  For those studies that included only the
  $5008$\,\AA\, line, the equivalent widths were multiplied by a
  factor of $1.34$ to account for the $4960$\,\AA\, emission line.
  The thick and thin solid lines indicate the average relationships
  between $W(\oiii)$ and stellar mass for MOSDEF galaxies (including
  those with upper limits in $W(\oiii)$) and SDSS galaxies,
  respectively.}
\label{fig:woiiivsmass}
\end{figure}

The tight correlation between rest-frame optical emission line
equivalent widths and sSFR (e.g., Figures~\ref{fig:ewssfrhaz},
\ref{fig:ewssfrmet}, and \ref{fig:haewssfrmet}), and the fact that the
former do not depend sensitively on model assumptions, have prompted
the use of equivalent width measurements to infer the redshift
evolution in the sSFR of galaxies, giving insights into the typical
star-formation histories of such galaxies (e.g., \citealt{papovich11,
  reddy12b, gonzalez12, bouwens12, stark13, gonzalez14, smit14,
  smit16, faisst16}).  In this context, we compare the \oiii\,
equivalent width measurements from the MOSDEF survey with others from
a subset of the literature at similar (and lower) redshifts in
Figure~\ref{fig:woiiivsmass}.  Note that some of these studies were
based on samples that targeted preferentially those galaxies with
large equivalent widths \citep{cardamone09, vanderwel11, maseda14} or
required the detection of the $\oiii\lambda 4364$ auroral line
\citep{ly15}.  The latter will tend to select those galaxies with the
lowest metallicities and brightest $\oiii$ lines.  As such, the
relationships between equivalent width and stellar mass derived in
these studies may not be reflective of the relationships for all
star-forming galaxies at the same redshifts.  Not shown in
Figure~\ref{fig:woiiivsmass} are combined $\oiii+\hb$ equivalent
widths derived either from narrowband-selected galaxies at $z\simeq
0.8-3.3$ (e.g., \citealt{khostovan16}) or from IRAC color excesses for
galaxies at $z\sim 6-8$ (e.g., \citealt{labbe13, smit14,
  castellano17}), but these studies generally point to rest-frame
$W(\oiii+\hb) \simeq 500-3000$\,\AA.

Regardless, when taking our results in context with those of other
investigations that targeted lower mass galaxies, we find some hint
that the $W(\oiii)$-versus-$M_{\ast}$ relation may flatten at
$M_{\ast}\la 10^{8.7}$\,$M_\odot$, with an absence of star-forming
galaxies having rest-frame $W(\oiii) \ga 3000$\,\AA.  This upper limit
appears to be consistent with the typical range of \oiii\, rest-frame
equivalent widths of extreme emission line star-forming galaxies found
from other studies at both low and high redshifts (e.g.,
\citealt{atek11, vanderwel11, maseda14, smit14, smit16, forrest17, tang18}).

While the lower redshift emission line samples of \citet{vanderwel11}
and \citet{maseda14} are unlikely to have missed a large population of
even more extreme emitters (i.e., $W(\oiii)\ga 3000$\,\AA) at the same
redshifts,
such emitters could
be present at even higher redshifts ($z\ga 4$) given the
aforementioned evolution toward higher $W(\oiii)$ with redshift at a
fixed stellar mass (Section~\ref{sec:residuals}).  Thus, the apparent
scarcity of galaxies with rest-frame $W(\oiii)\ga 3000$\,\AA\, may
simply reflect the current absence of line measurements for such
objects at $z\ga 4$.  With the exception of a few individual galaxies,
inferences of line equivalent widths at these redshifts are presently
limited to galaxies with $M_\ast \ga 10^{8.5}$\,$M_\odot$ (e.g.,
\citealt{labbe13, smit14, castellano17}).  These studies suggest
\oiii\, equivalent widths that are consistent with those of the
highest equivalent width galaxies in the $z\sim 2$ MOSDEF sample.  We
must keep in mind, however, that the equivalent widths inferred for
$z\sim 6-8$ studies are based on stacking photometry, a procedure that
can potentially mask the presence of individual strong line emitters.

Alternatively, the plateau in the relationship between $W(\oiii)$ and
$M_\ast$ at low masses ($M_\ast\la 10^9$\,$M_\odot$) may be physical
in nature.  The prompt enrichment of $\alpha$ elements such as oxygen
implies that a ``floor'' in gas-phase metallicity, and hence stellar
metallicity, may be reached over very short (less than tens of Myr)
timescales.\footnote{Given the apparent $\alpha$-enhancement present
  in high-redshift galaxies \citep{steidel16}, the actual {\em
    measured} stellar metallicity may lag the {\em measured} gas-phase
  metallicity, since the former is modulated by iron-peak elements,
  while the latter is sensitive to $\alpha$ elements (i.e., oxygen).}
Given the decrease in the ionizing photon production rate and
softening of the ionizing spectrum with increasing stellar
metallicity, one might naturally also expect a ``ceiling'' in the
\oiii\, equivalent widths for such low-mass galaxies.  Another
possibility is that the MZR would predict such galaxies to have very
low metallicities ($\la 0.2Z_\odot$) such that the collisional line
luminosities may be suppressed due to a low oxygen
abundance.\footnote{The critical density for \oiii\, is $\approx 3.5$
  orders of magnitude larger than the typical $n_{\rm e}$ of $z\sim 2$
  galaxies (e.g., \citealt{sanders16b, strom17}), so collisional
  de-excitation is unlikely to be responsible for the suppression of
  the \oiii\, doublet emission at low stellar masses.}  Indeed,
Figure~\ref{fig:cloudy} shows that the \oiii\, line intensity at a
fixed SFR reaches a maximum around $\frac{1}{3} Z_\odot$, below which
point the \oiii\, line intensity decreases.  This non-monotonic
behavior in $L(\oiii)$ versus $Z$ would translate to a change in slope
of the $W(\oiii)$ versus $M_\ast$ relation around a stellar mass of
$\approx 10^9$\,$M_\odot$, similar to the behavior seen in
Figure~\ref{fig:woiiivsmass}.  It is also possible that the stochastic
star formation believed to be present in these low-mass galaxies
\citep{weisz12, hopkins14, dominguez15, guo16, sparre17, faucher18}
may result in the efficient expulsion of metals via supernovae (SNe) and stellar
feedback.  This effect results in a decrease of the gas-phase
metallicity and potential reduction in the strengths of oxygen lines.
Future spectroscopic campaigns to measure the oxygen (and other metal
and \ion{H}{1} recombination) lines in low-mass galaxies at
high redshift may be able to distinguish between these various
possibilities.

\begin{figure*}
\epsscale{1.00}
\plotone{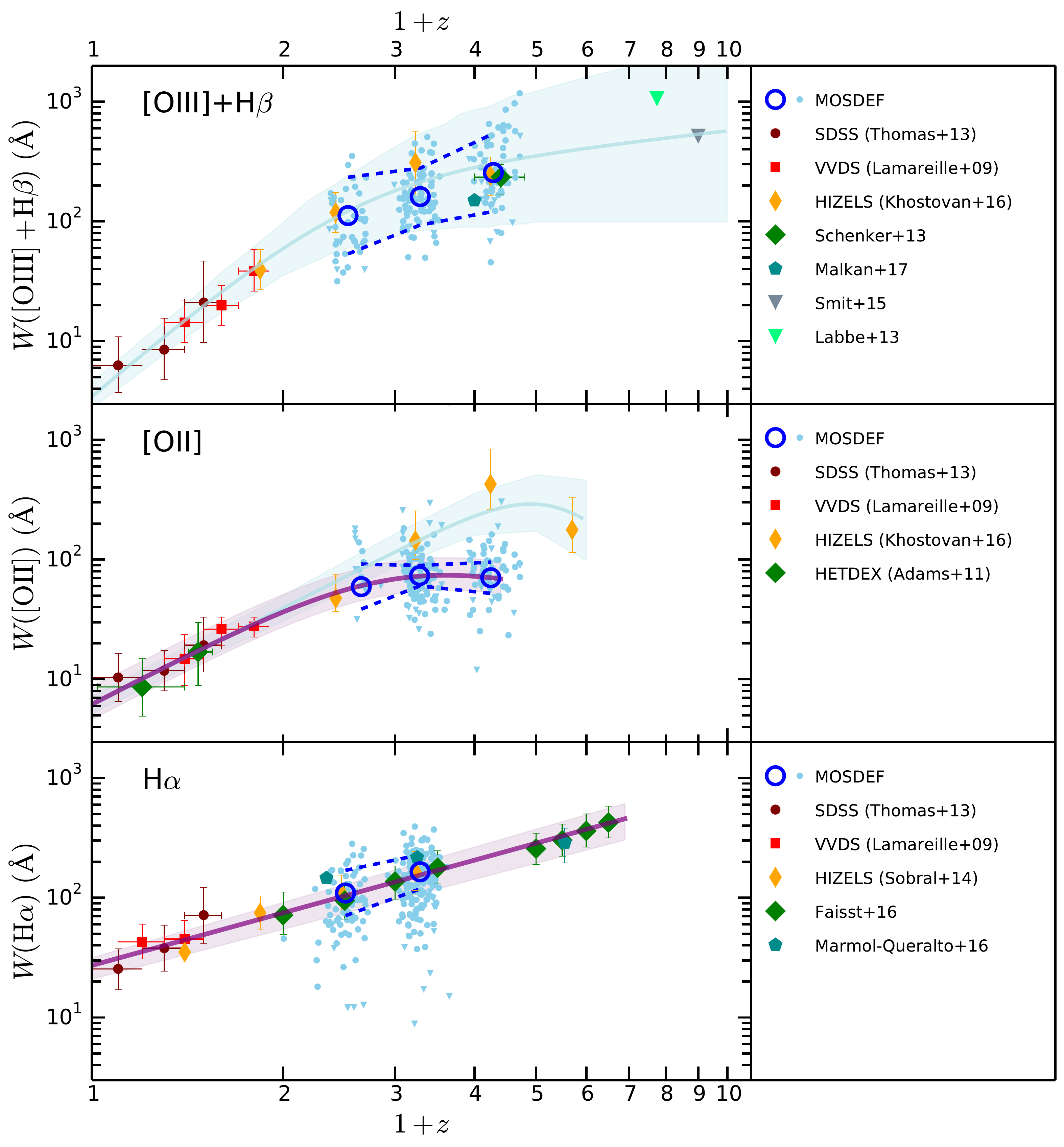}
\caption{Average rest-frame equivalent widths of $\oiii+\hb$ ({\em
    top}), $\oii$ ({\em middle}), and $\ha$ ({\em bottom}) as a
  function of redshift for star-forming galaxies with $M_\ast =
  10^{9.5}-10^{10.0}$\,$M_\odot$ in the MOSDEF sample, where
  individual objects with detected lines are indicated by the small
  cyan circles, objects with upper limits in line equivalent widths
  are shown by the small downward-pointing cyan triangles, and averages
  for each redshift subsample (including both detections and
  non-detections) are denoted by the large open blue circles.  Several
  other literature values are included for galaxies over the same
  aforementioned stellar mass range (e.g., see \citealt{khostovan16}).
  In each panel, the thick dashed blue lines indicate the average
  equivalent widths for galaxies in two other mass ranges, $M_\ast =
  10^{9.0}-10^{9.5}$\,$M_\odot$ (upper dashed lines) and $M_\ast =
  10^{10.0}-10^{10.5}$\,$M_\odot$ (lower dashed lines).  The light
  blue line and shaded regions in the upper two panels show double
  power-law fits and their $95\%$ confidence bands to the evolution in
  $W(\oiii+\hb)$ and $W(\oii)$ with redshift presented in
  \citet{khostovan16}.  The purple line and shaded region in the
  middle panel shows a double power-law fit and the $95\%$ confidence
  band to the evolution in $W(\oii)$ with redshift when including only
  the MOSDEF points above $z\ga 1$.  The purple line and shaded region
  in the bottom panel indicate the same for a single power-law fit to
  $W(\ha)$ versus redshift for all the data shown.}
\label{fig:wevolz}
\end{figure*}

\subsubsection{Redshift Evolution in $W$ at a Fixed Mass}

For another view of the redshift evolution in equivalent widths, and
one that can be used to more directly infer the redshift evolution of
the sSFRs and ionization parameters of galaxies, we show in
Figure~\ref{fig:wevolz} the average $\oiii+\hb$, $\oii$, and $\ha$
equivalent widths of $M_{\ast}=10^{9.5} - 10^{10.0}$\,$M_\odot$
galaxies as a function of redshift.\footnote{The redshift
  evolution in equivalent widths shown in Figure~\ref{fig:wevolz}
  should not be interpreted as indicating an evolution in equivalent
  widths for individual galaxies since, for example,
  $M_{\ast}=10^{9.5}$\,$M_\odot$ galaxies in the local Universe had
  progenitors at $z\sim 2$ that were significantly less massive than
  the $M_\ast=10^{9.5}$\,$M_\odot$ galaxies at $z \sim 2$.}  We
include data from the MOSDEF sample analyzed here, as well as several
other results from the literature \citep{thomas13, lamareille09,
  khostovan16, schenker13, malkan17, smit15, labbe13, adams11,
  sobral14, faisst16, marmol16}.  Focusing on the top panel, we find
that the average $W(\oiii+\hb)$ for the redshift subsamples of the
MOSDEF survey are consistent with previous measurements at similar
redshifts (e.g., \citealt{schenker13, khostovan16, malkan17}).
Moreover, we find that the average equivalent widths deduced for the
MOSDEF sample are consistent with the double power-law fit to
$W(\oiii+\hb)$ versus $1+z$ that includes data at both lower ($z\la
1$) and higher ($z\ga 5$) redshifts provided in \citet{khostovan16}.
This fit---which does not include the MOSDEF galaxies but is
nonetheless consistent with our data---indicates that the \oiii+\hb\,
equivalent widths do not continually increase with redshift, but reach
a plateau consistent with the absence of very high \oiii\, equivalent
width galaxies ($W(\oiii)\ga 3000$\,\AA) as suggested by
Figure~\ref{fig:woiiivsmass}.

The average $W(\oii)$ for galaxies with similar stellar masses also
increases with redshift, at least up to $z\sim 2$.  While
\citet{khostovan16} found that $W(\oii)$ continues to increase up to
$z\sim 3$, the MOSDEF data suggest an average $W(\oii)$ at $z\sim 3$
that is similar to that measured at $z\sim 2$.  There are a few
possible reasons for this difference in the inferred redshift
evolution of $W(\oii)$.  During the course of the MOSDEF survey, slits
were placed on galaxies according to their {\em H}-band continuum
centroids.  If the \oii\, line emission is significantly offset from
the continuum centroid of a galaxy, then the
spectroscopically measured $W(\oii)$ for such a galaxy may
underestimate the true value.

Another possible contributing reason for this disparity in the
inferred redshift evolution of $W(\oii)$ is that the
\citet{khostovan16} analysis was based on a narrowband selection of
line emitters that is sensitive to galaxies with $W(\oii)\ga 25$\,\AA.
If these narrowband samples miss populations of galaxies with
equivalent widths lower than this limit, then the average equivalent
widths deduced from these samples will be biased high relative to
those inferred from spectroscopic surveys such as MOSDEF.  Indeed, in
the MOSDEF sample, we find a small number of galaxies with either
measured $W(\oii)\la 25$\,\AA\, or upper limits in \oii\, that could
potentially lie below this equivalent width limit.  Consequently, one
should be cautious of using narrowband-selected samples of
high-redshift galaxies to deduce the evolution of equivalent widths of
weak lines such as \oii.

Finally, while the MOSDEF sample has been vetted for AGNs using a
number of different photometric, spectroscopic, and multi-wavelength
criteria \citep{coil15}, such detailed (and particularly
spectroscopic) information was only available for a subset of the
HIZELS sample \citep{sobral16}.  Contaminating AGNs with very
high \oii\, equivalent widths may result in a high average $W(\oii)$
for the narrowband sample.  In principle, one might expect such AGNs to
also exhibit high equivalent widths of \oiii\, as well, but we find
that insofar as \oiii+\hb\, is concerned, the mean equivalent widths
of this line complex are similar between the MOSDEF and HIZELS samples
(top panel of Figure~\ref{fig:wevolz}).  At any rate, complete
spectroscopy for the existing narrowband samples should help to
clarify the cause of the discrepancy in the evolution of $W(\oii)$
versus redshift deduced from these samples relative to that inferred
from the MOSDEF sample.  One obvious implication of the steep redshift
evolution of $W(\oiii+\hb)$ (and $W(\oiii)$) relative to that of
$W(\oii)$ for galaxies with similar masses is that the O32 ratio, or
ionization parameter, increases with redshift at a fixed stellar mass,
a result that has been noted previously (e.g., \citealt{khostovan16,
  sanders16a}).

Separately, we note that the nebular reddening of galaxies of a fixed stellar
mass does not appear to evolve strongly with redshift (e.g.,
\citealt{dominguez13}), in much the same way that the stellar
attenuation at a fixed stellar mass also appears to be invariant with
redshift (e.g., see Section~\ref{sec:betairxtrend}).  As such, the
redshift evolution in the dust-corrected $\sfrha$ versus $M_\ast$
relationship translates directly to a corresponding evolution in the
observed $L(\ha)$, or $W(\ha)$, versus $M_\ast$ relation, such that
galaxies at a fixed stellar mass have $W(\ha)$ that increase with
redshift.  This expectation is borne out by the bottom panel of
Figure~\ref{fig:wevolz}, which shows a steady increase in $W(\ha)$ for
galaxies of a fixed stellar mass of $M_\ast\simeq
10^{9.75}$\,$M_\odot$.  One of the primary results of our analysis is
that the redshift evolution of the equivalent widths is directly tied
to the redshift evolution in the SFR-$M_\ast$ and MZR relations
(Section~\ref{sec:residuals}).

\subsection{A Strategy for Identifying High-excitation and High-ionization Galaxies}

The relationships presented in Section~\ref{sec:calibrations} and
discussed in Section~\ref{sec:discussion} point to a clear strategy of
identifying galaxies that are likely typical of those at much higher
redshifts ($z\ga 5$), namely, those with bluer UV spectral slopes and
lower reddening, lower stellar masses, younger ages, higher sSFRs, ISM
line ratios indicative of higher ionization parameters and excitation
conditions, lower metallicities, and higher ionizing photon production
efficiencies ($\xi_{\rm ion}$).  Specifically, these characteristics
are typical of galaxies with the highest equivalent widths of the
\oiii\, and recombination lines.  Additionally, many of the trends
between equivalent width and other galaxy/ISM properties are dependent
on redshift, in the sense of increasing equivalent widths with
redshift at fixed values of reddening, stellar mass, SFR, and
metallicity.

\begin{figure}[!t]
\epsscale{1.15}
\plotone{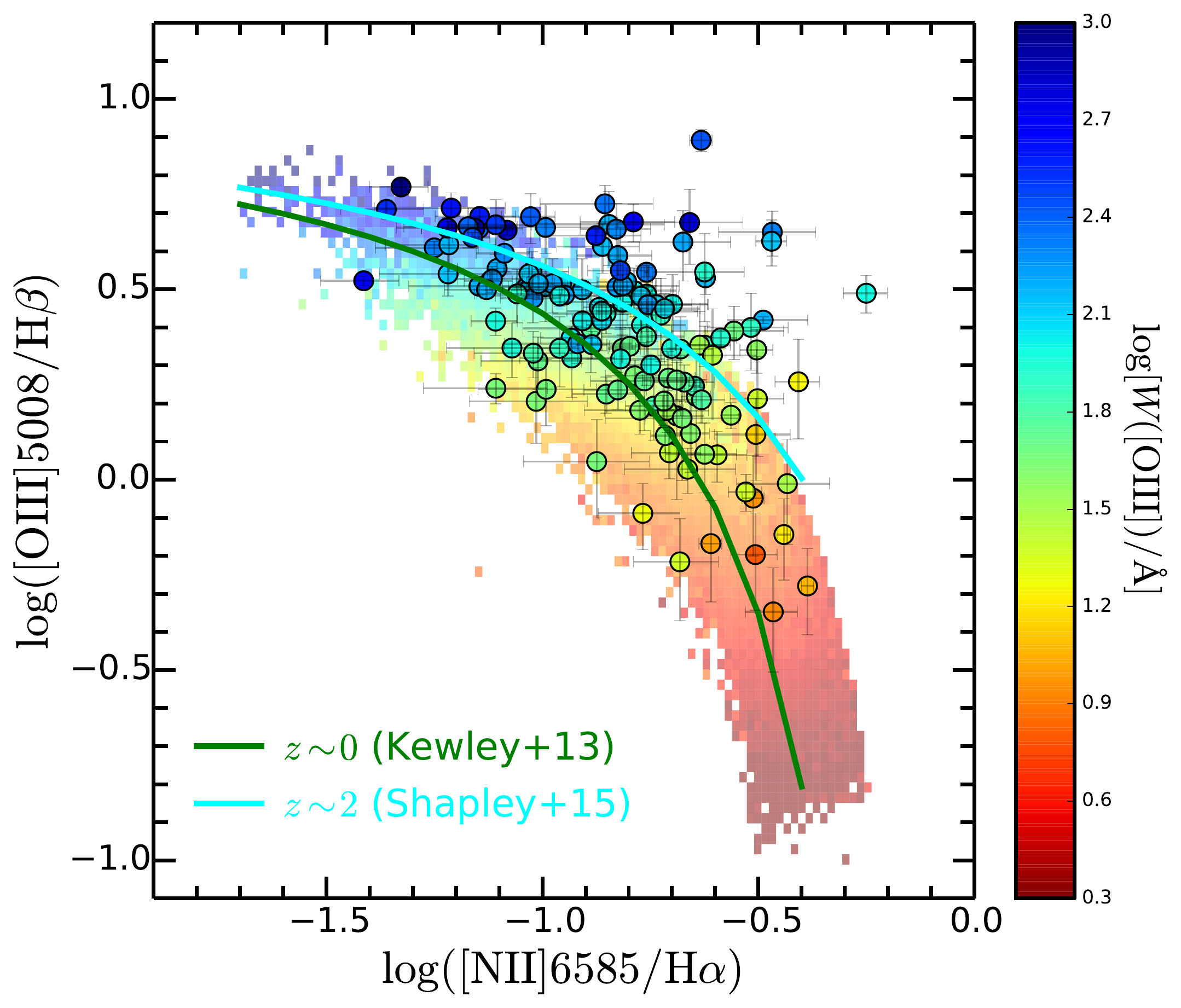}
\caption{$\log({\rm O3})$ versus $\log({\rm N2})$ (i.e., the BPT
  diagram) for galaxies in the MOSDEF sample where all four lines are
  detected, indicated by the circles.  The same is also shown for the
  SDSS sample---objects lying above the \citet{kauffmann03}
  demarcation between star-forming galaxies and AGNs are not shown as
  they were excluded from the local comparison sample (see
  Section~\ref{sec:sampleselection}).  To facilitate the comparison
  with previous studies (i.e., \citealt{kewley13} and
  \citet{shapley15}), the O3 and N2 line ratios shown here for the
  SDSS galaxies are not corrected for DIG emission.  Points have been
  color-coded by \oiii\, equivalent width.  The green and cyan lines
  indicate the average trend between $\log({\rm O3})$ and $\log({\rm
    N2})$ for the SDSS and $z\sim 2$ MOSDEF samples, respectively,
  from \citet{kewley13} and \citet{shapley15}.}
\label{fig:bpt}
\end{figure}

Of the emission lines explored in our study, the equivalent width of
\oiii\, correlates most strongly with many of the aforementioned
properties, while \oii\, shows the least significant correlations.
Fortuitously, the \oiii\, line can be the strongest rest-frame optical
emission line, particularly for low-mass and faint star-forming
galaxies where the recombination lines may be weak (e.g., see
\citealt{malkan17} and references therein).  By combining the
information shown in Figures~\ref{fig:ewo3hb} and \ref{fig:ewn2met},
we can assess how the equivalent width can be used to select galaxies
in certain regions of ``BPT'' space, traditionally defined as
$\log[{\rm O3}]$ versus $\log[{\rm N2}]$ \citep{baldwin81}.  A number
of studies have found that $z\ga 1$ galaxies are offset toward higher
$\log[{\rm O3}]$ at a fixed $\log[{\rm N2}]$ than local galaxies
(e.g., \citealt{shapley05, erb06b, liu08, masters14, steidel14,
  shapley15}), a result that is shown through the MOSDEF sample in
Figure~\ref{fig:bpt}.  Despite this offset, it is clear that both the
low- and high-redshift galaxies exhibit increasing $W(\oiii)$ with
decreasing N2 and increasing O3 (e.g., see also
Figures~\ref{fig:ewo3hb} and \ref{fig:ewn2met}).  Galaxies with the
highest equivalent widths of \oiii\, occupy the upper-left region of
the BPT diagram, indicative of them being the most ``extreme'' sources
in our sample in terms of their low metallicities and high
excitation-sensitive line ratios (e.g., O3).  This is a similar region
of BPT parameter space where local compact starbursting galaxies known
as ``Green Peas'' lie, many of which have low stellar masses ($M_\ast
\la 10^{10}$\,$M_\odot$; \citealt{cardamone09}), low metallicities
($12+\log({\rm O/H}) \simeq 8.05$; \citealt{amorin10, amorin12}), and
high sSFRs (typically between $10^{-10}$ and $10^{-8}$\,yr$^{-1}$;
\citealt{cardamone09,izotov11}).  The ease of detecting the strong
\oiii\, line and the sensitivity of its equivalent width to stellar
population and ISM properties---assuming that AGNs can be isolated
based on other imaging and/or spectroscopy---make the \oiii\, line an
ideal target for future narrowband and spectroscopic followup of
high-excitation, low-mass, and low-metallicity galaxies at high
redshift.

Such galaxies are also believed to be conducive to both the production
and escape of ionizing radiation.  In terms of production, for
example, galaxies with higher equivalent widths also tend to have
larger ionizing photon production efficiencies, $\xi_{\rm ion}$
(Figure~\ref{fig:ewxiion}).  The sensitivity of equivalent width to
$\xi_{\rm ion}$ is expected for simple physical reasons.  Higher
ionizing photon production rates can result from stellar populations
with harder ionizing spectra, which in turn can lead to higher
ionization parameters and higher \oiii\, and Balmer emission line
equivalent widths (e.g., Figure~\ref{fig:cloudy}).

In terms of the escape of ionizing radiation, the trend
of increasing neutral gas covering fraction with reddening
\citep{reddy16b} and the trend between equivalent width and reddening/UV slope
imply that galaxies with higher equivalent widths have
lower gas covering fractions, and hence larger ionizing escape
fractions, on average.  Moreover, galaxies that occupy the upper-left
region of the BPT diagram also show significantly larger Ly$\alpha$
equivalent widths \citep{erb14, trainor16} which may be indicative of
lower gas covering fractions \citep{shapley03, hayes11, jones13,
  rudie13, borthakur14, alexandroff15, trainor15, dijkstra16,
  reddy16b} and/or a higher intrinsic production rate of ionizing
photons \citep{erb14, trainor16, du18}.  While the gas covering
fraction may correlate inversely with $W(\oiii)$, we find that the
highest $W(\oiii)$ galaxies in our sample have average O32 ratios $\la
5$ (Figure~\ref{fig:ewo32}), below the values expected for
density-bounded nebulae (e.g., \citealt{nakajima13}).  Thus, one might
still expect significant Ly$\alpha$ opacity along the line-of-sight
for galaxies with high $W(\oiii)$, as would be the case for a patchy
ISM.  Regardless, local Lyman continuum emitters also appear to lie in
the region of the BPT diagram occupied by high $W(\oiii)$ galaxies
\citep{izotov16b, izotov17}.

Along these lines, several studies have suggested that local galaxies
with higher Balmer line equivalent widths have larger Ly$\alpha$
escape fractions \citep{cowie11, henry15, izotov16b}.  The connection
between Ly$\alpha$ and the escape of ionizing radiation in both local
and high-redshift galaxies (e.g., \citealt{kornei10, hayes11, jones13,
  wofford13, borthakur14, henry15, izotov16b, trainor15, dijkstra16,
  trainor16, reddy16b, steidel18}) then suggests that galaxies with
high Balmer emission line equivalent widths are likely to have larger
ionizing escape fractions, at least up to the point where the escape
fractions are not so high as to suppress the \ion{H}{1} recombination
line strengths.  Our results show that galaxies with high Balmer
line equivalent widths exhibit the highest \oiii\, equivalent widths
as well (e.g., see also \citealt{cowie11, nakajima13, henry15,
  izotov16b}).  Thus, galaxies with high \oiii\, (and Balmer emission
line) equivalent widths may be typical of those that leak ionizing
radiation and contribute significantly to cosmic reionization.  Given
the connection between escaping ionizing radiation and the intensity
and hardness of the ionizing radiation field, it is clear that
understanding the relationship between proxies for the latter (e.g.,
O32, O3, and $\xi_{\rm ion}$) and line equivalent widths will be
essential for identifying and characterizing galaxies that dominate
the ionizing photon budget at high redshift.

\section{CONCLUSIONS}
\label{sec:conclusions}

We use ground-based {\em YJHK} spectroscopy for 1,134 galaxies with
stellar masses $M_\ast \ga 10^9$\,$M_\odot$ from the MOSDEF survey to
investigate how strong rest-frame optical emission line equivalent
widths vary with a number of galaxy and ISM properties, including
stellar mass, UV slope, age, SFR, sSFR, ionization parameter,
gas-phase metallicity, and ionizing photon production efficiency.  Our
main results are summarized below.

\begin{itemize}

\item In general, rest-frame optical emission line equivalent widths are
significantly correlated with several galaxy properties.  We find that
at a fixed redshift, galaxies with larger equivalent widths have lower
stellar masses, bluer UV slopes, younger ages, higher specific SFRs,
higher ionization parameters, lower gas-phase metallicities, and
higher ionizing photon production efficiencies ($\xi_{\rm ion}$).  The
equivalent width of \oii\, generally exhibits the least significant
correlations with the aforementioned properties, while the equivalent
widths of \oiii\, and \oiii+\hb\, correlate most strongly with these
properties.

\item The redshift evolution of the trends between equivalent width and
stellar mass, UV slope, and SFR can be explained in the context of the
increasing SFR and decreasing metallicity with redshift for galaxies
of a fixed stellar mass.  At a fixed mass and metallicity, the \oiii\,
line luminosity and equivalent width increase with SFR due to a larger
ionizing photon rate, while at a fixed mass and SFR, the \oiii\, line
luminosity and equivalent width increase with decreasing metallicity
due to a harder ionizing spectrum.  The aforementioned trends imply
that when the SFR, stellar mass, and metallicity are fixed, galaxies
have similar equivalent widths irrespective of redshift.  It is
for this reason that the relationship between equivalent width and
specific SFR evolves very little with redshift between $z\sim 0$ and
$z\sim 4$, particularly when considering galaxies of a fixed
metallicity.

\item  The relationship between \oiii\, equivalent width and stellar mass
plateaus at stellar masses $M_\ast \la 10^{9}$\,$M_\odot$.  This
behavior could be physically explained by a rapid enrichment of
$\alpha$ elements, the suppression of the oxygen lines in galaxies
with metallicities below $\simeq \frac{1}{3}Z_\odot$, and/or the
efficient dilution of the gas-phase metallicity due to SNe and stellar
feedback in low-mass galaxies.

\item Of all the strong rest-frame optical lines, the equivalent width
  of \oiii\, correlates most significantly with many of the galaxy and
  ISM properties considered in this study, thus marking it as a
  powerful probe of low-mass, young, high-excitation, and
  low-metallicity galaxies at high redshift.  In particular, the
  \oiii\, (and Balmer emission line) equivalent widths correlate
  significantly with age, ionization parameter, and the ionizing
  photon production efficiency $\xi_{\rm ion}$, all factors that are
  believed to be conducive to the escape of ionizing radiation.  As a
  consequence, galaxies with high \oiii\, and Balmer emission line
  equivalent widths are likely to contribute significantly to cosmic
  reionization.

\end{itemize}

In Appendix~\ref{sec:ewcorrections}, we provide prescriptions for
translating the equivalent widths of line complexes (e.g.,
$\oiii+\hb$, $\ha+\nii$) obtained from narrowband or
low-spectral-resolution data to those of single ionic species.  Beyond
enabling such simple prescriptions and allowing one to evaluate the
impact of line emission on galaxy photometry, the MOSDEF sample
provides a robust foundation for evaluating the evolution of
equivalent widths over broader dynamic ranges in galaxy/ISM properties
and redshifts made possible with the next generation of ground- and
space-based observatories.  In the future, the {\em James Webb Space
  Telescope} ({\em JWST}) will allow spectroscopic access to many of
the same emission features for galaxies up to $z\simeq 10$.  Thus, it
is anticipated that within the next several years, we will have a
comprehensive view of how the strong rest-frame optical emission lines
and the properties inferred from them evolve over roughly $97\%$ of
cosmic history.

\acknowledgements

N.A.R. thanks Matt Malkan and Daniel Cohen for providing some of the data
shown in Figure~\ref{fig:woiiivsmass} in electronic format.  We
acknowledge support from NSF AAG grants AST-1312780, 1312547, 1312764,
and 1313171; archival grant AR-13907 provided by NASA through the
Space Telescope Science Institute; and grant NNX16AF54G from the NASA
ADAP program.  R.L.S. was supported by a UCLA Graduate Division
Dissertation Year Fellowship.  A.E.S. acknowledges a NASA contract
supporting the ``WFIRST Extragalactic Potential Observations (EXPO)
Science Investigation Team" (15-WFIRST15-0004), administered by GSFC.
We are grateful to the MOSFIRE instrument team for building this
powerful instrument, and to Marc Kassis at the Keck Observatory for
his many valuable contributions to the execution of the MOSDEF
survey. We also acknowledge the 3D-HST collaboration, who provided us
with spectroscopic and photometric catalogs used to select MOSDEF
targets and derive stellar population parameters. We also thank
I. McLean, K. Kulas, and G. Mace for taking observations for the
MOSDEF survey in 2013 May and June.  We wish to extend special thanks
to those of Hawaiian ancestry on whose sacred mountain we are
privileged to be guests.  Without their generous hospitality, the
observations presented herein would not have been possible.



\appendix

\section{Line Luminosity Completeness}
\label{sec:completeness}

Here, we present evidence that the MOSDEF sample is mostly complete in line
luminosity for galaxies above our ``mass-completeness'' threshold of
$\simeq 10^{9.0}$\,$M_\odot$.  Specifically,
limiting our sample to
galaxies that have \oiii\, line fluxes that are twice our nominal
detection threshold changes the normalization of the trend between
$W(\oiii)$ and $M_\ast$ by $\la 0.1$\,dex, implying that our sample
probes the tail of the $W(\oiii)$ distribution over the relevant range
of stellar mass.  Moreover, we achieve a high spectroscopic success
rate ($\approx 80\%$) of identifying redshifts for objects where prior
targeting information---i.e., photometric redshifts or external
grism/spectroscopic redshifts---indicate that the lines of interest
should fall in the wavelength ranges accessible to MOSFIRE,
independent of redshift \citep{kriek15}.  In other words, there would
only be an incremental increase in successful redshifts for
star-forming galaxies if we were to have increased the spectroscopic
integration time.  Finally, as shown in
Section~\ref{sec:calibrations}, the inclusion of objects with
successful redshifts, but where one or more lines may not have been
significantly detected, does little to alter the mean trends between
equivalent widths and the other galaxy properties analyzed here.
Further evidence of the line luminosity completeness of the MOSDEF
sample comes from a comparison of the line luminosity functions
inferred from the MOSDEF survey and those obtained from narrowband
selected samples of high-redshift galaxies, a topic that will be
discussed elsewhere.  As such, we assume that relative comparisons of
the trends between equivalent widths and other galaxy properties
between the MOSDEF and local SDSS samples remain valid, irrespective
of the depth of the MOSDEF spectroscopic observations.

\section{Composite Spectra}
\label{sec:composite}

\begin{figure}
\epsscale{1.00}
\plotone{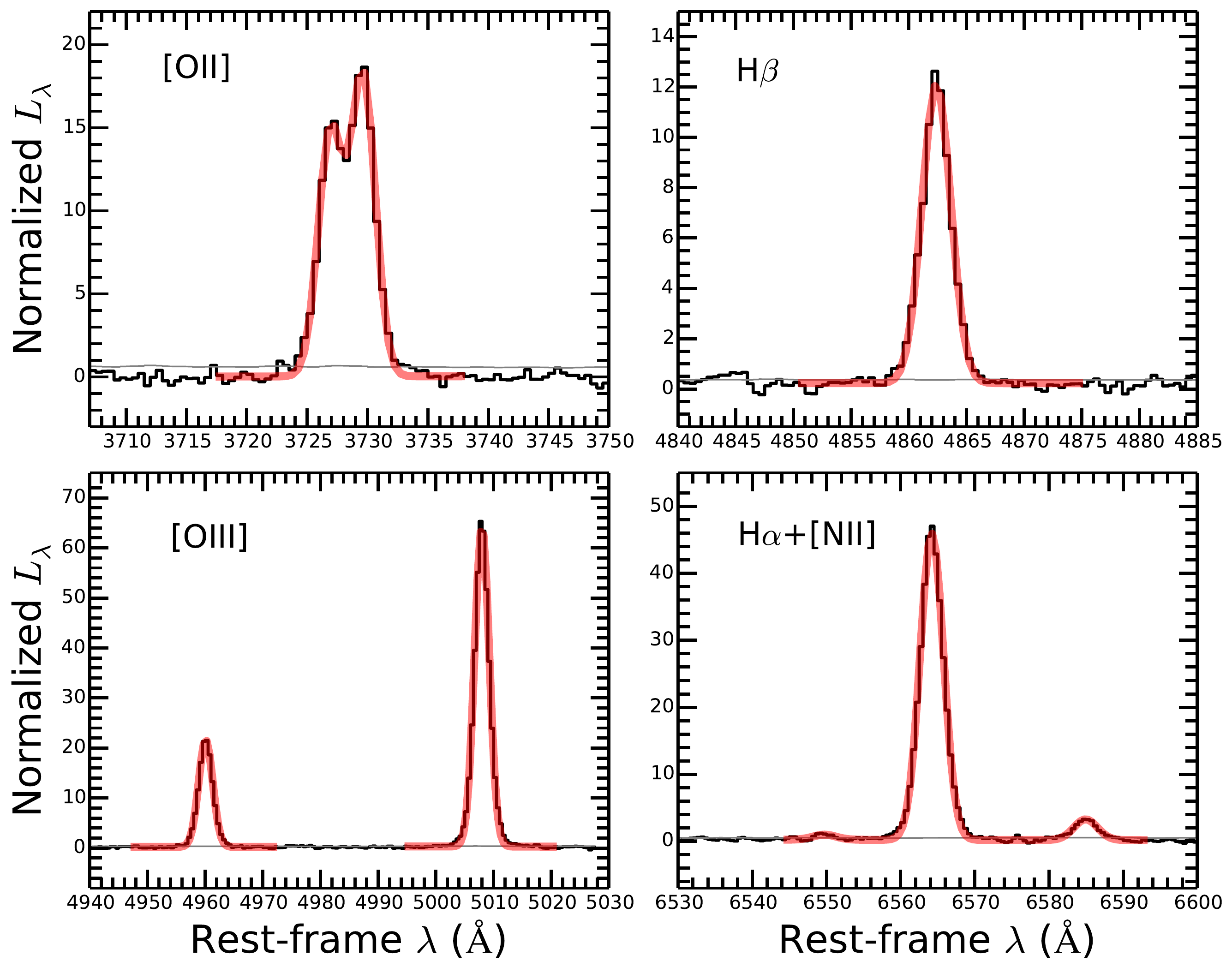}
\caption{Composite spectrum of 292 galaxies lying
  in the lowest quartile of the distribution of stellar mass ({\em
    black} curve), along with the associated error spectrum ({\em thin
    gray line}), and the best-fit line profile ({\em red}) to each of the lines
  considered in our analysis.}
\label{fig:example_comp}
\end{figure}

Some of the lines included in our analysis are formally undetected for
some galaxies, yielding only upper limits on $W$.  To include such
objects in our analysis, we combined the spectra of galaxies with
detected and undetected lines in order to create a composite spectrum
and compute mean equivalent widths.  A composite spectrum was computed
by shifting each galaxy's spectrum into the rest frame, converting to
luminosity density units, normalizing the individual spectra by the
continuum luminosity density at the center of the line for which we
were interested in computing the mean equivalent width for, and
averaging the spectra for all galaxies contributing to the composite.
Further details on how composite (and associated error) spectra were
calculated are provided in \citet{shivaei18}.
Figure~\ref{fig:example_comp} shows an example composite spectrum---in
this case, derived from 292 galaxies lying in the lowest quartile in
stellar mass---along with the best-fit line profiles used to calculate
the average equivalent widths.

The average equivalent width was then computed by simply integrating
the composite spectrum over the line of interest in the same way that
fluxes are computed for emission lines in individual galaxy spectra.
Also, analogous to how line flux errors are computed for individual
galaxies, the error in the average line flux was obtained by
perturbing the composite spectrum by its associated error spectrum,
remeasuring the line fluxes for each of these realizations, and then
taking the error as the $1\sigma$ width of the distribution of the
resulting line flux measurements.  The error in average line flux is
folded into the uncertainty on the average equivalent width.

\section{Effect of Changing the Assumed Stellar Metallicity and Dust
Obscuration Curve}
\label{sec:sedpop}

\begin{figure*}
\epsscale{1.00}
\plotone{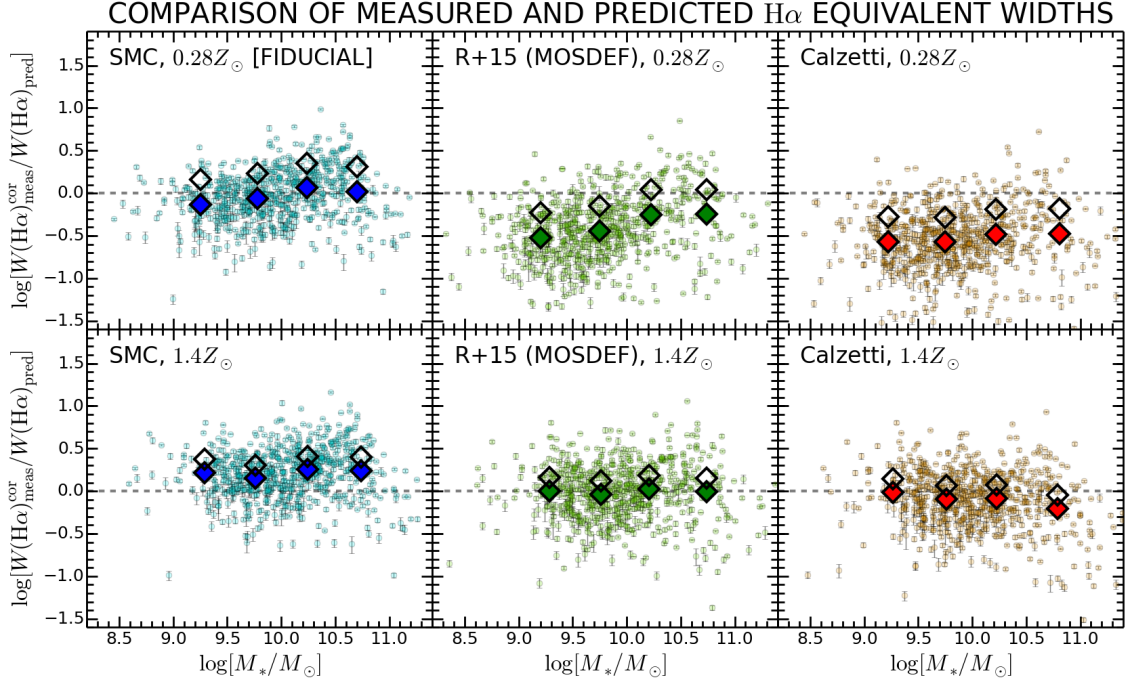}
\caption{Comparison of the measured and predicted \ha\, equivalent
  widths for MOSDEF galaxies with detected \ha\, for different
  assumptions of the stellar attenuation curve and stellar population
  model, as indicated at the top of each panel.  For comparison with
  the model predictions, the measured equivalent widths are corrected
  for dust attenuation ($W(\ha)_{\rm meas}^{\rm cor}$) based on the
  Balmer decrement and assuming the \citet{cardelli89} extinction
  curve.  For those galaxies where \hb\, is undetected, the Balmer
  decrement is computed assuming the mean relationship between Balmer
  decrement and stellar mass (see text for details).  The small points
  denote individual objects, where $\log[W(\ha)_{\rm meas}^{\rm cor}/
    W(\ha)_{\rm pred}]$ assumes the BPASS stellar population synthesis
  models.  The large filled and open diamonds in each panel correspond
  to the mean $\log[W(\ha)_{\rm meas}^{\rm cor}/ W(\ha)_{\rm pred}]$
  in four bins of stellar mass for the BPASS and BC03 stellar
  population models, respectively.}
\label{fig:modelhaews}
\end{figure*}

As noted in Section~\ref{sec:sedmodeling}, we assumed a fixed stellar
population model ($0.28Z_\odot$ BC03 model) and
attenuation curve (SMC) for all galaxies in our sample.  In principle,
a more realistic approach would allow for the stellar population to
increase in stellar metallicity with the stellar mass, and also allow
for grayer attenuation curves toward higher stellar masses (e.g.,
\citealt{reddy18}).
One key test that can be used to evaluate whether a more sophisticated
treatment is warranted is to directly compare the measured equivalent
widths of \ha\, with those predicted from the intrinsic stellar
population that best fits the observed photometry.  Specifically, the
ionizing flux per unit SFR increases with decreasing metallicity, among
other things (see below).  As
recombination line fluxes are directly related to the ionizing flux, a
comparison between the observed recombination lines (e.g., \ha) and
those predicted from various stellar population models can be used to
potentially distinguish between these models.  

The \ha\, line luminosity predicted from a given stellar population
model was computed as follows.  First, the intrinsic SED corresponding
to the SED that best fits the observed photometry was integrated in
the range $\lambda = 200-912$\,\AA\, to compute the H-ionizing photon
luminosity, $Q({\rm H})$.  The corresponding intrinsic \ha\,
luminosity, $L(\ha)$, was calculated assuming Case B recombination and
an electron temperature of $T_{\rm e} = 10,000$\,K.  We also assumed
no escape of ionizing photons ($f_{\rm esc}=0$) and that there is no
dust attenuation of ionizing photons, referred to as Lyman continuum
extinction \citep{inoue01a, inoue01b}.  The former is
justified as $f_{\rm esc}$ is on average $\la 10\%$ for typical
($L^{\ast}$) star-forming galaxies at $z\sim 2-3$ \citep{reddy16b,
  steidel18}.  The latter assumption is based on the high gas column
densities inferred for $z\sim 3$ galaxies, such that photoelectric
absorption, rather than dust attenuation, dominates the depletion of
ionizing photons \citep{reddy16b}.  With these assumptions, the
predicted equivalent width of \ha\, ($W(\ha)_{\rm pred}$) was then
computed by dividing $L(\ha)$ by the continuum luminosity density from
the best-fit SED model.

To compare $W(\ha)_{\rm pred}$ with the measured $W(\ha)$, the latter
must be corrected for the attenuation of the \ha\, line.  For those
galaxies where both \ha\, and \hb\, were significantly detected, the
Balmer decrement was computed directly from these lines.  For those
galaxies where only \ha\, was significantly detected, and \hb\, was
not, the Balmer decrement was derived as follows.  First, we
constructed composite spectra of all galaxies where \ha\, was
detected, without regard to whether \hb\, was detected, in four bins
of stellar mass.  The average Balmer decrement was computed from each
of these composite spectra, and then we fit a linear function to these
mean Balmer decrements as a function of stellar mass.  Based on this
linear function, a Balmer decrement was assigned to each galaxy for
which \hb\, was undetected.  Balmer decrements were then used to
correct the measured \ha\, equivalent widths, $W(\ha)_{\rm meas}$, for
the attenuation of \ha\, using the procedure outlined in
Section~\ref{sec:calcsfrha}, to yield $W(\ha)_{\rm meas}^{\rm cor}$.
Note that the dust corrections are only applied to the \ha\, emission;
the continuum flux density is not corrected for dust in either the
model predictions or the measured values.

There are large systematic offsets in the ionizing flux per unit SFR
predicted by various models, particularly those that include binaries
and/or rotating stars (e.g., \citealt{eldridge09, brott11, levesque12,
  leitherer14, eldridge17}).  Given the apparent success of models
that include binaries in reproducing simultaneously the rest-frame
far-UV continuum, stellar photospheric lines, and nebular lines (both
in the rest-frame far-UV and optical) of the average galaxy at $z\sim
2$ (e.g., \citealt{steidel16}), we examined the $W(\ha)$ predictions
for the Binary Population and Spectral Synthesis (BPASS) models
\citep{eldridge17} that match the metallicity assumed for our fiducial
model ($0.28Z_\odot$) and a metallicity of $1.4Z_\odot$.

The log of the ratio of the measured and predicted \ha\, equivalent
widths for galaxies with detected $\ha$ is shown as a function of
stellar mass in Figure~\ref{fig:modelhaews} for these two different
metallicities and for three different dust curves: the SMC extinction
curve, the \citet{reddy15} attenuation curve derived for MOSDEF
galaxies at $z\sim 2$, and the \citet{calzetti00} attenuation
curve.\footnote{Similar to the SMC curve used here, the dust curves
  given in \citet{reddy15} and \citet{calzetti00} were updated in the
  far-UV ($\lambda\simeq 950-1250$\,\AA) based on the analysis of
  \citet{reddy16a}.}  The attenuation curve used for modeling the SEDs
is relevant in this comparison because the shape of the dust curve
affects $\sfrsed$: for a given observed SED, a steeper (SMC-like)
attenuation curve yields lower $\ebmvstars$, lower $\sfrsed$, and hence
lower $Q(H)$, relative to those derived with a grayer attenuation
curve.  Thus, modeling that includes an SMC-like dust curve results in
$W(\ha)_{\rm pred}$ that are systematically lower, and hence
$\log[W(\ha)_{\rm meas}^{\rm cor}/W(\ha)_{\rm pred}]$ that are
systematically higher, than those derived with either the
\citet{reddy15} or \citet{calzetti00} attenuation curves.  These
offsets that result from different assumptions of the attenuation
curve increase as metallicity decreases.

The average $\log[W(\ha)_{\rm meas}^{\rm cor}/W(\ha)_{\rm pred}]$ in
four bins of stellar mass are indicated by the large filled diamonds
in each panel of Figure~\ref{fig:modelhaews}.  For comparison---and to
illustrate the relative offsets in the predicted $L(\ha)$ between
single and binary evolution stellar synthesis models---the large open
diamonds indicate the average $\log[W(\ha)_{\rm meas}^{\rm
    cor}/W(\ha)_{\rm pred}]$ assuming the BC03 models when computing
the intrinsic ionizing flux.  The single stellar evolution models
predict lower ionizing fluxes per unit SFR than the binary models,
resulting in larger $\log[W(\ha)_{\rm meas}^{\rm cor}/W(\ha)_{\rm
    pred}]$ for the former.

These comparisons indicate that models that include sub-solar
metallicity stellar populations and gray attenuation curves, or solar
metallicity models with steep attenuation curves, predict \ha\,
equivalent widths that are systematically offset by up to a factor of
three from the measured \ha\, equivalent widths.  On the other hand,
our fiducial model, as well as models that include grayer attenuation
curves with solar metallicity stellar populations, are generally able
to reproduce the measured \ha\, equivalent widths.  As noted earlier,
sub-solar metallicity models with steeper attenuation curves are able
to better-reproduce the dust obscuration (as measured from far-IR
data) at a given UV slope for $z\sim 2$ galaxies \citep{reddy18}.
Furthermore, modeling of far-UV stellar photospheric absorption lines
for typical star-forming galaxies at $z\sim 2$ also favors sub-solar
metallicity models \citep{steidel16}.  It is for these reasons that we
adopted the assumptions corresponding to our fiducial modeling.
However, given that the combination of solar metallicity models with
gray attenuation curves is able to reproduce the measured \ha\,
equivalent widths, it is useful to examine how adopting these
assumptions may alter some of the aforementioned trends seen between
equivalent width and SED-derived parameters.

\begin{figure}
\epsscale{0.85}
\plotone{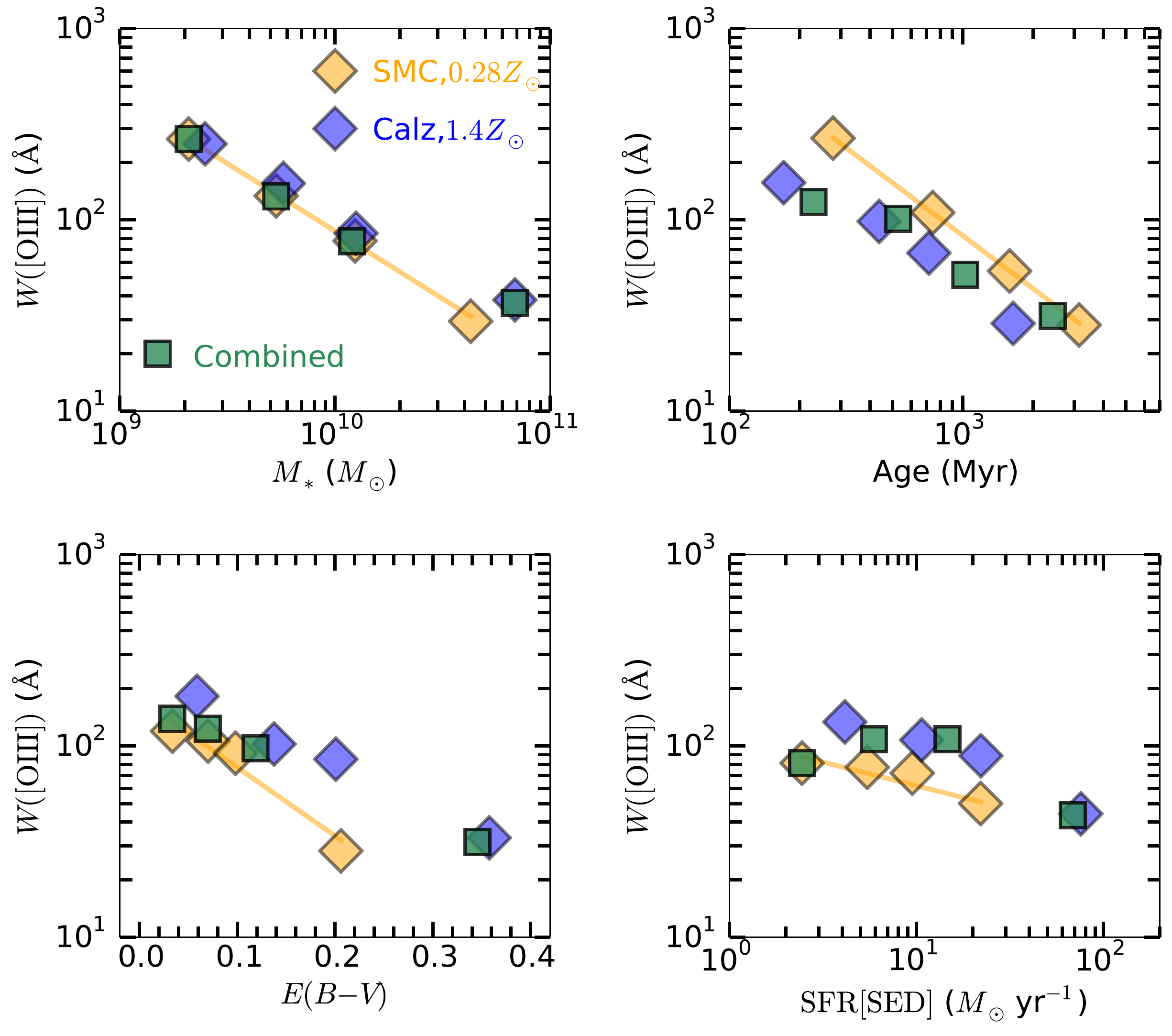}
\caption{$W(\oiii)$ versus $M_{\ast}$, age, $\ebmvstars$, and $\sfrsed$.
  Orange and diamonds and the solid lines indicate the average values
  and best-fit relations, respectively, assuming our fiducial modeling
  with a \citet{bruzual03} $0.28Z_\odot$ metallicity stellar
  population and the SMC extinction curve.  Blue diamonds indicate the
  same for a $1.4Z_\odot$ metallicity and the \citet{calzetti00}
  attenuation curve.  The green squares show the results when assuming
  the latter set of assumptions for galaxies with $M_{\ast}\ge 3\times
  10^{10}$\,$M_\odot$), and the fiducial modeling for lower mass
  galaxies.}
\label{fig:ewsedpop}
\end{figure}

In Figure~\ref{fig:ewsedpop}, we show how changing our initial
assumptions for the metallicity of the stellar population and
reddening curve affects the relationship between line equivalent width
and SED parameters, focusing on $W(\oiii)$.  For this demonstration,
we compared the trends obtained with the fiducial modeling (i.e., SMC
extinction curve and a $0.28Z_\odot$ stellar population) and those
obtained with a \citet{calzetti00} attenuation curve and a
$1.4Z_\odot$ stellar population.  The latter set of assumptions yields
stellar masses that are roughly $10\%$ smaller and ages that are a factor
of $\simeq 2$ younger than those obtained with the fiducial modeling.
The $\ebmvstars$ are on average larger for the \citet{calzetti00}
curve---as this is a grayer curve than the SMC extinction curve, a
larger reddening is required to reproduce an observed UV spectral
shape.  Because of this, $\sfrsed$ is also systematically larger.

If we assume that all galaxies with $M_\ast>3\times
10^{10}$\,$M_\odot$ are better described by a $1.4Z_\odot$ BC03 model
and the \citet{calzetti00} curve, and assume that galaxies below this
mass follow a $0.28Z_\odot$ model and the SMC curve, then we obtain
the mean trends of $W(\oiii)$ versus $M_\ast$, age, $\ebmvstars$, and
$\sfrsed$ indicated by the green squares in Figure~\ref{fig:ewsedpop}.
Not surprisingly, the resultant trends lie somewhere between those
obtained when assuming either model for all of the galaxies in the
sample.  The slopes of the $W(\oiii)$ versus $M_\ast$ and $W(\oiii)$
versus $\ebmvstars$ relations become slightly less negative than those
obtained with the fiducial modeling.  On the other hand, the slopes of
$W(\oiii)$ versus age and $W(\oiii)$ versus $\sfrsed$ relations are
not significantly affected while their normalizations are slightly
lower and higher by $\simeq 0.2$\,dex, respectively, than those
obtained with the fiducial modeling.  Similar trends are seen for
lines other than \oiii.  These results underscore the importance of
keeping in mind the assumptions of the stellar population model and
attenuation curve when examining trends between equivalent width and
stellar population parameters.  In the context of even higher redshift
galaxies (e.g., $z\ga 3$), the results obtained with our fiducial
modeling are likely to be closer to reality given that such galaxies
will be less metal-rich than more massive galaxies at lower redshifts.

\section{EMISSION LINE EQUIVALENT WIDTH CORRECTIONS}
\label{sec:ewcorrections}

\begin{figure}
\epsscale{0.80}
\plotone{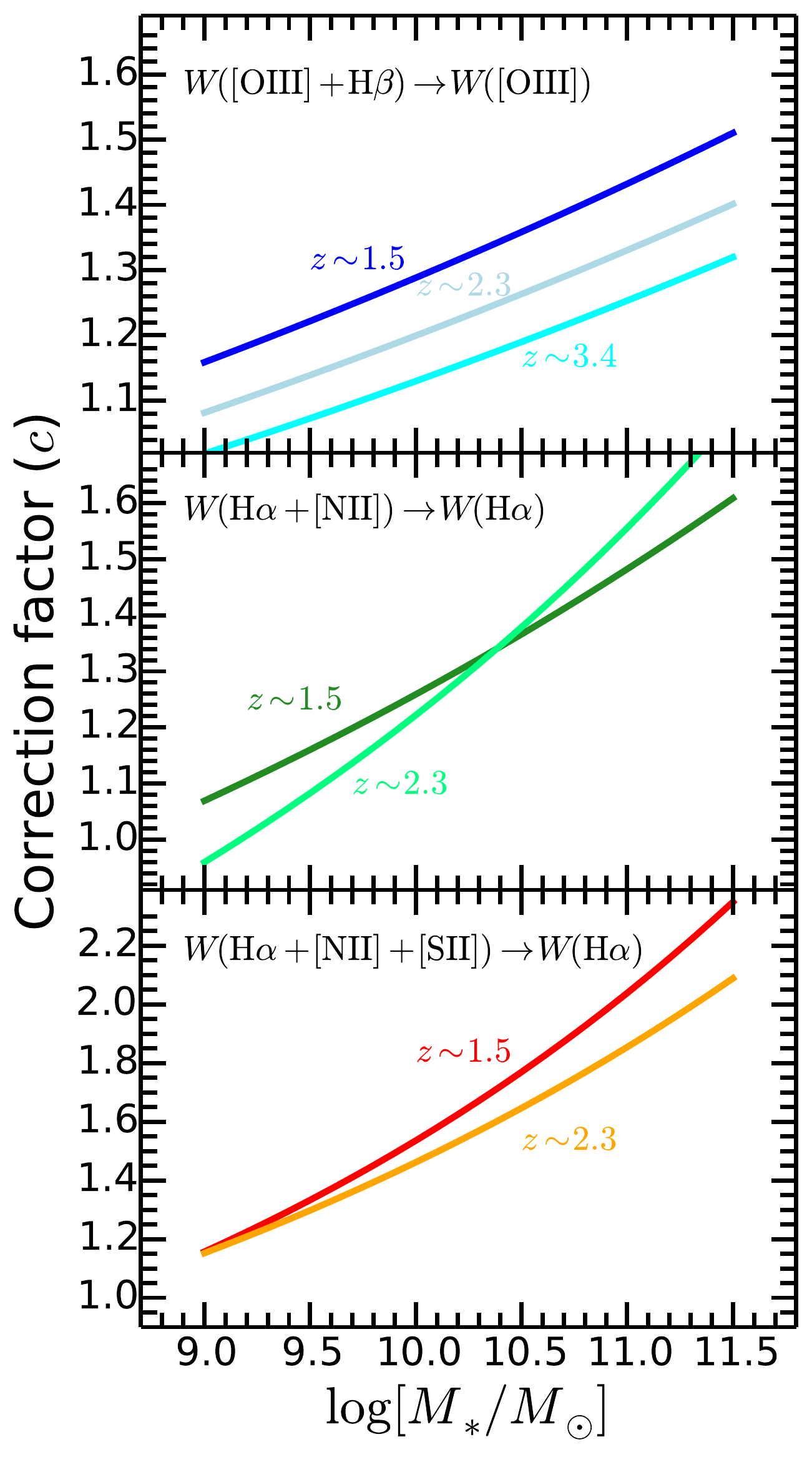}
\caption{Correction factors ($c$) by which $W(\oiii+\hb)$,
  $W(\ha+\nii)$, and $W(\ha+\nii+\sii)$ must be {\em divided} by to
  yield $W(\oiii)$, $W(\ha)$, and $W(\ha)$, respectively, as a
  function of stellar mass for different redshift subsamples.
  Functional forms for these corrections are given in
  Table~\ref{tab:ewcorrections}.}
\label{fig:ewcorrections}
\end{figure}

As is often the case, various combinations of emission lines may enter
into a narrowband or broadband filter depending on the redshifts of
the galaxies of interest.  It is common practice to use color excesses
between photometric filters that contain and surround such emission
lines in order to select strong line emitters.  In this context, it is
useful to assess the corrections required to recover the equivalent
widths of single emission lines, or the emission associated with a
single ionic species, from the equivalent widths inferred from such
color excesses, or from low-resolution spectral data.  The most common
corrections involve translating $W(\oiii+\hb)$ to $W(\oiii)$,
$W(\ha+\nii)$ to $W(\ha)$, and $W(\ha+\nii+\sii)$ to $W(\ha)$.  The
relations discussed in Section~\ref{sec:calibrations} and tabulated in
Table~\ref{tab:relations_sedparms} allow us to compute these
corrections as a function of various properties.  Here, we focus on
the corrections as a function of stellar mass computed for each
redshift subsample, and shown in Figure~\ref{fig:ewcorrections} and
tabulated in Table~\ref{tab:ewcorrections}.

\begin{deluxetable*}{lcrrr}
\tabletypesize{\footnotesize}
\tablewidth{0pc}
\tablecaption{Parameters for the Equivalent Width Corrections\tablenotemark{a}}
\tablehead{
\colhead{Correction} &
\colhead{Redshift Subsample} &
\colhead{$c_0$} &
\colhead{$c_1$} &
\colhead{$c_2$}}
\startdata
$W(\oiii+\hb) \rightarrow W(\oiii)$ & $z\sim 1.5$ & 0.665 & -0.0119 & $7.430\times 10^{-3}$ \\
                                    & $z\sim 2.3$ & 0.616 & -0.0079 & $6.616\times 10^{-3}$ \\
                                    & $z\sim 3.4$ & 0.581 & -0.0074 & $6.232\times 10^{-3}$ \\
\hline
$W(\ha+\nii) \rightarrow W(\ha)$ & $z\sim 1.5$ & 0.951 & -0.1451 & $1.758\times 10^{-2}$ \\
                                 & $z\sim 2.3$ & 2.063 & -0.4662 & $3.820\times 10^{-2}$ \\
\hline
$W(\ha+\nii+\sii) \rightarrow W(\ha)$ & $z\sim 1.5$ & 3.812 & -0.8955 & $6.677\times 10^{-2}$ \\
                                     & $z\sim 2.3$ & 2.376 & -0.5334 & $4.419\times 10^{-2}$ 
\enddata
\tablenotetext{a}{This table gives the correction factors ($c$) by which $W(\oiii+\hb)$ must be divided
by to compute $W(\oiii)$, and by which $W(\ha+\nii)$ and $W(\ha+\nii+\sii)$ must be divided by
to compute $W(\ha)$.  The correction factors are expressed as $c=c_0 + c_1x + c_2x^2$ where $x=\log[M_\ast/M_\odot]$,
and are valid over the range $9.0\le \log[M_\ast/M_\odot] \le 11.5$.}
\label{tab:ewcorrections}
\end{deluxetable*}

Focusing on the translation between $W(\oiii+\hb)$ and $W(\oiii)$, the
associated corrections decrease with decreasing stellar mass at a
fixed redshift, reflecting the fact that the O3 ratio decreases with
increasing stellar mass (e.g., \citealt{sanders16a}).  This conclusion is
supported by the results shown in Figure~\ref{fig:ewsedmassz}, where
the $\oiii$ line luminosity is effectively constant with mass above
$10^{9}$\,$M_\odot$, but the $\hb$ line luminosity decreases with
decreasing stellar mass over the same range in stellar mass.  We also
note that the correction factor at fixed mass decreases with redshift,
implying that the $W(\oiii)$ evolves more strongly with redshift than
$W(\hb)$, another result that is evident in
Figure~\ref{fig:ewsedmassz}, and one that can be explained in terms of
the redshift evolution of the SFR-$M_\ast$ and mass-metallicity
relations (Section~\ref{sec:residuals}).

The corrections required to translate $W(\ha+\nii)$ to $W(\ha)$ also
increase with stellar mass.  In this case, as the ratio of \nii-to-\ha
(i.e., the N2 index) is sensitive to oxygen abundance, the increasing
strength of \nii\, relative to \ha\, with stellar mass simply reflects
the higher metallicities of more massive galaxies (e.g., see also
\citealt{faisst18}).  While there is little discernible redshift
evolution in the aforementioned corrections between the $z\sim 1.5$
and $z\sim 2.3$ subsamples, the redshift evolution of the MZR implies
that the corrections to $W(\ha+\nii)$ should decrease with increasing
redshift at a fixed stellar mass.

The correction factors associated with $W(\ha+\nii+\sii)$ depend on
stellar mass in a way that implies that $\sii/\ha$ is not constant
with stellar mass for the lower redshift galaxies.  Comparing the
slopes of the corrections for $W(\ha+\nii)$ and $W(\ha+\nii+\sii)$
implies that that $\sii/\ha$ increases with stellar mass for the
$z\sim 1.5$ sample, but is essentially constant with stellar mass for
the $z\sim 2.3$ sample.  These results may suggest the varying
contribution of shock-heated gas or diffuse ionized gas to the
measured line ratios for galaxies of different stellar masses, a topic
that is beyond the scope of this paper and will be discussed
elsewhere.


\LongTables

\begin{deluxetable*}{llllrrcrc}
\tabletypesize{\footnotesize}
\tablewidth{0pc}
\tablecaption{Dependence of Equivalent Widths on Stellar Populations}
\tablehead{
\colhead{Attribute\tablenotemark{a}} &
\colhead{Line\tablenotemark{a}} &
\colhead{$z$-Range ($\langle z\rangle$)\tablenotemark{b}} &
\colhead{$N$ (det/undet)\tablenotemark{c}} &
\colhead{$\rho$\tablenotemark{d}} &
\colhead{$\sigma_{\rm P}$\tablenotemark{d}} &
\colhead{Intercept\tablenotemark{e}} &
\colhead{Slope\tablenotemark{e}} &
\colhead{RMS\tablenotemark{e}}}
\startdata
$\log\left[\frac{M_{\ast}}{M_\odot}\right]$ & \oii & All: $1.6013-3.7152$ (2.4970) & 542 (407/135) & -0.50 & 10.1 & $3.814\pm 0.070$ & $-0.202 \pm 0.007$ & 0.17 \\
             &      & Low-$z$: $1.6013-1.7112$ (1.6545) & 54 (25/29) & -0.53 & 2.6 & $5.440\pm 0.279$ & $-0.376\pm 0.029$ & 0.21 \\
             &      & Mid-$z$: $1.8456-2.6196$ (2.2792) & 344 (274/70) & -0.53 & 8.8 & $3.533\pm 0.080$ & $-0.171\pm0.008$ & 0.17 \\
             &      & High-$z$: $2.9286-3.7152$ (3.2446) & 144 (108/36) & -0.39 & 4.0 & $4.393\pm 0.240$ & $-0.261\pm 0.025$ & 0.17 \\
\cline{2-9}
             & \hb & All: $1.3685-3.7152$ (2.3641) & 632 (515/117) & -0.68 & 15.4 & $5.337\pm 0.138$ & $-0.407 \pm 0.014$ & 0.21 \\
             &     & Low-$z$: $1.3685-1.7338$ (1.5326) & 160 (121/39) & -0.66 & 7.2 & $5.290\pm 0.160$ & $-0.406\pm 0.016$ & 0.18 \\
             &      & Mid-$z$: $2.0461-2.6541$ (2.3036) & 308 (267/41) & -0.63 & 10.2 & $4.145\pm0.164$ & $-0.283\pm 0.016$ & 0.22 \\
             &      & High-$z$: $2.9750-3.7152$ (3.2838) & 164 (127/37) & -0.73 & 8.2 & $6.351\pm 0.334$ & $-0.492\pm 0.034$ & 0.17 \\
\cline{2-9}
             & \oiii & All: $1.2467-3.7152$ (2.3540) & 925 (830/95) & -0.67 & 19.1 & $9.007\pm 0.046$ &  $-0.707\pm0.005$ & 0.32 \\
             &        & Low-$z$: $1.2467-1.7292$ (1.5293) & 232 (189/43) & -0.70 & 9.6 & $8.639\pm 0.183$ & $-0.686\pm 0.018$ & 0.30 \\
             &      & Mid-$z$: $1.9233-2.5855$ (2.2659) & 454 (424/30) & -0.66 & 13.5 & $7.210\pm 0.057$ & $-0.520\pm 0.006$ & 0.29 \\
             &      & High-$z$: $2.9262-3.7152$ (3.2444) & 239 (217/22) & -0.66 & 9.6 & $9.143\pm 0.168$ & $-0.695\pm 0.017$ & 0.28 \\
\cline{2-9}
             & \oiii+ & All: $1.3685-3.7152$ (2.3956) & 555 (448/107) & -0.70 & 14.8 & $8.632\pm 0.051$ &  $-0.661\pm0.005$ & 0.28 \\
             & \hb       & Low-$z$: $1.3685-1.7292$ (1.5305) & 141 (103/38) & -0.73 & 7.3 & $8.289\pm 0.152$ & $-0.640\pm 0.015$ & 0.24 \\
             &      & Mid-$z$: $2.0461-2.5855$ (2.2869) & 252 (218/34) & -0.67 & 9.8 & $6.839 \pm 0.057$ & $-0.475\pm 0.006$ & 0.25 \\
             &      & High-$z$: $2.9750-3.7152$ (3.2838) & 162 (127/35) & -0.72 & 8.1 & $8.746\pm 0.169$ & $-0.650\pm 0.017$ & 0.23 \\
\cline{2-9}
             & \ha & All: $1.0048-2.6541$ (2.0122) & 771 (704/67) & -0.50 & 13.2 & $5.991\pm 0.025$ & $-0.398\pm 0.002$ & 0.26 \\
             &        & Low-$z$: $1.0048-1.7338$ (1.5067) & 264 (247/17) & -0.54 & 8.4 & $5.725\pm 0.036$ & $-0.378\pm 0.004$ & 0.25 \\
             &      & Mid-$z$: $1.9233-2.6541$ (2.2854) & 507 (457/50) & -0.47 & 9.9 & $5.005\pm 0.033$ & $-0.286\pm 0.003$ & 0.23 \\
\cline{2-9}
             & \ha+ & All: $1.0048-2.5805$ (2.0188) & 683 (384/299) & -0.50 & 9.7 & $5.233\pm 0.032$ & $-0.314\pm 0.003$ & 0.23 \\
             & \nii & Low-$z$: $1.0048-1.7338$ (1.5104) & 205 (133/72) & -0.41 & 4.7 & $5.115\pm 0.048$ & $-0.307\pm 0.005$ & 0.20 \\
             &      & Mid-$z$: $1.9727-2.5805$ (2.2883) & 478 (251/227) & -0.54 & 8.6 & $4.042\pm 0.043$ & $-0.181\pm 0.004$ & 0.20 \\
\hline
$\beta$ & \oii & All: $1.6013-3.7152$ (2.4970) & 542 (407/135) & -0.20 & 4.0 & $1.666\pm 0.009$ & $-0.065\pm 0.006$ & 0.20 \\
        &     & Low-$z$: $1.6013-1.7112$ (1.6545) & 54 (25/29) & -0.39 & 3.8 & $1.532\pm 0.047$ & $-0.105\pm 0.028$ & 0.21 \\
        &       & Mid-$z$: $1.8456-2.6196$ (2.2792) & 344 (274/70) & -0.21 & 3.5 & $1.705\pm 0.012$ & $-0.049\pm 0.008$ & 0.20 \\
        &       & High-$z$: $2.9286-3.7152$ (3.2446) & 144 (108/36) & -0.17 & 1.8 & $1.581\pm 0.026$ & $-0.154\pm 0.018$ & 0.20 \\
\cline{2-9}
        & \hb & All: $1.3685-3.7152$ (2.3641) & 632 (515/117) & -0.32 & 7.2 & $0.889\pm 0.017$ & $-0.223\pm 0.012$ & 0.33 \\
        &       & Low-$z$: $1.3685-1.7338$ (1.5326) & 160 (121/39) & -0.35 & 3.8 & $0.929\pm 0.019$ & $-0.134\pm 0.013$ & 0.27 \\
        &       & Mid-$z$: $2.0461-2.6541$ (2.3036) & 308 (267/41) & -0.29 & 4.8 & $0.865\pm 0.022$ & $-0.254\pm 0.013$ & 0.29 \\
        &       & High-$z$: $2.9750-3.7152$ (3.2838) & 164 (127/37) & -0.53 & 5.9 & $1.037\pm 0.055$ & $-0.321\pm 0.038$ & 0.22 \\
\cline{2-9}
        & \oiii & All: $1.2467-3.7152$ (2.3540) & 925 (830/95) & -0.29 & 8.4 & $1.413\pm 0.006$ & $-0.282\pm 0.003$ & 0.46 \\
        &       & Low-$z$: $1.2467-1.7292$ (1.5293) & 232 (189/43) & -0.44 & 6.1 & $1.279\pm 0.012$ & $-0.231\pm 0.007$ & 0.42 \\
        &       & Mid-$z$: $1.9233-2.5855$ (2.2659) & 454 (424/30) & -0.32 & 6.6 & $1.275\pm 0.007$ & $-0.451\pm 0.004$ & 0.40 \\
        &       & High-$z$: $2.9262-3.7152$ (3.2444) & 239 (217/22) & -0.51 & 7.5 & $1.586\pm 0.017$ & $-0.459\pm 0.010$ & 0.35 \\
\cline{2-9}
        & \oiii+ & All: $1.3685-3.7152$ (2.3956) & 555 (448/107) & -0.27 & 5.8 & $1.529\pm 0.006$ & $-0.267\pm 0.004$ & 0.46 \\
        & \hb          & Low-$z$: $1.3685-1.7292$ (1.5305) & 141 (103/38) & -0.31 & 3.1 & $1.418\pm 0.011$ & $-0.219\pm 0.006$ & 0.39 \\
        &           & Mid-$z$: $2.0461-2.5855$ (2.2869) & 252 (218/34) & -0.29 & 4.3 & $1.420\pm 0.007$ & $-0.414\pm 0.004$ & 0.37 \\
        &           & High-$z$: $2.9750-3.7152$ (3.2838) & 162 (127/35) & -0.56 & 6.3 & $1.690\pm 0.017$ & $-0.437\pm 0.010$ & 0.31 \\
\cline{2-9}
        & \ha & All: $1.0048-2.6541$ (2.0122) & 771 (704/67) & -0.24 & 6.4 & $1.746\pm 0.004$ & $-0.104\pm 0.003$ & 0.34 \\
        &     & Low-$z$: $1.0048-1.7338$ (1.5067) & 264 (247/17) & -0.31 & 4.8 & $1.638\pm 0.005$ & $-0.125\pm 0.003$ & 0.33 \\
        &     & Mid-$z$: $1.9233-2.6541$ (2.2854) & 507 (457/50) & 0.22 & 4.8 & $1.879\pm 0.005$ & $-0.131\pm 0.004$ & 0.27 \\
\cline{2-9}
        & \ha+ & All: $1.0048-2.5805$ (2.0188) & 683 (384/299) & -0.21 & 4.0 & $1.893\pm 0.004$ & $-0.077\pm 0.003$ & 0.30 \\
        & \nii         & Low-$z$: $1.0048-1.7338$ (1.5104) & 205 (133/72) & -0.25 & 2.9 & $1.806\pm 0.005$ & $-0.087\pm 0.003$ & 0.24 \\
        &          & Mid-$z$: $1.9727-2.5805$ (2.2883) & 478 (251/227) & -0.20 & 3.1 & $2.047\pm 0.006$ & $-0.080\pm 0.004$ & 0.24 \\ 
\hline
$\log\left[\frac{\rm Age}{\rm Myr}\right]$ & \oii & All: $1.6013-3.7152$ (2.4970) & 542 (407/135) & -0.32 & 6.5 & $2.418\pm 0.026$ & $-0.220 \pm 0.009$ & 0.19 \\
             &      & Low-$z$: $1.6013-1.7112$ (1.6545) & 54 (25/29) & -0.22 & 1.1 & $3.175\pm 0.137$ & $-0.481\pm 0.046$ & 0.27 \\
             &      & Mid-$z$: $1.8456-2.6196$ (2.2792) & 344 (274/70) & -0.40 & 6.6 & $2.533\pm 0.033$ & $-0.246\pm 0.011$ & 0.19 \\
             &      & High-$z$: $2.9286-3.7152$ (3.2446) & 144 (108/36) & -0.22 & 2.2 & $2.687\pm 0.077$ & $-0.340\pm 0.030$ & 0.20 \\
\cline{2-9}
             & \hb & All: $1.3685-3.7152$ (2.3641) & 632 (515/117) & -0.64 & 14.4 & $3.253\pm 0.122$ & $-0.657 \pm 0.036$ & 0.22 \\
             &     & Low-$z$: $1.3685-1.7338$ (1.5326) & 160 (121/39) & -0.62 & 6.8 & $3.099\pm 0.086$ & $-0.613\pm 0.028$ & 0.21 \\
             &      & Mid-$z$: $2.0461-2.6541$ (2.3036) & 308 (267/41) & -0.62 & 10.1 & $2.692\pm 0.051$ & $-0.461\pm 0.018$ & 0.23 \\
             &      & High-$z$: $2.9750-3.7152$ (3.2838) & 164 (127/37) & -0.62 & 7.0 & $2.960\pm 0.102$ & $-0.562\pm 0.040$ & 0.21 \\
\cline{2-9}
             & \oiii & All: $1.2467-3.7152$ (2.3540) & 925 (830/95) & -0.68 & 19.7 & $4.673\pm 0.010$ &  $-0.919\pm0.004$ & 0.33 \\
             &        & Low-$z$: $1.2467-1.7292$ (1.5293) & 232 (189/43) & -0.70 & 9.5 & $5.427\pm 0.089$ & $-1.138\pm 0.028$ & 0.30 \\
             &      & Mid-$z$: $1.9233-2.5855$ (2.2659) & 454 (424/30) & -0.64 & 13.1 & $3.901\pm 0.086$ & $-0.677\pm 0.026$ & 0.32 \\
             &      & High-$z$: $2.9262-3.7152$ (3.2444) & 239 (217/22) & -0.55 & 8.1 & $4.240\pm 0.105$ & $-0.762\pm 0.035$ & 0.33 \\
\cline{2-9}
             & \oiii+ & All: $1.3685-3.7152$ (2.3956) & 555 (448/107) & -0.72 & 15.2 & $4.605\pm 0.022$ &  $-0.866\pm0.007$ & 0.29 \\
             & \hb       & Low-$z$: $1.3685-1.7292$ (1.5305) & 141 (103/38) & -0.72 & 7.3 & $5.085\pm 0.075$ & $-1.016\pm 0.023$ & 0.25 \\
             &      & Mid-$z$: $2.0461-2.5855$ (2.2869) & 252 (218/34) & -0.67 & 9.9 & $4.261\pm 0.083$ & $-0.753\pm 0.025$ & 0.29 \\
             &      & High-$z$: $2.9750-3.7152$ (3.2838) & 162 (127/35) & -0.64 & 7.1 & $4.184\pm 0.094$ & $-0.714\pm 0.031$ & 0.30 \\
\cline{2-9}
             & \ha & All: $1.0048-2.6541$ (2.0122) & 771 (704/67) & -0.60 & 15.8 & $3.860\pm 0.009$ & $-0.600\pm 0.003$ & 0.24 \\
             &        & Low-$z$: $1.0048-1.7338$ (1.5067) & 264 (247/17) & -0.60 & 9.4 & $4.067\pm 0.017$ & $-0.666\pm 0.005$ & 0.24 \\
             &      & Mid-$z$: $1.9233-2.6541$ (2.2854) & 507 (457/50) & -0.54 & 11.6 & $3.555\pm 0.014$ & $-0.485\pm 0.005$ & 0.22 \\
\cline{2-9}
             & \ha+ & All: $1.0048-2.5805$ (2.0188) & 683 (384/299) & -0.55 & 10.7 & $3.708\pm 0.011$ & $-0.517\pm 0.004$ & 0.23 \\
             & \nii       & Low-$z$: $1.0048-1.7338$ (1.5104) & 205 (133/72) & -0.49 & 5.6 & $3.843\pm 0.021$ & $-0.563\pm 0.006$ & 0.22 \\
             &      & Mid-$z$: $1.9727-2.5805$ (2.2883) & 478 (251/227) & -0.53 & 8.3 & $3.343\pm 0.018$ & $-0.381\pm 0.006$ & 0.21 \\
\hline
$\log\left[\frac{\rm SFR[H\alpha]}{M_\odot \,{\rm yr}^{-1}}\right]$  & \oii & All: $1.6013-2.6196$ (2.2431) & 231 (203/28) & -0.09 & 1.2 & $1.730\pm 0.014$ & $0.073 \pm 0.010$ & 0.21 \\
 &      & Low-$z$: $1.6013-1.7112$ (1.6605) & 25 (15/10) & -0.27 & 1.0 & $1.786\pm0.052$ & $0.075\pm0.044$ & 0.24 \\
             &      & Mid-$z$: $2.0824-2.6196$ (2.2896) & 206 (188/18) & -0.07 & 0.9 & $1.727\pm 0.015$ & $0.076\pm 0.010$ & 0.21 \\
\cline{2-9}
             & \hb & All: $1.3685-2.6541$ (2.0584) & 354 (354/0) & -0.12 & 2.2 & $1.229\pm 0.021$ & $0.027 \pm 0.018$ & 0.28 \\
             &     & Low-$z$: $1.3685-1.7338$ (1.5267) & 112 (112/0) & -0.26 & 2.8 & $1.302\pm 0.030$ & $-0.092\pm 0.022$ & 0.23 \\
             &      & Mid-$z$: $2.0461-2.6541$ (2.3023) & 243 (243/0) & -0.07 & 1.0 & $1.214\pm 0.020$ & $0.087\pm 0.013$ & 0.29 \\
\cline{2-9}
             & \oiii & All: $1.2467-2.5855$ (2.0595) & 388 (373/15) & -0.07 & 1.4 & $1.873\pm 0.005$ &  $0.046\pm0.004$ & 0.42 \\
             &        & Low-$z$: $1.2467-1.7292$ (1.5323) & 123 (111/12) & -0.27 & 2.9 & $1.884\pm 0.012$ & $-0.121\pm 0.010$ & 0.40 \\
             &      & Mid-$z$: $2.0151-2.5855$ (2.2828) & 265 (262/3) & -0.05 & 0.8 & $1.986\pm 0.008$ & $0.045\pm 0.005$ & 0.39 \\
\cline{2-9}
             & \oiii+ & All: $1.3685-2.5855$ (2.0411) & 301 (290/11) & 0.28 & 5.9 & $1.929\pm 0.006$ &  $0.066\pm0.005$ & 0.39 \\
             & \hb       & Low-$z$: $1.3685-1.7292$ (1.5232) & 102 (93/9) & -0.30 & 2.9 & $1.958\pm 0.013$ & $-0.102\pm 0.009$ & 0.35 \\
             &      & Mid-$z$: $2.0461-2.5855$ (2.2856) & 199 (197/2) & -0.01 & 0.1 & $2.036\pm 0.008$ & $0.060\pm 0.005$ & 0.36 \\
\cline{2-9}
             & \ha & All: $1.2467-2.6541$ (2.0897) & 434 (434/0) & 0.28 & 5.9 & $1.876\pm 0.004$ & $0.134\pm 0.003$ & 0.25 \\
             &        & Low-$z$: $1.2467-1.7338$ (1.5361) & 121 (121/0) & 0.22 & 2.4 &  $1.954\pm 0.006$ & $0.015\pm 0.005$ & 0.24 \\
             &      & Mid-$z$: $2.0151-2.6541$ (2.3020) & 314 (314/0) & 0.28 & 5.0 & $1.866\pm 0.007$ & $0.216\pm 0.004$ & 0.25 \\
\cline{2-9}
             & \ha+ & All: $1.3896-2.5605$ (2.0728) & 398 (267/131) & 0.29 & 4.8 & $1.925\pm 0.005$ & $0.181\pm 0.004$ & 0.23 \\
             & \nii       & Low-$z$: $1.3896-1.7338$ (1.5366) & 104 (79/25) & 0.21 & 1.9 & $1.995\pm 0.008$ & $0.082\pm 0.006$ & 0.18 \\
             &      & Mid-$z$: $2.0151-2.5605$ (2.2981) & 294 (188/106) & 0.28 & 3.9 & $1.893\pm 0.008$ & $0.268\pm 0.005$ & 0.22 \\
\hline
$\log\left[\frac{\rm sSFR[H\alpha]}{{\rm yr}^{-1}}\right]$ & \oii & All: $1.6013-2.6196$ (2.2431) & 231 (203/28) & 0.51 & 7.2 & $3.876\pm 0.080$ & $0.234\pm 0.009$ & 0.17 \\
 &      & Low-$z$: $1.6013-1.7112$ (1.6605) & 25 (15/10) & 0.27 & 1.0 & $4.688\pm 0.367$ & $0.325\pm 0.042$ & 0.20 \\
             &      & Mid-$z$: $2.0824-2.6196$ (2.2896) & 206 (188/18) & 0.54 & 7.3 & $3.751\pm 0.084$ & $0.220\pm 0.010$ & 0.17 \\
\cline{2-9}
             & \hb & All: $1.3685-2.6541$ (2.0584) & 354 (354/0) & 0.56 & 10.5 & $4.129\pm 0.099$ & $0.323 \pm 0.012$ & 0.23 \\
             &     & Low-$z$: $1.3685-1.7338$ (1.5267) & 112 (112/0) & 0.50 & 5.2 & $3.605\pm 0.255$ & $0.266 \pm 0.028$ & 0.24 \\
             &      & Mid-$z$: $2.0461-2.6541$ (2.3023) & 243 (243/0) & 0.63 & 9.8 & $4.292\pm 0.196$ & $0.334 \pm 0.022$ & 0.22 \\
\cline{2-9}
             & \oiii & All: $1.2467-2.5855$ (2.0595) & 388 (373/15) & 0.67 & 12.9 & $6.654\pm 0.032$ &  $0.537\pm0.004$ & 0.32 \\
             &        & Low-$z$: $1.2467-1.7292$ (1.5323) & 123 (111/12) & 0.54 & 5.6 & $5.193\pm 0.091$ & $0.386\pm 0.010$ & 0.38 \\
             &      & Mid-$z$: $2.0151-2.5855$ (2.2828) & 265 (262/3) & 0.69 & 11.1 & $6.112\pm 0.036$ & $0.467\pm 0.004$ & 0.28 \\
\cline{2-9}
             & \oiii+ & All: $1.3685-2.5855$ (2.0411) & 301 (290/11) & 0.69 & 11.7 & $6.393\pm 0.034$ &  $0.497\pm0.004$ & 0.30 \\
             & \hb       & Low-$z$: $1.3685-1.7292$ (1.5232) & 102 (93/9) & 0.50 & 4.8 & $5.053\pm 0.090$ & $0.357\pm 0.010$ & 0.34 \\
             &      & Mid-$z$: $2.0461-2.5855$ (2.2856) & 199 (197/2) & 0.73 & 10.2 & $5.829\pm 0.044$ & $0.423\pm 0.005$ & 0.27 \\
\cline{2-9}
             & \ha & All: $1.2467-2.6541$ (2.0897) & 434 (434/0) & 0.85 & 17.8 & $5.808\pm 0.022$ & $0.419\pm 0.003$ & 0.13 \\
             &        & Low-$z$: $1.2467-1.7338$ (1.5361) & 121 (121/0) & 0.86 & 9.3 & $5.565\pm 0.037$ & $0.394\pm 0.004$ & 0.13 \\
             &      & Mid-$z$: $2.0151-2.6541$ (2.3020) & 314 (314/0) & 0.87 & 15.3 & $5.440\pm 0.034$ & $0.374\pm 0.004$ & 0.14 \\
\cline{2-9}
             & \ha+ & All: $1.3896-2.5605$ (2.0728) & 398 (267/131) & 0.82 & 13.4 & $5.408\pm 0.027$ & $0.361\pm 0.003$ & 0.12 \\
             & \nii       & Low-$z$: $1.3896-1.7338$ (1.5366) & 104 (79/25) & 0.81 & 7.1 & $5.307\pm 0.044$ & $0.353\pm 0.005$ & 0.11 \\
             &      & Mid-$z$: $2.0151-2.5605$ (2.2981) & 294 (188/106) & 0.82 & 11.2 & $5.085\pm 0.040$ & $0.320\pm 0.005$ & 0.13
\enddata
\tablenotetext{a}{Statistics are presented for the relationship between $\log[W/{\rm \AA}]$ for the line (or lines)
listed under column heading ``Line'' and the property listed under column heading ``Attribute.''}
\tablenotetext{b}{Redshift range and mean redshift of objects in this subsample.}
\tablenotetext{c}{Total number of objects and the number of detections and non-detections of the line (or lines)
listed under column heading ``Line.''}
\tablenotetext{d}{Spearman rank correlation coefficient and the number of standard deviations by which the correlation
deviates from the null hypothesis of no correlation.}
\tablenotetext{e}{Intercept and slope of the best-fit linear function to the composite averages and the rms of the data points about this best-fit linear function.}
\label{tab:relations_sedparms}
\end{deluxetable*}

\end{document}